\documentclass[10pt,journal,compsoc, cspaper, twoside]{IEEEtran}

\ifCLASSOPTIONcompsoc
  \usepackage[nocompress]{cite}
\else
  \usepackage{cite}
\fi

\ifCLASSINFOpdf
  \usepackage{graphicx}
  \graphicspath{ {Figures/} }
\else
\fi

\usepackage{amsmath}
\usepackage{amssymb}
\usepackage{graphicx}
\usepackage{algorithm}
\usepackage{algpseudocode}
\usepackage{multirow}
\usepackage{balance}

\usepackage{array}
\usepackage{lipsum}

\ifCLASSOPTIONcompsoc
  \usepackage[caption=false,font=footnotesize,labelfont=sf,textfont=sf]{subfig}
\else
  \usepackage[caption=false,font=footnotesize]{subfig}
\fi

\usepackage{url}

\newcommand\MYhyperrefoptions{bookmarks=true,bookmarksnumbered=true,
pdfpagemode={UseOutlines},plainpages=false,pdfpagelabels=true,
colorlinks=true,linkcolor={black},citecolor={black},urlcolor={black},
pdftitle={An Approach for Automatic Generation of System Test Cases from Use Case Specifications},%
pdfsubject={Automatic Test Case Generation},%
pdfauthor={Chunhui Wang},%
pdfkeywords={System Test Case Generation; Use Case Specifications; Natural Language Processing; Semantic Labeling}}%
\ifCLASSINFOpdf
\usepackage[\MYhyperrefoptions,pdftex]{hyperref}
\else
\usepackage[\MYhyperrefoptions,breaklinks=true,dvips]{hyperref}
\usepackage{breakurl}
\fi

\usepackage{array,booktabs} %
\usepackage{tabularx}
\usepackage{float}
\usepackage{comment}
\usepackage{balance}
\usepackage{wrapfig}

\makeatletter
\newcommand{\manuallabel}[2]{\def\@currentlabel{#2}\label{#1}}
\makeatother

\usepackage{tikz}
\usetikzlibrary{calc}
\usetikzlibrary{trees}
\usetikzlibrary{arrows}
\usetikzlibrary{graphs}
\usetikzlibrary{positioning}
\usetikzlibrary{shapes.geometric}
\usetikzlibrary{shapes.misc}
\tikzstyle{block}=[draw opacity=0.7,line width=1.0cm]

\tikzset{
modal/.style={>=stealth',shorten >=1pt,shorten <=1pt,auto,node distance=1.5cm, semithick},
world/.style={circle,draw,minimum size=0.5cm,fill=gray!15},
ghost/.style={rectangle,minimum size=0.5cm,fill=white, fill opacity=0},
atom/.style={rectangle,minimum size=0.5cm,fill=green!50!black!20,draw=green!50!black},
structure/.style={=>stealth’,shorten >=1pt,shorten <=1pt,auto,node distance=25mm, thick, minimum size=5mm},
set/.style={rectangle, minimum size=0.5cm,fill=green!50!black!20,draw=green!50!black, font=\sffamily\scriptsize},
lattice/.style={=>stealth’,shorten >=1pt,shorten <=1pt, auto, node distance=1cm, semithick, minimum size=3mm},
every node/.style={font=\sffamily\footnotesize},
point/.style={circle,draw,inner sep=0.5mm,fill=black},
scriptsize/.style={font=\sffamily\scriptsize},
tiny/.style={font=\sffamily\tiny},
highlight/.style={fill=red!50!black!20,draw=red!50!black},
highlightgreen/.style={fill=green!50!black!20,draw=green!50!black},
comment/.style={rectangle, fill=white,text opacity=1,fill opacity=0, draw opacity=0, minimum size=0.5cm,font=\itshape\sffamily\scriptsize},
reflexive above/.style={->,loop,looseness=7,in=120,out=60},
reflexive below/.style={->,loop,looseness=7,in=240,out=300},
reflexive left/.style={->,loop,looseness=7,in=150,out=210},
reflexive right/.style={->,loop,looseness=7,in=30,out=330}
}



\usepackage{color}
\usepackage{xspace}

\usepackage{pdfcomment}
\usepackage{textpos}

\algrenewcommand\algorithmicindent{0.5em}%

\newcommand{\EquationsSize}{\footnotesize}

\newcommand{\SNA}{S1}
\newcommand{\SA}{S2}
\newcommand{\SB}{S3}
\newcommand{\SC}{S4}
\newcommand{\SE}{S5}
\newcommand{\SF}{S6}
\newcommand{\SG}{S7}
\newcommand{\SH}{S8}
\newcommand{\SL}{S9}
\newcommand{\SM}{S10}
\newcommand{\SN}{S11}
\newcommand{\SO}{S12}
\newcommand{\SP}{S13}
\newcommand{\SQ}{S14}

\newcommand{\BigO}[0]{\mathcal{O}}

\newcounter{cLINE}
\newcommand{\LLRESET}{\setcounter{cLINE}{0}}

\newcommand{\LL}{\addtocounter{cLINE}{1}\arabic{cLINE}}
\newcommand{\LINE}{\arabic{cLINE}}

\newcommand{\PTWO}{A1.1\xspace}
\newcommand{\PTHREE}{A1.2\xspace}

\newcommand{\PSIX}{A1.3\xspace}

\newcommand{\PONE}{A1.4\xspace}

\newcommand{\PFOUR}{A1.5\xspace}
\newcommand{\PFIVE}{A1.6\xspace}

\newcommand{\inst}{\emph{node}\xspace}

\newcommand{\MM}{Use Case Modeling for System-level, Acceptance Tests Generation}

\newcommand{\UMTG}{UMTG\xspace}
\newcommand{\M}{\UMTG}
\newcommand{\UCTM}{UCTM\xspace}
\newcommand{\BodySense}{\emph{BodySense}\xspace}

\newcommand{\CNP}{CNP}

\newcommand{\CHANGED}[1]{\textcolor{black}{#1}}%

\usepackage{marginnote}
\usepackage[backgroundcolor=white,bordercolor=blue,linecolor=blue]{todonotes}

\newcommand{\MMREVISION}[2]{#2}

\newcommand{\MINREVT}[2]{\textcolor{black}{#2}}

\newcommand{\MREVISION}[2]{#2}

\newcommand{\MINREV}[2]{\textcolor{black}{#2}}

\newcommand{\HOD}{\emph{HOD}\xspace}

\newcolumntype{P}[1]{>{\centering\arraybackslash}p{#1}}

\begin{document}
\title{Automatic Generation of Acceptance Test Cases from Use Case Specifications: an NLP-based Approach}

\newif\ifArxiv
\Arxivtrue

\ifArxiv
\author{Chunhui~Wang, %
        Fabrizio~Pastore,
        Arda~Goknil,
        and~Lionel~C.~Briand%

\IEEEcompsocitemizethanks{
\IEEEcompsocthanksitem C. Wang, F. Pastore, A. Goknil and L.C. Briand are with the SnT Centre for Security, Reliability and Trust, University of Luxembourg, Luxembourg. L.C. Briand is also affiliated with the school of EECS, University of Ottawa.\protect \\
 E-mail: wangchunhui@me.com fabrizio.pastore@uni.lu ar.goknil@gmail.com lionel.briand@uni.lu lbriand@uottawa.ca
}%

\thanks{Manuscript submitted to TSE}}
\else
\author{Chunhui~Wang, %
        Fabrizio~Pastore,~\IEEEmembership{Member,~IEEE,}
        Arda~Goknil,
        and~Lionel~C.~Briand,~\IEEEmembership{Fellow,~IEEE}%

\IEEEcompsocitemizethanks{
\IEEEcompsocthanksitem C. Wang, F. Pastore, A. Goknil and L.C. Briand are with the SnT Centre for Security, Reliability and Trust, University of Luxembourg, Luxembourg. L.C. Briand is also affiliated with University of Ottawa.\protect \\
 E-mail: wangchunhui@me.com fabrizio.pastore@uni.lu ar.goknil@gmail.com lionel.briand@uni.lu lbriand@uottawa.ca
}%

\thanks{Manuscript received July 15th, 2019; revised March 30th, 2020.}}
\fi

\ifArxiv
\else
\markboth{IEEE Transactions on Software Engineering,~Vol.~00, No.~0, MONTH~YEAR}%
{Wang \MakeLowercase{\textit{et al.}}: Automatic Generation of System Test Cases from Use Case Specifications: an NLP-based Approach} 
\fi

\newcommand{\Lionel}[1]{{ \color{green} Lionel: {#1}} }
\newcommand{\Fabrizio}[1]{{ \color{orange} Fabrizio: {#1}} }

\IEEEtitleabstractindextext{%
\begin{abstract}

\MMREVISION{R1.60}{Acceptance testing is a validation activity performed to ensure the conformance of software systems with respect to their functional requirements.
In safety critical systems, it plays a crucial role since it is enforced by software standards, which mandate that each requirement be validated by such testing in a clearly traceable manner.}
Test engineers need to identify all the representative test execution scenarios from requirements, determine the runtime conditions that trigger these scenarios, and finally provide the input data that satisfy these conditions. %
Given that requirements specifications are typically large and often provided in natural language (e.g., use case specifications), the generation of acceptance test cases tends to be expensive and error-prone.

In this paper, we present \MM{} (\UMTG), an approach that supports the generation of \CHANGED{executable, system-level, acceptance test cases} from requirements specifications in natural language, with the goal of reducing the manual effort required to generate test cases and ensuring requirements coverage. 
More specifically, \UMTG automates the generation of \CHANGED{acceptance test cases} based on use case specifications and a domain model for the system under test, which are commonly produced in many development environments. %
Unlike existing approaches, %
it does not impose strong restrictions on the expressiveness of use case specifications. %
We rely on recent advances in natural language processing to automatically identify test scenarios and to generate formal constraints that capture conditions triggering the execution of the scenarios, thus enabling the generation of test data. 
In two industrial case studies, \UMTG automatically and correctly translated 95\% of the use case specification steps into formal constraints required for test data generation; furthermore, it generated test cases that exercise not only all the test scenarios manually implemented by experts, but also some critical scenarios not previously considered.

\end{abstract}

}

\maketitle

\IEEEdisplaynontitleabstractindextext

\IEEEpeerreviewmaketitle

\ifCLASSOPTIONcompsoc
\IEEEraisesectionheading{\section{Introduction}\label{sec:introduction}}
\else
\section{Introduction}
\label{sec:introduction}
\fi

\IEEEPARstart{T}{he complexity} of embedded software in safety critical domains, e.g., automotive and avionics, has significantly increased over the years. 
\MREVISION{R1.60}{In such contexts, software testing plays a fundamental role to ensure that the system, as a whole, meets its functional requirements. 
Such system testing activity is commonly referred to as \emph{acceptance testing}~\cite{ieee24765}.
As opposed to verification, which aims at detecting faults, acceptance testing is a validation activity performed at the very end of the  development lifecycle to demonstrate compliance with requirements. Acceptance testing is enforced and regulated by international standards. 
For example, traceability between requirements and acceptance test cases is enforced by standards for safety-critical embedded systems~\cite{DO178C,ISO26262}.
}

\MREVISION{R3.1}{Functional requirements in safety critical domains  are often expressed in natural language (NL), and acceptance test cases are either manually derived from requirements specifications or -- a much less frequent practice -- generated from models produced for testing purposes only~\cite{ANDA2009,Nebut-AutomaticTestGenerationUseCaseDriven-TSE-2006,Arora:16,Provenzano2017}.}
Automatic test generation from requirements specifications in NL achieves the best of both worlds as it does not require additional test modeling, reduces the cost of testing and also guarantees that testing is systematic and properly covers all requirements.

The benefits of automatic test generation are widely acknowledged today and there are many proposed approaches in the literature~\cite{Escalona-JSS-2011}. 
\MREVISION{R3.2}{However, existing approaches are often not applicable in industrial context ~\cite{Shirole13} since they typically 
require that system specifications be captured as UML behavioral models such as activity diagrams~\cite{Linzhang2004}, statecharts~\cite{Ryeser-Scenario-ICSSEA-1999}, and sequence diagrams~\cite{Nebut-AutomaticTestGenerationUseCaseDriven-TSE-2006}.} 
In modern industrial systems, these behavioral models tend to be complex and expensive if they are to be precise and complete enough to support test automation, and are thus often not part of development practice. There are techniques~\cite{Yue-RUCM-TSE-2013}~\cite{Yue-RUCMtoState-ECMFA-2011}~\cite{Yue-UseCasesToActivity-MFA-2010} that generate test models from NL requirements, but the generated models need to be manually edited to enable test automation, thus creating scalability issues. 
In approaches generating test cases directly from NL requirements~\cite{Zhang-SAM-2014}~\cite{Sarmiento2016}~\cite{Sarmiento14}~\cite{Gutierrez08}, test cases are not executable and often require significant manual intervention to %
provide test input data %
(e.g., they need additional formal specifications~\cite{Gutierrez08}).
A few approaches can generate executable test cases including test input data directly from NL requirements specifications~\cite{deFigueiredo-2006}~\cite{Carvalho-NAT2TEST-SCP-2014}, but they require that requirements specifications be written according to a controlled natural language (CNL). %
The CNL specifications are translated into formal specifications which are then used to automatically generate test input data (e.g., using constraint solving). The CNL language supported by these approaches is typically very focused and therefore limited (e.g., it enables the use of only a few verbs in requirements specifications), thus reducing their usability.

Our goal in this paper is to enable automated generation of \CHANGED{executable, system-level, acceptance test cases} from NL requirements, with no additional behavioral modeling. Our motivation is to rely, to the largest extent possible, on practices that are already in place in many companies developing embedded systems, including our industry partner, i.e., IEE S.A. (in the following ``IEE'')~\cite{IEE}, with whom we performed the case studies reported in this paper. In many environments like IEE, development processes are use case-driven and this strongly influences their requirements engineering and \CHANGED{acceptance testing} practices. Use case specifications are widely used for communicating requirements among stakeholders and, in particular, facilitating communication with customers. A domain model typically complements use cases by specifying the terminology and concepts shared among all stakeholders and thus helps avoid misunderstandings. 

In this paper, we propose, apply and assess \textit{\MM{} (\M{})}, an approach that generates executable \CHANGED{system test cases for acceptance testing} by exploiting behavioral information in use case specifications. 
\M{} requires a domain model (\MREVISION{R1.1}{i.e.,} a class diagram) of the system, which enables the generation of constraints that are used to generate test input data. 
Use case specifications and domain models are common in requirements engineering practice~\cite{Larman-Applying-2002}, such as our industry partner's organisation in our case studies. Consistent with the objectives stated above, we avoid behavioral modeling (e.g., activity and sequence diagrams) by applying Natural Language Processing (NLP) to a more structured and analysable form of use case specifications, i.e., Restricted Use Case Modeling (RUCM)~\cite{Yue-RUCM-TSE-2013}. Without limiting expressiveness, RUCM introduces a template with keywords and restriction rules to reduce ambiguity in requirements and enables automated analysis of use case specifications. It enables the extraction of behavioral information by reducing imprecision and incompleteness in use case specifications. RUCM has been successfully applied in many domains (e.g.,~\cite{Zhou2014, Yue-RTCM-ISSTA-2015, Hajri2015, Hajri2016pumconf, Hajri2016, Hajri2018, Hajri2017a, Hajri2017b, Mai2018, Hajri2019, Mai2018b, Mai2019, Zhang2018}). It has been previously evaluated through controlled experiments and showed to be usable and beneficial with respect to making use case specifications less ambiguous and more amenable to precise analysis and design~\cite{Yue-RUCM-TSE-2013}. 
\MREVISION{R2.1}{By relying on RUCM, \UMTG, differently from approaches relying on CNL, enables engineers to write specifications using the entire English vocabulary. \UMTG puts limits only on the complexity of the use case specification sentences, not on the terminology in use.
Keywords are only used to define a use case specification template, which is a common practice, and to introduce conditional sentences which are written in free form.}
In short, \M{} attempts to strike a balance among several objectives: use cases legible by all stakeholders, sufficient information for automated acceptance test cases generation, and minimal modeling.

\UMTG employs NLP to build \textit{Use Case Test Models} (\UCTM{s}) from RUCM specifications. A UCTM captures the control flow implicitly described in an RUCM specification and enables the model-based identification of use case scenarios (i.e, the sequences of use case steps in the model).
\UMTG includes three model-based, coverage strategies for the generation of use case scenarios from UCTMs: branch, def-use, and subtype coverages.
A list of textual pre, post and guard conditions in each use case specification is extracted during NLP. 
The extracted conditions enable \UMTG to determine the constraints that test inputs need to satisfy to cover a test scenario.
To automatically generate test input data for testing, \UMTG automatically translates each extracted condition in NL into a constraint in the Object Constraint Language (OCL)~\cite{OCL} that describes the condition in terms of the entities in the domain model.
\UMTG relies on OCL since it is the natural choice for constraints in UML class diagrams.
To generate OCL constraints, it exploits the capabilities of advanced NLP techniques (e.g., Semantic Role Labeling~\cite{PunyakanokRoYi08}).
The generated OCL constraints are then used to automatically generate test input data via constraint solving using Alloy~\cite{Jackson2006}. Test oracles are generated by processing the postconditions.

Engineers are expected to manually inspect the automatically generated OCL constraints, possibly make corrections and write new constraints when needed. Note that the required manual effort is very limited since, according to our industrial case studies, \UMTG can automatically and correctly generate 95\% of the OCL constraints. The accuracy of the OCL constraint generation is very high, since 99\% of the generated constraints are correct. 
Executable test cases are then generated by identifying -- using a mapping table -- the test driver API functions to be used to provide the generated test input data to the system under test.

This paper extends our previous conference papers concerning the automatic generation of UCTMs~\cite{Wang--UMTG-ISSTA-2015} and the automatic generation of OCL constraints from specifications in NL~\cite{Wang2018} published at the International Symposium on Software Testing and Analysis (ISSTA'15) and at the 11th IEEE Conference on Software Testing, Validation and Verification (ICST'18). An earlier version of our tool was demonstrated~\cite{Wang2015} at the 10th Joint meeting of the European Software Engineering Conference and the ACM SIGSOFT Symposium on the Foundations of Software Engineering (ESEC/FSE'15). This paper brings together, refines, and extends the ideas from the above papers. %
Most importantly, we extend the expressiveness of the automatically generated OCL constraints \MREVISION{R1.2 R3.7}{by supporting existential quantifiers and size operators in addition to universal quantifiers}, \CHANGED{we} introduce an Alloy-based constraint solving algorithm that solves the path conditions in OCL, and \CHANGED{we} integrate the def-use and subtype coverage strategies not presented in our previous work.
Finally, the paper further provides substantial new empirical evidence to support the scalability of our approach, and demonstrates its effectiveness using two industrial case studies (i.e., automotive embedded systems sold in the US and EU markets). 
Our contributions include:

\begin{itemize} 

\item \M{}, an approach for the automatic generation of executable acceptance test cases from use case specifications and a domain model, without resorting to behavioral modeling;

\item an NLP technique generating test models (UCTMs) from use case specifications expressed with RUCM;

\item an NLP technique generating OCL constraints from use case specifications for test input data generation;

\item an algorithm combining UCTMs and constraint solving to automatically generate test input data, based on three different coverage criteria;

\item a publicly available tool integrated as a plug-in for IBM DOORS and Eclipse, which generates executable acceptance test cases from use case specifications; 

\item two industrial case studies from which we provide credible empirical evidence demonstrating the applicability, scalability and benefits of our approach.

\end{itemize}

\MREVISION{R2.2}{In this paper we focus on embedded systems. Although \UMTG is a generic approach that relies on artefacts that may be available for any type of software system, 
it comprises solutions that have been designed specifically for embedded systems. These includes RUCM, which has only been partially evaluated with other types of systems (e.g., Web systems~\cite{Mai2018}), the NLP technique for generating OCL constraints which relies on patterns defined based on the characteristics of embedded systems, and the coverage criteria, which reflect the stringent testing thoroughness typical of safety-critical embedded systems.}

This paper is structured as follows. Section~\ref{sec:background} provides the background on the NLP techniques on which this paper builds the proposed test case generation approach. Section~\ref{sec:context} introduces the industrial context of our case study to illustrate the practical motivations for our approach. Section~\ref{sec:related} discusses the related work in light of our industrial needs. In Section~\ref{sec:overview}, we provide an overview of the approach. From Section~\ref{sec:rucm} to Section~\ref{sec:generationOfTestCases}, we provide the details of the core technical parts of our approach. Section~\ref{sec:tool} presents our tool support for test case generation. Section~\ref{sec:evaluation} reports on the results of the empirical validation conducted with two industrial case studies. We conclude the paper in Section~\ref{sec:conclusion}.

\section{Background and Glossary}
\label{sec:background}
In this section, we present the background regarding the Natural Language Processing (NLP) techniques which we employ in \M{}, as well as a glossary defining the terminology used in the paper. 
\MREVISION{R3.3}{The background on model-based testing is not presented because it is widely known to readers interested in software testing, who may further refer to books and surveys on the topic~\cite{Pretschner05,WenbinMBT,Utting:2006:PMT}.}

NLP refers to a set of procedures that extract structured information from documents written in NL. They are implemented as a pipeline that executes multiple analyses, e.g., tokenization, morphology analysis, and syntax analysis~\cite{Jurafsky-SLP-BOOK}. %

\M{} relies on five different NLP analyses: tokenization, named entity recognition, part-of-speech tagging, semantic role labeling (SRL), and semantic similarity detection.
Tokenization splits a sentence into tokens based on a predefined set of rules (e.g., the identification of whitespaces and punctuation).
Named entity recognition identifies and classifies named entities in a text into predefined categories (e.g., the names of cities). 
Part-of-speech (POS) tagging assigns parts of speech to each word in a text (e.g., noun, verb, pronoun, and adjective).
SRL automatically determines the roles played by the phrases\footnote{The term \emph{phrase} indicates a word or a group of consecutive words.} in a sentence~\cite{Jurafsky-SLP-BOOK}, e.g., the actor performing an activity. Semantic similarity detection determines the similarity between two given phrases.

Tokenization, named entity recognition, and POS tagging are well known in the software engineering community since they have been adopted by several approaches integrating NLP \cite{Zhang:ASE:2015,Sinha2009,Bajwa-NL2OCL-IEDOCC-2010,Bajwa-NL2Alloy-2010, Arora2015a, Arora2015b}. However, none of the existing software testing approaches relies on SRL or combines SRL with semantic similarity detection.

Section~\ref{subsec:nlpsrl} provides a brief description of SRL, while we present the basics of semantic similarity detection in Section~\ref{subsec:similarity}. \MREVISION{R3.6}{In Section~\ref{subsec:glossary}, we present a glossary for some of the terms and concepts frequently used in the paper.}

\subsection{Semantic Role Labeling}
\label{subsec:nlpsrl}

SRL techniques are capable of  automatically determining the roles played by words in a sentence.
For the sentences \emph{The system starts} and \emph{The system starts the database}, SRL can determine that the actors affected by the actions are \emph{the system} and \emph{the database}, respectively. The component that is started coincides with the subject in the first sentence and with the object in the second sentence although the verb \emph{to start} is used with active voice in both. This information cannot be captured by other NLP techniques like POS tagging or dependency parsing.

There are few SRL tools~\cite{SEMAFORtool,SHALtool,COGCOMPtool}. 
They are different in terms of models they adopt to capture roles.
Semafor \cite{SEMAFORtool,Das-SEMAFOR-2014} and Shalmaneser \cite{SHALtool} are based on the FrameNet model, while the CogComp NLP pipeline (hereafter \CNP~\cite{COGCOMPtool}) 
uses the PropBank~\cite{Palmer2005} and NomBank models~\cite{Meyers-NomBank-HLTNAACL-2004,Gerber-NomBankImplicitArgumentation-NAACL-2009}.
To the best of our knowledge, \CNP~is the only tool under active development, and is thus used in \M{}. %

The tools using PropBank 
tag the words in a sentence with keywords (e.g., \emph{A0}, \emph{A1}, \emph{A2}, \emph{AN}) to indicate their roles.
\emph{A0} indicates who performs an action, while \emph{A1} indicates the actor most directly affected by the action. For instance, the term \emph{The system} is tagged with \emph{A1} in the sentence \emph{The system starts}, while the term \emph{the database} is tagged with \emph{A1} in the sentence \emph{The system starts the database}. %
The other roles are verb-specific despite some commonalities, e.g., \emph{A2} which is often used for the end state of an action.

PropBank includes additional roles which are not verb-specific (see Table~\ref{tab:propBankAdditionalRoles}). They are labeled with general keywords and match adjunct information in different sentences, e.g., \emph{AM-NEG} indicating negative verbs. 
NomBank, instead, captures the roles of nouns, adverbs, and adjectives in noun phrases. It uses the same keywords adopted by PropBank. 
For instance, using ProbBank, we identify that the noun phrase \emph{the watchdog counter} plays the role \emph{A1} in the sentence
\emph{The system resets the watchdog counter}. 
Using NomBank, we obtain complementary information indicating the term \emph{counter} is the main noun (tagged with \emph{A0}), and the term \emph{watchdog} is an attributive noun (tagged with \emph{A1}).

PropBank does not help identify two different sentences describing similar concepts. 
In the sentences \emph{The system stopped the database}, \emph{The system halted the database} and \emph{The system terminated the database}, an SRL tool using PropBank tags `\emph{the database}' with \emph{A1}, indicating the database is the actor affected by the action. However, \emph{A1} does not indicate that the three sentences have similar meanings (i.e., the verbs are synonyms). To identify similar sentences, \M{} employs semantic similarity detection techniques. %

\begin{table}[tb]
\scriptsize
\caption{PropBank Additional Semantic Roles used in the paper. }
\begin{center}
\begin{tabular}{|@{\hspace{0.5mm}}p{1.3cm}|p{6.8cm}|}
    \hline
    \multicolumn{2}{|c|}{\textbf{Verb-specific semantic roles}}\\
        \hline
    \textbf{Identifier} & \textbf{Definition} \\
    \hline
    A0& Usually indicates who performs an action.\\
    A1& Usually indicates the actor most directly affected by the action.\\
    A2& With motion verbs, indicates a final state or a location.\\
            \hline
        \multicolumn{2}{|c|}{\textbf{Generic semantic roles}}\\
            \hline
                \textbf{Identifier} & \textbf{Definition} \\
    \hline
    AM-ADV & Adverbial modification.\\
    AM-LOC & Indicates a location.\\
    AM-MNR & Captures the manner in which an activity is performed.\\
    AM-MOD & Indicates a modal verb.\\
    AM-NEG & Indicates a negation, e.g. 'no'.\\
    AM-TMP & Provides temporal information.\\
    AM-PRD & Secondary predicate with additional information about A1. \\
    \hline
\end{tabular}
\end{center}
\label{tab:propBankAdditionalRoles}
\end{table}%

\subsection{Semantic Similarity Detection}
\label{subsec:similarity}

For semantic similarity detection, we use the VerbNet lexicon~\cite{Kipper-ClassBasedCOnstructionOfVerbLexicon-NCAI-2000}, which clusters 
verbs that \emph{have a common semantics and share a common set of semantic roles}
into a total of 326 verb classes~\cite{VerbnetWebClasses}.
Each verb class is provided with a set of \emph{role patterns}. %
For example, \emph{$\langle{}$A1,V$\rangle{}$} and \emph{$\langle{}$A0,V,A1$\rangle{}$} are two \emph{role patterns} for the VerbNet class \emph{stop-55.4}, which includes, among others, the verbs \emph{to stop}, \emph{to halt} and \emph{to terminate}. 
In \emph{$\langle{}$A1,V$\rangle{}$}, the sentence contains only the verb (\emph{V}), and the actor whose state is altered (\emph{A1}). %
In \emph{$\langle{}$A0,V, A1$\rangle{}$}, the sentence contains the actor performing the action (\emph{A0}), the verb (\emph{V}), and the actor affected by the action (\emph{A1}).
Examples of these two patterns are \emph{the database stops} and \emph{the system stops the database}, respectively. \M{} uses VerbNet version 3.2~\cite{VerbnetWebClasses}, which includes 272 verb classes and 214 subclasses where %
a class may have more than one subclass.

VerbNet uses a model different than PropBank. There is a mapping between PropBank and the model in VerbNet~\cite{Palmer-SemLink-GenLex-2009}. 
For simplification, we use only PropBank role labels in the paper.
All the verbs in a VerbNet class are guaranteed to have a common set of role patterns, %
but are not guaranteed to be synonyms (e.g., the verbs \emph{repeat} and \emph{halt} in the VerbNet class \emph{stop-55.4}). We employ WordNet~\cite{Miller-WordNet-CACM-1995}, a database of lexical relations, %
to cluster verbs with similar meaning. 
\MREVISION{R2.3}{Further, we use WordNet to identify synonyms and antonyms of phrases in use case specifications (see Section~\ref{subsection:ocl_algorithm}).}

\subsection{Glossary}
\label{subsec:glossary}

\MREVISION{R3.6}{An \emph{actor} specifies a type of role played by an entity interacting with a system (e.g., by exchanging signals and data), but which is external to the system (See Section~\ref{sec:rucm}).}

\CHANGED{A \emph{use case} is a list of actions or event steps typically defining the interactions between an actor and a system to achieve a goal. It is generally named with a phrase that denotes the goal, e.g., \emph{Identify Occupancy Status} (See Section~\ref{sec:rucm}).}

\CHANGED{A \emph{use case specification} is a textual document that captures the specific details of a use case. Use case specifications provide a way to document the functional requirements of a system. They generally follow a template (See Section~\ref{sec:rucm}).}

\CHANGED{A \emph{use case flow} is a possible sequence of interactions between actors and the system captured by a use case specification. A use case specification may include multiple alternative use case flows (See Section~\ref{sec:rucm}).}

\CHANGED{A \emph{use case scenario} is a sequence of interactions between actors and the system. It represents a single use case execution. It is a possible path through a use case specification. It may include multiple use case flows (See Section~\ref{sec:testGeneration}).}

\CHANGED{An \emph{abstract test case} is a human-readable description of the interactions between actors and the system under test that should be exercised during testing. It exercises one use case
scenario. It includes one or more abstract test oracles (See Section~\ref{sec:generationOfTestCases}).}

\CHANGED{An \emph{abstract test oracle} is a human-readable description of a function that verifies if the behavior of the system meets its requirements. It captures the characteristics that the
outputs generated by the system should have (See Section~\ref{sec:generationOfTestCases}).}

\CHANGED{An \emph{executable test case} is a sequence of executable instructions (i.e., invocations of test driver functions) that trigger the system under test, thus simulating the interactions between one or more actors and the system.
It includes one or more executable test oracles (See Section~\ref{sec:generationOfTestCases}).}

\CHANGED{An \emph{executable test oracle} is an executable instruction that returns true when the systems behaves according to its requirements, false otherwise. 
An executable test oracle usually consists of a set of assertions verifying if a set of outputs generated by the system match a set of expected outputs (See Section~\ref{sec:generationOfTestCases}).}

\CHANGED{A \emph{test driver function} is a software module used to invoke the software under test. A test driver typically provides test inputs, controls and monitors execution, and reports test results~\cite{ieee24765}. See Section~\ref{sec:generationOfTestCases}.}

\CHANGED{A \emph{test verdict} (or \emph{test result}) is an indication of whether or not an executed test case has passed or failed, i.e., if the actual output corresponds to the expected result or if deviations were observed~\cite{ieee24765}. A test verdict can be either \emph{Pass} or \emph{Fail}.}

\CHANGED{An \emph{input equivalence partition} (also referenced as \emph{equivalence partition} in this paper) is a subset of the range of values for an input variable, or set of input variables, for the software under test, such that all the values in the partition can reasonably be expected to be treated similarly by the software under test (i.e., they are considered equivalent)~\cite{ieee24765}.}

\section{Motivation and Context} 
\label{sec:context}
The context for which we developed \M{} is that of safety-critical 
embedded software in the automotive domain.
The automotive domain is a representative example of the many domains for which compliance with requirements should be demonstrated through documented test cases. For instance, ISO-26262~\cite{ISO26262}, an automotive safety standard, states that all system requirements should be properly tested by corresponding system test cases.

In this paper, we use the system \emph{BodySense}$^{TM}$ as one of the case studies and also to motivate and illustrate \M{}. \BodySense is a safety-critical automotive software developed by IEE~\cite{IEE}, a leading supplier of embedded software and hardware systems in the automotive domain. \BodySense provides automatic airbag deactivation for child seats. It classifies vehicle occupants for smart airbag deployment. Using a capacitive sensor in the vehicle's passenger seat, it monitors whether the seat is occupied, as well as classifying the occupant. If the passenger seat has a child in a child seat or is unoccupied, the system disables the airbag. For seats occupied by adult passengers, it ensures the airbag is deployed in the event of an accident. \BodySense also provides occupant detection for the seat belt reminder function.

Table~\ref{tab:testCase} gives a simplified version of a real test case for \emph{BodySense}. 
Lines 1, 3, 5, 7, and 9 provide high-level operation descriptions, i.e., informal descriptions of the operations to be performed on the system. These lines are followed by the name of the functions that should be executed by the test driver along with the corresponding input and expected output values. For instance, Line 4 invokes the function \emph{SetBus} with a value indicating that the test driver should simulate the presence of an adult on the seat (for simplicity assume that, when an adult is seated, the capacitance sensor positioned on a seat sends a value above 600 on the bus).

\begin{table}[h]
\scriptsize
\caption{An example test case for \BodySense.}
\begin{center}
\begin{tabular}{|p{0.35cm}|p{4.6cm}|p{2.4cm}|}
\hline
\textbf{Line}&\textbf{Operation}&\textbf{Inputs/Expectations}\\
\hline
1&	\textit{Reset power and wait}&\\
\hline
2&	ResetPower&	Time=INIT\_TIME\\
\hline
3&	\textit{Set occupant status - Adult}&\\
\hline
4&	SetBus& Channel = {RELAY}\\
&&Capacitance  = {601}\\
\hline
5&\textit{Simulate a nominal temperature}&\\

\hline
6&	SetBus&Channel=RELAY\\
&&Temperature = 20\\

\hline
7&	\textit{Check that and Adult has been detected on the seat, i.e. SeatBeltReminder status is Occupied and AirBagControl status is Occupied.}&	\\

\hline
8&    ReadAndCheckBus&D0=OCCUPIED\\
&   &D1=OCCUPIED\\
\hline
9&	\textit{Check that the AirBagControl has received new data.}&	\\
\hline
10&    CheckAirbagPin&0x010\\
\hline
\end{tabular}
\end{center}
\label{tab:testCase}
\end{table}%

Exhaustive test cases needed to validate safety-critical, 
embedded software are difficult both to derive and maintain because requirements are often updated during the software lifecycle (e.g., when \BodySense needs to be customized for new car models).
For instance, the functional test suite for \emph{BodySense} is made of 192 test cases which include a total of 4,707 calls to test driver functions and around 21,000 variable assignments.
The effort required to specify test cases for \emph{BodySense} is overwhelming. 
Without automated test case generation, such testing activity is not only expensive but also error prone.

Within the context of testing safety-critical, 
embedded software such as \emph{BodySense}, we identify three challenges that need to be considered for the automatic generation of \CHANGED{system-level, acceptance} test cases from functional requirements:

\textbf{\textit{Challenge 1:} Feasible Modeling.} Most of the existing automatic system test generation approaches are model-based and rely upon behavioral models such as state, sequence or activity diagrams (e.g.,~\cite{Linzhang2004, Bandyopadhyay2009, Briand-UMLbasedApproachToSystemTesting-2001, Offutt1999, Abdurazik2000}). In complex industrial systems, behavioral models that are precise enough to enable test automation are so complex that their specification cost is prohibitive and the task is often perceived as overwhelming by engineers.
To evaluate the applicability of behavioral modeling on \emph{Body\-Sense}, we asked the IEE engineers to specify system sequence diagrams (SSDs) for some of the use cases of \emph{BodySense}. For example, the SSD for the use case \textit{Identify initial occupancy status of a seat} included 74 messages, 19 nested blocks, and 24 references to other SSDs that had to be derived. This was considered too complex for the engineers and required significant help from the authors of this paper, and many iterations and meetings. 
Our conclusion is that the adoption of behavioral modeling, at the level of detail required for automated testing, is not a practical option for \CHANGED{acceptance testing} automation unless detailed behavioral models are already used by engineers for other purposes, e.g., software design.

\textbf{\textit{Challenge 2:} Automated Generation of Test Data.} 
Without behavioral modeling, test generation can be driven only by existing requirements specifications in NL, which complicates the identification of the test data (e.g., the input values to send to the system under test). Because of this, most of the existing approaches focus on the identification of test scenarios (i.e., the sequence of activities to perform during testing), and ask engineers to manually produce the test data.
Given the complexity of the test cases to be generated (recall that the BodySense test suite includes 21000 variable assignments), it is extremely important to automatically generate test data, and not just test scenarios.

\textbf{\textit{Challenge 3:} Deployment and Execution of the Test Suite.} Execution of test cases for a system like \emph{BodySense} entails the deployment of software under test on the target environment. 
To speed up testing, test case execution is typically automated through test scripts invoking test driver functions. These functions simulate sensor values and read computed results from a communication bus. Any test generation approach should generate appropriate function calls and test data in a processable format for the test driver. For instance, the test drivers in  \emph{BodySense} need to invoke driver functions (e.g., \emph{SetBus}) to simulate seat occupancy. %

In the rest of this paper, we focus on how to best address these challenges in a practical manner, in the context of use case-driven development of embedded systems.

\section{Related Work}
\label{sec:related}

In this section, we cover the related work across three categories in terms of the challenges we presented in Section~\ref{sec:context}.

\textit{\textbf{Feasible Modeling.}} Most of the system test case generation approaches require that system requirements be given in UML behavioral models such as activity diagrams
(e.g.,~\cite{Linzhang2004, Hasling2008model, Nayak2011, Samuel2009}), statecharts (e.g.,~\cite{Bandyopadhyay2009, Sarma2009, Ryeser-Scenario-ICSSEA-1999, Kim1999}), and sequence diagrams (e.g.,~\cite{Briand-UMLbasedApproachToSystemTesting-2001, Pelliccione2005, Nebut-AutomaticTestGenerationUseCaseDriven-TSE-2006, Basanieri2002}). For instance, Nebut et al.~\cite{Nebut-AutomaticTestGenerationUseCaseDriven-TSE-2006} propose a use case driven test generation approach based on system sequence diagrams. Gutierrez et al.~\cite{Gutierrez2015} introduce a systematic process based on model-driven engineering paradigm to automate the generation of system cases from functional requirements given in activity diagrams. Briand and Labiche~\cite{Briand-UMLbasedApproachToSystemTesting-2001} use both activity and sequence diagrams to generate system test cases. While sequential dependencies between use cases are extracted from an activity diagram, sequences in a use case are derived from a system sequence diagram. In contrast, \M{} needs only use case specifications complemented by a domain model and OCL constraints. In addition, \M{} is able to automatically generate most of the OCL constraints from use case specifications. %

There are techniques generating behavioral models from NL requirements~\cite{Gutierrez2008, Yue-RUCMtoState-ECMFA-2011, Yue-RUCM-TSE-2013, Yue-UseCasesToActivity-MFA-2010, Yue2015, Ding2016}. %
Some approaches employ similar techniques %
in the context of test case generation. For instance, Frohlich and Link~\cite{frohlich2000} generate test cases from UML statecharts that are automatically derived from use cases. De Santiago et al.~\cite{de2012generating} provide a similar approach to generate test cases from statecharts derived from NL scenario specifications. Riebisch et al.~\cite{riebisch2003uml} describe a test case generation approach based on the semi-automated generation of state diagrams from use cases. Katara and Kervinen~\cite{katara2007making} propose an approach which generates test cases from labeled transition systems that are derived from use case specifications. \MREVISION{R1.3}{Nogueira et al.~\cite{Nogueira-TestGenerationFromStateBased-FAC-2014} provide a test case generation approach using labelled transition systems derived from use case specifications. The formal testing theory the approach is built upon is proved to be sound: if a generated test case fails, it necessarily means that the system under test does not conform to the specification (according to a formally defined conformance relation).} 
Sarmiento et al.~\cite{Sarmiento2016, Sarmiento14} propose another approach to generate test scenarios from a restricted form of NL requirements. The approach automatically translates restricted NL requirements into executable Petri-Net models; the generated Petri-Nets are used as input for test scenario generation. 
Soeken et al.~\cite{Soeken-BDD-TOOLS-2012} employ a statistical parser~\cite{StanfordParser} and a lexical database~\cite{Miller-WordNet-CACM-1995} to semi-automatically generate sequence diagrams from NL scenarios, which are later used to semi-automatically generate test cases. Hartmann et al.~\cite{Hartmann2005} provide a test-generation tool that creates a set of test cases from UML models that are manually annotated and semi-automatically extracted from use case specifications. All these approaches mentioned above have two major drawbacks in terms of feasible modeling: (i) generated test sequences have to be edited, corrected, and/or refined and (ii) test data have to be manually provided in the generated test models. In contrast, \M{} not only generates sequences of function calls that do not need to be modified but also generates test data for function calls.

Kesserwan et al.~\cite{Kesserwan2017} provide a model-driven testing methodology that supports test automation based on system requirements in NL. Using the methodology, the engineer first specifies system requirements according to Cockburn use case notation~\cite{Cockburn-WritingEffectiveUseCases-Book-2000} and then manually refines them into Use Case Map (UCM) scenario models~\cite{Buhr1998}. In addition, test input data need to be manually extracted from system requirements and modelled in a data model. \M{} requires that system requirements be specified in RUCM without any further refinement. Text2Test~\cite{Sinha-Text2Test-ICST-2010, Sinha2009} extracts control flow implicitly described in use case specifications, which can be used to automatically generate system test cases. The adaptation of such an approach in the context of test case generation has not been investigated.

\textit{\textbf{Automated Generation of Test Data.}} The ability to generate test data, and not just abstract test scenarios, is an integral part of automated test case generation~\cite{Stocks1996}. However, many existing NL-based test case generation approaches require manual intervention to derive test data for executable test cases (e.g.,~\cite{Zhang-SAM-2014, Kesserwan2017, Sarmiento2016, Sarmiento14}), while some other approaches focus only on generating test data (e.g.,~\cite{Weyuker1994, Offutt1999b, Offutt2003, Benattou2002, Soltana2017, Ali-GeneratingTestsFromOCL-TSE-2013}). For instance, Zhang et al.~\cite{Zhang-SAM-2014} generate test cases from RUCM use cases. The generated test cases cannot be executed automatically because they do not include test data. Sarmiento et al.~\cite{Sarmiento2016} generate test scenarios without test data from a restricted form of NL requirements specifications.

Similar to \M{}, Kaplan et al.~\cite{Kaplan-LessIsMore-ICST-2008} propose another approach, i.e., Archetest, which generates test sequences and test inputs from a domain model and use case specifications together with invariants, guardconditions and postconditions. Yue et al.~\cite{Yue-RTCM-ISSTA-2015} propose a test case generation tool (aToucan4Test), which takes RUCM use case specifications annotated with OCL constraints as input and generates automatically executable test cases. These two test generation approaches require that conditions and constraints be provided by engineers to automatically generate test data. In contrast, \M{} can automatically generate, from use case specifications, most of the OCL constraits that are needed for the automated generation of test data.

In some contexts, test data might be simple and consist of sequences of system events without any associated additional parameter value. 
This is the case of interaction test cases for smartphone systems, which can be automatically generated by the approach proposed by De Figueiredo et al.~\cite{deFigueiredo-2006}. The approach
processes use case specifications in a custom use case format to derive sequences of system operations and events.
\M{} complements this approach with the generation of parameter values, which is instead needed to perform functional testing at the system level.

Carvalho et al.~\cite{Carvalho-NAT2TEST-SCP-2014, Carvalho-SCR-SAC-2013} generate executable test cases for reactive systems from requirements written according to a restricted grammar and dictionary. The proposed approach effectively generates test data but has two main limitations: (i) the underlying dictionary may change from project to project (e.g., the current version supports only seven verbs of the English language), and (ii) the restricted grammar may not be suitable to express some system requirements (e.g., the approach does not tackle the problem of processing transitive and intransitive forms of the same verb).
In contrast, \M{} does not impose any restricted dictionary or grammar but simply relies on a use case format, RUCM, which can be used to express use cases for different kinds of systems.
RUCM does not restrict the use of verbs or nouns in use case steps and thus does not limit the expressiveness of use case specifications. Furthermore, the RUCM keywords are used to specify input and output steps but do not constraint internal steps or condition sentences (see Section~\ref{sec:rucm}). 
Finally, by relying on SRL and VerbNet, \UMTG \MREVISION{R1.6}{aims to ensure the correct generation of OCL constraints} (see Section~\ref{sec:oclGeneration}), without restricting the writing of sentences (e.g., it supports the use of both transitive and intransitive forms).

Other approaches focus on the generation of class invariants and method pre/postconditions, from NL requirements, which, in principle, could be used for test data generation (e.g.,~\cite{Pandita-ICSE-2012,Bajwa-NL2OCL-IEDOCC-2010,Bajwa-NL2Alloy-2010}). 
Pandita et al.~\cite{Pandita-ICSE-2012} focus only on API descriptions written according to a CNL.
NL2OCL~\cite{Bajwa-NL2OCL-IEDOCC-2010} and NL2Alloy~\cite{Bajwa-NL2Alloy-2010}, instead, process a UML class diagram and NL requirements to derive class invariants and method pre/postconditions. 
These two approaches rely on an ad-hoc semantic analysis algorithm that uses information in the UML class diagram (e.g., class and attribute names) to identify the roles of words in sentences. They rely on the presence of specific keywords to determine passive voices and to identify the operators to be used in the generated invariants and conditions. 
Their constraint generation is rule-based, but they do not provide a solution to ease the processing of a large number of verbs with a reasonable number of rules.
Thanks to the use of Wordnet synsets and VerbNet classes (see Section~\ref{sec:oclGeneration}), \M{} can process a large set of verbs with few rules to generate OCL constraints.

Though NL2OCL~\cite{Bajwa-NL2OCL-IEDOCC-2010} and NL2Alloy~\cite{Bajwa-NL2Alloy-2010} are no longer available for comparison, they seem more useful for deriving class invariants including simple comparison operators (i.e., the focus of the evaluation in~\cite{Bajwa-NL2OCL-IEDOCC-2010}), rather than for generating pre/postconditions of the actions performed by the system (i.e., the focus of \M{}). Pre/postconditions are necessary for deriving test data in our context. 

\begin{table*}[tb]
\scriptsize
\caption{Summary and comparison of the related work.}
\label{table:RelatedWork}
\begin{tabular}{
|@{\hspace{0.02cm}}p{2.7cm}
|@{\hspace{0.05cm}}p{1.23cm} 
|@{\hspace{0.05cm}}p{1.70cm} 
|@{\hspace{0.05cm}}p{1.7cm} 
|@{\hspace{0.05cm}}p{1.9cm} 
| @{\hspace{0.05cm}}p{1.8cm} 
|@{\hspace{0.05cm}}p{1.7cm} 
|@{\hspace{0.05cm}}p{1.2cm} 
|@{\hspace{0.05cm}}p{1.8cm}|}
\hline
&\textbf{No need for behavioral models}&\textbf{Automated extraction of control flow from NL text}&\textbf{No need for restricted grammar or dictionary}&\textbf{Automated generation of abstract test input sequences}&\textbf{No need for editing the generated test input sequences}&\textbf{Automated generation of test data}
&\textbf{No need for formal data constraints}
&\textbf{Automated generation of data constraints from NL text}\\
\hline

\textbf{\UMTG}& $+$ & $+$ & $+$ & $+$ & $+$ & $+$ & $+$ & $+$\\
\hline

Nebut et al.~\cite{Nebut-AutomaticTestGenerationUseCaseDriven-TSE-2006}& $-$ & $-$ & $\mathit{NA}$ & $+$ & $-$ & $-$ & $\mathit{NA}$ & $\mathit{NA}$\\
\hline
Gutierrez et al.~\cite{Gutierrez2015}& $-$ & $-$ & $\mathit{NA}$ & $+$ &$-$ & $-$ & $\mathit{NA}$ & $\mathit{NA}$\\
\hline

Briand et al.~\cite{Briand-UMLbasedApproachToSystemTesting-2001}& $-$ & $-$ & $\mathit{NA}$ & $+$ & $-$ & $-$ & $\mathit{NA}$ & $\mathit{NA}$\\
\hline

Yue et al.~\cite{Yue2015}& $+$ & $+$ & $+$ & $\mathit{NA}$ & $\mathit{NA}$ & $\mathit{NA}$ & $\mathit{NA}$ & $\mathit{NA}$\\
\hline

Gutierrez et al.~\cite{Gutierrez2008}& $+$ & $+$ & $+$ & $\mathit{NA}$ & $\mathit{NA}$ & $\mathit{NA}$ & $\mathit{NA}$ & $\mathit{NA}$\\
\hline

Ding et al.~\cite{Ding2016}& $+$ & $+$ & $+$ & $\mathit{NA}$ & $\mathit{NA}$ & $\mathit{NA}$ & $\mathit{NA}$ & $\mathit{NA}$\\
\hline

Frohlich and Link~\cite{frohlich2000}& $+$ & $+$ & $+$ & $+$ & $-$ & $-$ & $\mathit{NA}$ & $\mathit{NA}$\\
\hline

De Santiago et al.~\cite{de2012generating}& $+$ & $+$ & $+$ & $+$ & $-$ & $-$ & $\mathit{NA}$ & $\mathit{NA}$\\
\hline

Katara et al.~\cite{katara2007making}& $+$ & $+$ & $-$ & $+$ & $-$ & $-$ & $\mathit{NA}$ & $\mathit{NA}$\\
\hline

Sarmiento et al.~\cite{Sarmiento2016, Sarmiento14}& $+$ & $+$ & $-$ & $+$ & $-$ & $-$ & $\mathit{NA}$ & $\mathit{NA}$\\
\hline

Soeken et al.~\cite{Soeken-BDD-TOOLS-2012}& $+$ & $+$ & $+$ & $+$ & $-$ & $-$ & $\mathit{NA}$ & $\mathit{NA}$\\
\hline

Kesserwan et al.~\cite{Kesserwan2017}& $-$ & $-$ & $\mathit{NA}$ & $+$ & $-$ & $-$ & $\mathit{NA}$ &$\mathit{NA}$\\
\hline

Text2Test~\cite{Sinha-Text2Test-ICST-2010, Sinha2009}& $+$ & $+$ & $+$ & $\mathit{NA}$ & $\mathit{NA}$ & $\mathit{NA}$ & $\mathit{NA}$ & $\mathit{NA}$\\
\hline

Zhang et al.~\cite{Zhang-SAM-2014}& $+$ & $+$ & $+$ & $+$ & $-$ & $-$ & $\mathit{NA}$ & $\mathit{NA}$\\
\hline

Weyuker et al.~\cite{Weyuker1994}& $\mathit{NA}$ & $\mathit{NA}$ & $\mathit{NA}$ & $\mathit{NA}$ & $\mathit{NA}$ & $+$ & $-$ & $-$\\
\hline

Offutt et al.~\cite{Offutt1999b}& $\mathit{NA}$ & $\mathit{NA}$ & $\mathit{NA}$ & $\mathit{NA}$ & $\mathit{NA}$ & $+$ & $-$ & $-$\\
\hline

Offutt et al.~\cite{Offutt2003}& $\mathit{NA}$ & $\mathit{NA}$ & $\mathit{NA}$ & $\mathit{NA}$ & $\mathit{NA}$ & $+$ & $-$ & $-$\\
\hline

Benattou et al.~\cite{Benattou2002}& $\mathit{NA}$ & $\mathit{NA}$ & $\mathit{NA}$ & $\mathit{NA}$ & $\mathit{NA}$ & $+$ & $-$ & $-$\\
\hline

Soltana et al.~\cite{Soltana2017}& $\mathit{NA}$ & $\mathit{NA}$ & $\mathit{NA}$ & $\mathit{NA}$ & $\mathit{NA}$ & $+$ & $-$ & $-$\\
\hline

Ali et al.~\cite{Ali-GeneratingTestsFromOCL-TSE-2013}& $\mathit{NA}$ & $\mathit{NA}$ & $\mathit{NA}$ & $\mathit{NA}$ & $\mathit{NA}$ & $+$ & $-$ & $-$\\
\hline

Carvalho et al.~\cite{Carvalho-NAT2TEST-SCP-2014, Carvalho-SCR-SAC-2013}& $+$ & $+$ & $-$ & $+$ & $+$ & $+$ & $+$ & $+$\\
\hline

Kaplan et al.~\cite{Kaplan-LessIsMore-ICST-2008}& $+$ & $+$ & $+$ & $+$ & $+$ & $+$ & $-$ & $-$\\
\hline

Yue et al.~\cite{Yue-RTCM-ISSTA-2015}& $+$ & $+$ & $+$ & $+$ & $+$ & $+$ & $-$ & $-$\\
\hline

Pandita et al.~\cite{Pandita-ICSE-2012}& $\mathit{NA}$ & $\mathit{NA}$ & $\mathit{NA}$ & $\mathit{NA}$ & $\mathit{NA}$ & $\mathit{NA}$ & $\mathit{NA}$ & $+$\\
\hline

NL2OCL~\cite{Bajwa-NL2OCL-IEDOCC-2010}& $\mathit{NA}$ & $\mathit{NA}$ & $\mathit{NA}$ & $\mathit{NA}$ & $\mathit{NA}$ & $\mathit{NA}$ & $\mathit{NA}$ & $+$\\
\hline

NL2Alloy~\cite{Bajwa-NL2Alloy-2010}& $\mathit{NA}$ & $\mathit{NA}$ & $\mathit{NA}$ & $\mathit{NA}$ & $\mathit{NA}$ & $\mathit{NA}$ & $\mathit{NA}$ & $+$\\
\hline

\end{tabular}
\\
\\

\end{table*}%

\textit{\textbf{Deployment and Execution of the Test Suite.}} 
The generation of executable test cases impacts on the usability of test generation techniques. In code-based approaches (e.g.,~\cite{Fraser-EvoSuite-FSE-2011,KLEE}), %
the generation of executable test cases is facilitated by the fact that it is based on processing the interfaces used during the test execution (e.g., test driver API). %

In model-based testing, %
the artefacts used to drive test generation are %
software abstractions (e.g., UML models). In this context, the generation of executable test cases is usually based on adaptation and transformation approaches~\cite{Utting:2006:PMT}. The adaptation approaches require the implementation of a software layer that, at runtime, matches high-level operations to software interfaces. They support the execution of complex system interactions (e.g., they enable feedback-driven, model-based test input generation~\cite{Larsen-UPPAAL_TRON-2005}). The transformation approaches, instead, translate an abstract test case into an executable test case by using a mapping table containing regular expressions for the translation process. They
require only abstract test cases and a mapping table, 
while the adaptation approaches need communication channels between the software under test and the adaptation layer, which might not be possible for many embedded systems. Therefore, \UMTG uses a mapping table that matches abstract test inputs to test driver function calls.

Model Transformation by Example (MTBE) approaches aim to learn transformation programs from source and target model pairs supplied as examples (e.g.,~\cite{Varro2016,Kessentini2012,Balogh2009}). These approaches search for a model transformation in a space whose boundaries are defined by a model transformation language and the source and target metamodels~\cite{Kappel2012}. Given the metamodels of abstract and executable test cases, MTBE can be applied to automatically generate part of the mapping table as a transformation program. However, this solution can be considered only when there are already some example abstract and executable test cases, which is not the case in our context, and we leave it for future work.

\MREVISION{R1.5}{In Table~\ref{table:RelatedWork}, based on a set of features necessary for the automatic generation of system test cases for acceptance testing, we summarize the differences between UMTG and prominent related work.
For each approach, the symbol '+' indicates that the approach provides the feature, the symbol '-' indicates that it does not provide the feature,
and 'NA' indicates that the feature is not applicable because it is out of its scope. For instance, the approach by Yue et al.~\cite{Yue2015} automatically generates UML analysis models, not test cases. Therefore, all the features related to the generation of test input sequences and test data are not considered for Yue et al.~\cite{Yue2015} in Table~\ref{table:RelatedWork}.
Most of the existing approaches extract behavioral models from NL specifications~\cite{Yue2015,Gutierrez2008,Ding2016}, generate abstract test inputs sequences~\cite{Nebut-AutomaticTestGenerationUseCaseDriven-TSE-2006,Gutierrez2015,Briand-UMLbasedApproachToSystemTesting-2001,frohlich2000,de2012generating,katara2007making,Sarmiento2016,Sarmiento14,Soeken-BDD-TOOLS-2012,Kesserwan2017,Zhang-SAM-2014} or derive test input data~\cite{Weyuker1994,Offutt1999b,Offutt2003,Benattou2002,Soltana2017,Ali-GeneratingTestsFromOCL-TSE-2013}.
The few approaches generating both abstract test input sequences and test data~\cite{Carvalho-NAT2TEST-SCP-2014, Carvalho-SCR-SAC-2013,Kaplan-LessIsMore-ICST-2008,Yue-RTCM-ISSTA-2015} either require the specification of data constraints using a formal language~\cite{Kaplan-LessIsMore-ICST-2008,Yue-RTCM-ISSTA-2015} or rely on a restricted grammar or dictionary~\cite{Carvalho-NAT2TEST-SCP-2014, Carvalho-SCR-SAC-2013}. Finally, a few approaches focus only on the automated generation of data constraints from text in NL~\cite{Pandita-ICSE-2012,Bajwa-NL2OCL-IEDOCC-2010,Bajwa-NL2Alloy-2010}.
\MREVISION{R2.1}{\UMTG is the only approach that automates the generation of both test input sequences and test input data, without requiring neither behavioral models nor a very restricted language. In the latter case, CNLs are limited to a restricted vocabulary while UMTG only requires the use of simple sentences and keywords but otherwise allows the use of the full English vocabulary.} 
Test input generation is enabled by the capability of automatically extracting data constraints from NL, a feature that has not so far been integrated by existing techniques.}

\section{Overview of the Approach} 
\label{sec:overview}

\begin{figure}[t]
  \centering
    \includegraphics{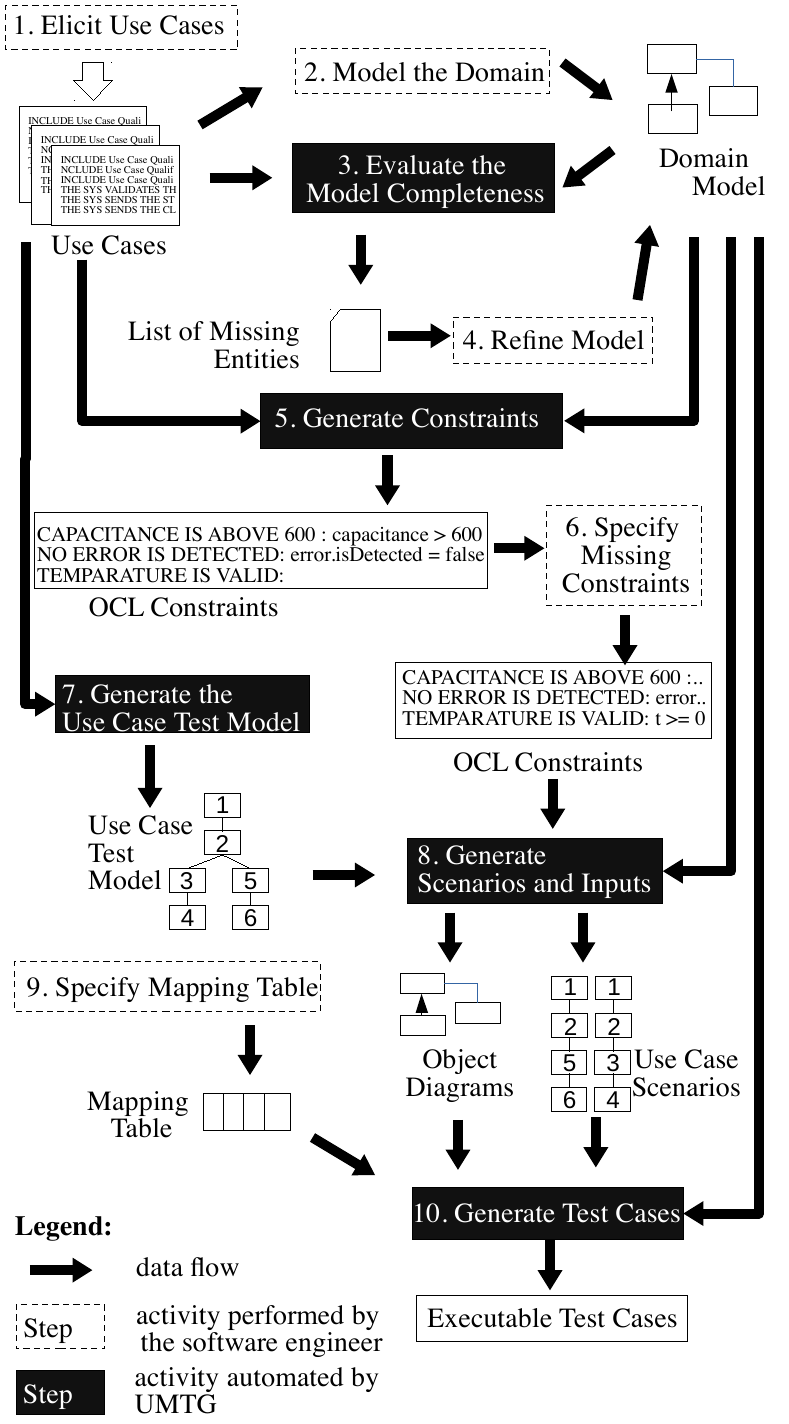}
      \caption{Overview of the \M{} approach.}
      \label{fig:approach}
\end{figure}

The process in Fig.~\ref{fig:approach} presents an overview of our approach.
In \M{}, behavioral information and high-level operation descriptions are extracted from use case specifications (\textit{Challenge 1}). \M{} generates OCL constraints from the use case specifications, while test inputs are generated from the OCL constraints through constraint solving (\textit{Challenge 2}). Test driver functions corresponding to the high-level operation descriptions and oracles implementing the postconditions in the use case specifications are generated through the mapping tables provided by the engineer (\textit{Challenge 3}). 

The engineer elicits requirements with RUCM (Step 1). The domain model is manually created as a UML class diagram (Step 2). \M{} automatically 
checks if the domain model includes all the entities mentioned in the use cases (Step 3). %
NLP is used to extract domain entities from the use cases. Missing entities are shown to the engineer who refines the domain model (Step 4).
Steps 3 and 4 are iterative: the domain model is refined until it is complete.

Once the domain model is complete, %
most of the OCL constraints are automatically generated from the extracted conditions (Step 5). 
\MREVISION{R1.4b R3.15}{The generated constraints are manually inspected by engineers to ensure they capture the meaning of the use case conditions and, as a result, guarantee the soundness of the generated test cases (i.e., to ensure that correct implementations are never rejected). Based on experience in different testing contexts, we conjecture that the manual validation of constraints is significantly less expensive than their manual definitions~\cite{PastoreICST13,PastoreICSE15}.}

The engineer manually writes the few OCL constraints that cannot be automatically generated (Step 6). \M{} further processes the use cases with the OCL constraints to generate a use case test model for each use case specification (Step 7). A use case test model is a directed graph that explicitly captures the implicit behavioral information in a use case specification.  

\M{} employs constraint solving for OCL constraints to generate test inputs associated with use case scenarios (Step 8). We use the term use case scenario for a sequence of use case steps that starts with a use case precondition and ends with a postcondition of either a basic or alternative flow. Test inputs cover all the paths in the testing model, and therefore all possible use case scenarios.

The engineer provides a mapping table that maps high-level operation descriptions and test inputs to the concrete driver functions and inputs that should be executed by the test cases (Step 9). Executable test cases are automatically generated through the mapping table (Step 10). If the test infrastructure and hardware drivers change in the course of the system lifespan, then only this table needs to change.

The rest of the paper explains the details of each step in Fig.~\ref{fig:approach}, with a focus on how we achieved our automation objectives.

\section{Elicitation of Requirements}
\label{sec:rucm}

Our approach starts with the elicitation of requirements in RUCM (Step 1 in Fig.~\ref{fig:approach}). RUCM has a template with keywords and restriction rules to reduce ambiguity in use case specifications~\cite{Yue-RUCM-TSE-2013}.
Since it was not originally designed for test generation, we introduce some extensions to RUCM. %

Table~\ref{useCaseInitialOccupancy} provides a simplified version of three \emph{BodySense} use case specifications in RUCM (i.e., \emph{Identify Occupancy Status}, \emph{Self Diagnosis}, and \emph{Classify Occupancy Status}). 
We omit some basic information such as actors and dependencies. 

The use cases contain basic and alternative flows.
A basic flow describes a main successful scenario that satisfies stakeholder interests. It contains a sequence of steps and a postcondition (Lines \ref{uc1:bf:start}-\ref{uc1:bf:post}). 
A step can describe one of the following activities: an actor sends data to the system (Lines \ref{uc1:bf:input} and \ref{uc2:requestTemperature}); the system validates some data (Line \ref{uc1:bf:validate}); the system replies to an actor with a result (Line \ref{uc1:bf:output}); the system alters its internal state (Line \ref{uc1:af:reset}).
The inclusion of another use case is specified as a step using the keyword \textit{INCLUDE USE CASE} (Line \ref{uc1:bf:include}). 
All keywords are written in capital letters for readability.

\begin{table}[h!]
\tiny
\caption{Part of \BodySense Use Case Specifications}
\begin{tabular}{|@{\hspace{0.1cm}}p{0.2cm} | @{\hspace{0.05cm}}p{8.0cm} |}
\LLRESET
\LL	& \textbf{1. Use Case} Identify Occupancy Status\\
\LL	& \textbf{1.1 Precondition}\\
\LL	& The system has been initialized. \\
\LL 	& \textbf{1.2 Basic Flow}\\
\LL 	& 1. The SeatSensor SENDS capacitance TO the system. \manuallabel{uc1:bf:input}{\LINE}  \manuallabel{uc1:bf:start}{\LINE} \\
\LL	& 2. INCLUDE USE CASE Self Diagnosis. \manuallabel{uc1:bf:include}{\LINE}\\
\LL	& 3. The system VALIDATES THAT no error is detected and no error is qualified. \manuallabel{uc1:bf:validate}{\LINE}\\
\LL	& 4. INCLUDE USE CASE Classify Occupancy Status. \manuallabel{uc1:bf:includeClassify}{\LINE}\\
\LL	& 5. The system SENDS the occupant class for airbag control TO AirbagControlUnit. \manuallabel{uc1:bf:output}{\LINE}\\
\LL	& 6. The system SENDS the occupant class for seat belt reminder TO SeatBeltControlUnit. \manuallabel{uc1:bf:end}{\LINE}\\
\LL	& Postcondition: The occupant class for airbag control has been sent to AirbagControlUnit. The occupant class for seat belt reminder has been sent to SeatBeltControlUnit.\manuallabel{uc1:bf:post}{\LINE}\\
\LL	& \textbf{1.3 Bounded Alternative Flow}\\
\LL	& RFS 2-4 \manuallabel{uc1:sat:ref:bound}{\LINE}\\
\LL	& 1. IF voltage error is detected THEN \manuallabel{uc1:sat:ref:boundIf}{\LINE}\\
\LL	& 2. The system resets the occupant class for airbag control to error.\manuallabel{uc1:af:reset}{\LINE}\\
\LL	& 3. The system resets the occupant class for seat belt reminder to error.\\
\LL	& 4. ABORT\manuallabel{uc1:af:abort}{\LINE}\\
\LL	& 5. ENDIF\\
\LL	& Postcondition: The occupant classes have been reset to error.\\

\LL 	& \textbf{1.4 Specific Alternative Flow}\manuallabel{uc1:sat:start}{\LINE}\\
\LL	& RFS 3 \manuallabel{uc1:sat:ref}{\LINE}\\
\LL	& 1. IF some error has been qualified THEN \manuallabel{uc1:af:occ2invalid}{\LINE}\\
\LL	& 2. The system SENDS the error occupant class TO AirbagControlUnit.\\
\LL	& 3. The system SENDS the error occupant class TO SeatBeltControlUnit.\\ 
\LL	& 4. ABORT\\
\LL	& 5. ENDIF\\
\LL	& Postcondition: The error occupant class has been sent to AirbagControlUnit. The error occupant class has been sent to SeatBeltControlUnit.\manuallabel{uc1:sat:post}{\LINE}\\

\LL	& \textbf{1.5 Specific Alternative Flow}\manuallabel{uc1:sat:startTwo}{\LINE}\\
\LL	& RFS 3\\
\LL	& 1. The system SENDS the previous occupant class for airbag control TO AirbagControlUnit.\\
\LL	& 2. The system SENDS the previous occupant class for seat belt reminder TO SeatBeltControlUnit.\\
\LL	& 3. ABORT\\
\LL	& Postcondition: The previous occupant class for airbag control has been sent to AirbagControlUnit. The previous occupant class for seat bet reminder has been set to SeatBeltControlUnit.\manuallabel{uc1:sat:postTwo}{\LINE}\\

\LL	& \textbf{2. Use Case} Self Diagnosis\\
\LL	& \textbf{2.1 Precondition}\\
\LL	& The system has been initialized.\\
\LL 	& \textbf{2.2 Basic Flow}\\
\LL 	& 1. The system sets temperature errors to not detected. \manuallabel{uc2:af:setTempErrors}{\LINE}\\
\LL 	& 2. The system sets memory errors to not detected. \manuallabel{uc2:af:setMemoryErrors}{\LINE}\\
\LL 	& 3. The system VALIDATES THAT the NVM is accessible. \manuallabel{uc2:sd:memory}{\LINE}\\
\LL 	& 4. The system REQUESTS the temperature FROM the SeatSensor. \manuallabel{uc2:requestTemperature}{\LINE} \\
\LL 	& 5. The system VALIDATES THAT the temperature is above -10 degrees. \manuallabel{uc2:TCOND}{\LINE}\\
\LL 	& 6. The system VALIDATES THAT the temperature is below 50 degrees. \manuallabel{uc2:bf:internal}{\LINE}\\
\LL 	& 7. The system sets self diagnosis as completed. \\
\LL	& Postcondition: Error conditions have been examined.\\
\LL	& \textbf{2.3 Specific Alternative Flow} \manuallabel{uc2:af3N:start}{\LINE}\\
\LL	& RFS 3\\
\LL 	& 1. The System sets MemoryError to detected. \manuallabel{uc2:af3N:resetMemoryError}{\LINE}\\
\LL 	& 2. RESUME STEP 4 \manuallabel{uc2:af3N:resume1}{\LINE}\\
\LL	& Postcondition: The system has detected a MemoryError. \manuallabel{uc2:af3N:post}{\LINE}\\
\LL	& \textbf{2.4 Specific Alternative Flow} \manuallabel{uc2:af3:start}{\LINE}\\
\LL	& RFS 5\\
\LL 	& 1. The System sets TemperatureLowError to detected. \manuallabel{uc2:af:resetTempLowError}{\LINE}\\
\LL 	& 2. RESUME STEP 7 \manuallabel{uc2:af:resume1}{\LINE}\\
\LL	& Postcondition: The system has detected a TemperatureLowError. \manuallabel{uc2:af3:post}{\LINE}\\
\LL	& \textbf{2.5 Specific Alternative Flow}\\
\LL	& RFS 6\\
\LL 	& 1. The System sets TemperatureHighError to detected. \\
\LL 	& 2. RESUME STEP 7 \manuallabel{uc2:af:resume2}{\LINE}\\
\LL	& Postcondition: The system has detected a TemperatureHighError.\\

\LL	& \textbf{3. Use Case} Classify Occupancy Status\\
\LL	& \textbf{3.1 Precondition}\\
\LL	& The system has been initialized.\\
\LL 	& \textbf{3.2 Basic Flow} \manuallabel{uc3:bf}{\LINE}\\
\LL 	& 1. The system sets the occupant class for airbag control to Init. \\
\LL 	& 2. The system sets the occupant class for seatbelt reminder to Init. \\
\LL 	& 4. The system VALIDATES THAT the capacitance is above 600. \manuallabel{uc3:bf:capacitance}{\LINE} \\
\LL 	& 5. The system sets the occupant class for airbag control to Occupied. \manuallabel{uc3:bf:setAirbagOccupied}{\LINE} \\
\LL 	& 6. The system sets the occupant class for seatbelt reminder to Occupied. \\
\LL	& Postcondition: An adult has been detected on the seat. \manuallabel{uc3:bf:post}{\LINE} \\
\LL	& \textbf{3.3 Specific Alternative Flow} \manuallabel{uc3:af:ifthen}{\LINE} \\
\LL	& RFS 4\\
\LL 	& 1. IF capacitance is above 200 THEN\\
\LL 	& 2. The system sets the occupant class for airbag control to Empty. \\
\LL 	& 3. The system sets the occupant class for seatbelt reminder to Occupied. \\
\LL 	& 4. EXIT \manuallabel{uc3:af:exit}{\LINE}\\
\LL 	& 5. ENDIF\\
\LL	& Postcondition: A child has been detected on the seat.\\
\LL	& \textbf{3.4 Specific Alternative Flow}\\
\LL	& RFS 4\\
\LL 	& 1. The system sets the occupant class for airbag control to Empty. \\
\LL 	& 2. The system sets the occupant class for seatbelt reminder to Empty. \\
\LL	& 3. EXIT \\
\LL	& Postcondition: The seat has been recognized as being empty.\\
\end{tabular}
\label{useCaseInitialOccupancy}
\label{useCaseClassify}
\end{table}%

The keyword \textit{VALIDATES THAT} indicates a condition that must be true to take the next step; otherwise an alternative flow is taken. %
\MREVISION{R1.9}{For example, the condition in Line \ref{uc1:bf:validate} indicates that no error condition should be detected or qualified to proceed with the step in Line \ref{uc1:bf:includeClassify} and, otherwise, the alternative flow in Line \ref{uc1:sat:start} is taken.
In \BodySense, an error is qualified (i.e., confirmed), when it remains detected for 3400 ms.}

Alternative flows describe other scenarios than the main one, both success and failure. An alternative flow always depends on a condition. %
In RUCM, there are three types of alternative flows: \textit{specific}, \textit{bounded} and \textit{global}. 
For specific and bounded alternative flows, the keyword \textit{RFS} is used~to refer to one or more reference flow steps (e.g., Lines \ref{uc1:sat:ref} and \ref{uc1:sat:ref:bound}).
A specific alternative flow refers to a step in its reference flow (Line \ref{uc1:sat:ref}). A bounded alternative flow refers to more than one step in the reference flow (Line \ref{uc1:sat:ref:bound}) while a global alternative flow refers to any step in the reference flow.

Bounded and global alternative flows begin with the keyword \textit{IF .. THEN} for the condition under which the alternative flow is taken (Line \ref{uc1:sat:ref:boundIf}). Specific alternative flows do not necessarily begin with \textit{IF .. THEN} since a condition may already be indicated in its reference flow step (Line \ref{uc1:bf:validate}). 
In Line~\ref{uc3:af:ifthen}, we have an example of a specific alternative flow beginning with an \textit{IF .. THEN} condition.
The alternative flows are evaluated in the order they appear in the specification.   

UMTG  introduces  extensions  into  RUCM regarding  the IF conditions,  the  keyword EXIT,  and  the  way  input/output messages are expressed
\MREVISION{R1.10}{\M{} prevents the definition of use case flows containing branches~\cite{Larman-Applying-2002}}, thus enforcing the adoption of \textit{IF} conditions only as a means to specify guard conditions for alternative flows. \M{} introduces the keywords \textit{SENDS ... TO} and \textit{REQUESTS ... FROM} for the system-actor interactions. Depending on the subject of the sentence, the former indicates either that an actor provides an input to the system (Line \ref{uc1:bf:input}) or that the system provides an output to an actor (Line \ref{uc1:bf:output}). The latter is used only for inputs, and indicates that the input provided by the actor has been requested by the system (Line~\ref{uc2:requestTemperature}). 
\UMTG introduces the keyword \emph{EXIT} to indicate use case termination under alternative valid execution conditions (Line~\ref{uc3:af:exit} describing the case of a child being detected on a seat). The keyword \textit{EXIT} complements the keyword \textit{ABORT}, which is used to indicate the abnormal use case termination (Line~\ref{uc1:af:abort}).

\section{\CHANGED{Tagging of Use Case Specifications}}
\label{sec:nlpPipeline}

We implemented an NLP application to extract the information required for three \M{} steps in Fig.~\ref{fig:approach}: \textit{evaluate the model completeness} (Step 3 in Fig.~\ref{fig:approach}), \textit{generate OCL constraints} (Step 5), and \textit{generate the use case test model} (Step 7). %
\MREVISION{R1.12}{This application annotates the phrases in the use case specification sentences to enable the generation of use case models and to identify the steps for which an OCL constraint should be generated. It does not include the procedures for generating OCL constraints (i.e., SRL and semantic similarity detection) because these are integrated in a dedicated algorithm described in Section~\ref{sec:oclGeneration}}.

\begin{figure}[h]
          \includegraphics{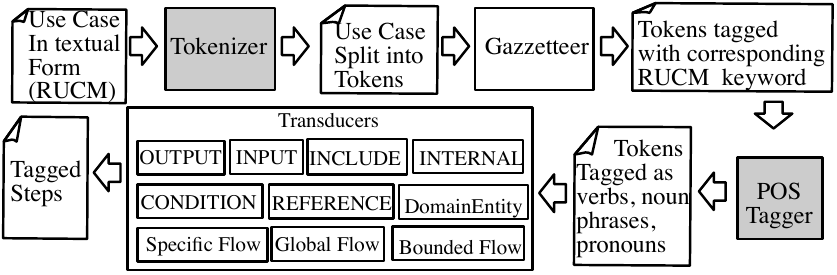}
    \caption{NLP pipeline applied to extract data used by \UMTG Steps 3, 5, and 7.}
      \label{fig:NLPpipeline}
\end{figure}

The NLP application is based on the GATE workbench~\cite{Cunningham2002}, an open source NLP framework, and implements the NLP pipeline in Fig.~\ref{fig:NLPpipeline}. The pipeline includes both default NLP components (grey) and components built to process use case specifications in RUCM (white). 
The \emph{Tokenizer} splits the use cases into tokens. The \emph{Gazetteer} identifies 
the RUCM keywords. 
\MREVISION{R1.14}{For example, according to RUCM, the system under test is expected to be referred to with the keyword \emph{The system}.}
The \emph{POS Tagger} tags tokens according to their nature: \textit{verb}, \textit{noun}, and 
\textit{pronoun}. 
The pipeline is terminated by a set of \emph{transducers} that tag blocks of words with additional information required by the three \M{} steps. 
The transducers integrated in \UMTG (1) identify the kinds of 
RUCM steps (i.e., output, input, include, condition and internal steps), (2)
 distinguish alternative flows, and (3) detect RUCM references (i.e., the RFS keyword), conditions, and domain entities in the use case steps.

\begin{figure}[t]
              \includegraphics{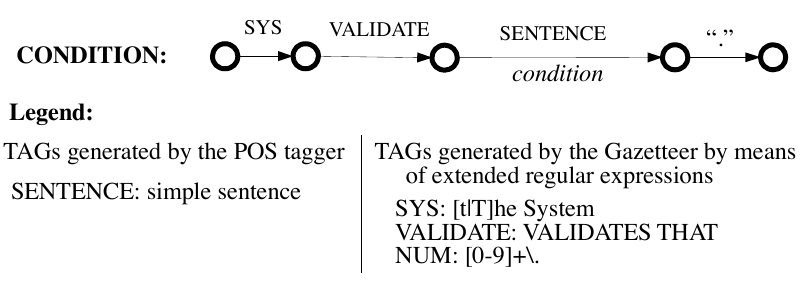}
      \caption{Part of the transducer that identifies conditions.}
      \label{fig:transExample}
\end{figure}

\begin{figure}[t]
    \includegraphics{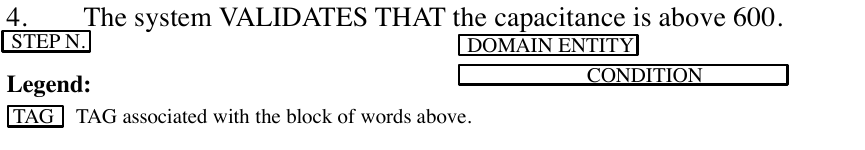}
      \caption{Tags associated with the use case step in Line~\ref{uc3:bf:capacitance} of Table~\ref{useCaseInitialOccupancy}.}
      \label{fig:tagsExample}
\end{figure}

Fig.~\ref{fig:transExample} gives an example transducer for condition steps. The arrow labels in higher case represent the transducer's inputs, i.e., tags previously identified by the POS tagger, the gazzetteer or other transducers. The italic labels show the tags assigned by the transducer to the words representing the transducer's input. Fig.~\ref{fig:tagsExample} gives the tags associated with the use case step in Line \ref{uc3:bf:capacitance} of Table~\ref{useCaseInitialOccupancy} after the execution of the transducer in Fig.~\ref{fig:transExample}. In Fig.~\ref{fig:tagsExample}, multiple tags are assigned to the same blocks of words. For example, the noun phrase `\emph{the capacitance}' is tagged both as a \emph{domain entity} and as part of a \emph{condition}.

\begin{table}[t]
\center
\scriptsize
\caption{List of UMTG steps in Fig~\ref{fig:approach} that process the tags generated by transducers.}
\begin{tabular}{|p{3cm}|p{4cm}|}
\hline
\textbf{Transducer Tag}&\textbf{UMTG Step(s)}\\
\hline
OUTPUT STEP& Step 7.\\
INPUT STEP& Step 7.\\
INCLUDE STEP& Step 7.\\
CONDITION STEP& Step 5, Step 7.\\
INTERNAL STEP& Step 7.\\
BASIC FLOW& Step 7.\\
ALTERNATIVE FLOW& Step 7.\\
RFS& Step 7.\\
CONDITION& Step 5.\\
DOMAIN ENTITY& Step 3, Step 7.\\
\hline
\end{tabular}
\label{table:tags}
\end{table}%

\MREVISION{R1.15}{The tags generated by transducers are used in Steps 3, 5, and 7 of Fig.~\ref{fig:approach}. 
Table~\ref{table:tags} provides, for each generated tag, %
the list of UMTG steps that process the associated phrases.
For example, the noun phrases annotated with the tag \emph{domain entity} are used in Step 3 to determine if the domain model is complete and, in Step 7, to identify inputs. Further details are provided in the following sections.}

\section{Evaluation of the Domain Model Completeness}
\label{sec:completeness}
The completeness of the domain model is important to generate correct and complete test inputs. 
\M{} automatically identifies missing domain entities to evaluate the model completeness (Step 3 in Fig.~\ref{fig:approach}). This is done by checking correspondences between the domain model elements and the domain entities identified by the NLP application. %

\begin{figure}[h]
          \includegraphics{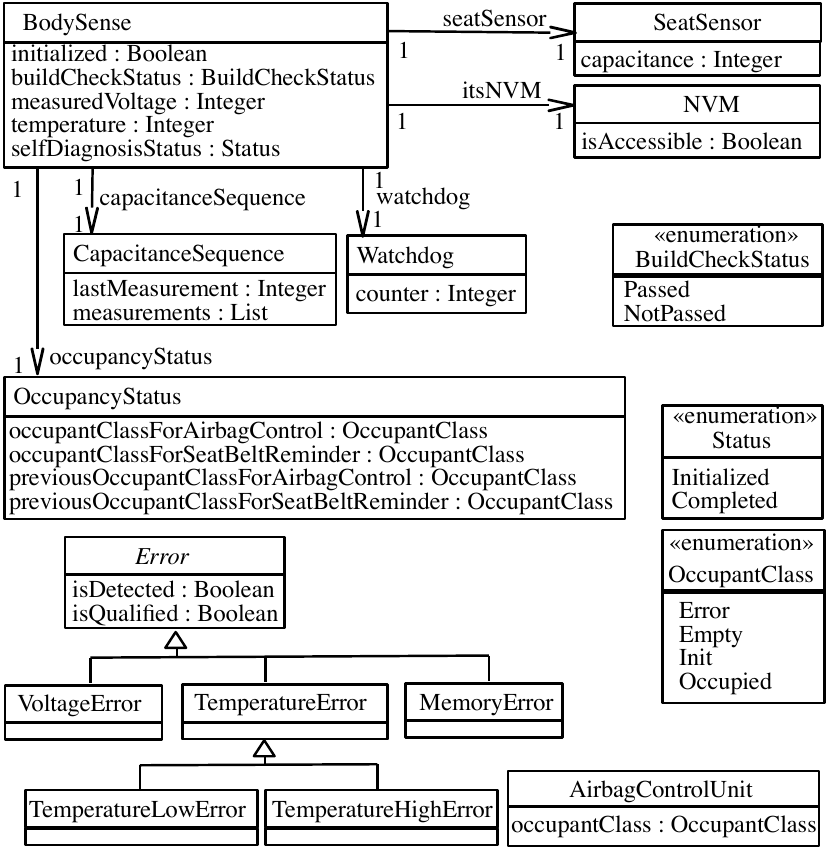}
      \caption{Part of the domain model for $BodySense$.}
      \label{fig:domainModel}
\end{figure}

Domain entities in a use case may not be modelled as classes but as attributes.
Fig.~\ref{fig:domainModel} shows a simplified excerpt of the domain model for \emph{BodySense} where 
the domain entities `\emph{occupant class for airbag control}' and `\emph{occupant class for seat belt reminder}' 
are modelled as attributes of the class \emph{OccupancyStatus}. \M{} follows a simple yet effective solution to check entity and attribute names. For each domain entity identified through NLP, \M{} generates an entity name by removing all white spaces and by putting all first letters following white spaces in capital. For instance, the domain entity `\emph{occupant class for airbag control}' becomes `\emph{OccupantClassForAirbagControl}'. \M{} checks the string similarity between the generated entity names and the domain model elements. Engineers are asked to correct their domain model and use case specifications in the presence of confirmed mismatches.

\section{Generation of OCL Constraints}
\label{sec:oclGeneration}

To identify test inputs via constraint solving, \M{} needs to derive OCL constraints that capture %
(1) the effect that the internal steps have on the system state (i.e., the postconditions of the internal steps), (2) the use case preconditions, and (3) the conditions in the condition steps. 
For instance, for the basic flow of the use case \emph{Classify Occupancy Status} (Lines \ref{uc3:bf} to \ref{uc3:bf:post} in Table~\ref{useCaseClassify}), we need a test input that satisfies the condition \emph{`the capacitance is above 600'} (Line~\ref{uc3:bf:capacitance}). %

As part of \M{}, we automate the generation of OCL constraints (Step 5 in Fig.~\ref{fig:approach}). Using some predefined constraint generation rules (hereafter transformation rules),
\M{} automatically generates an OCL constraint for each precondition, internal step and condition step identified by the transducers in Fig.~\ref{fig:NLPpipeline}. 
The generated constraint captures the meaning of the NL sentence in terms of the concepts in the domain model. 
Table~\ref{table:oclExamples} shows some of the OCL constraints generated from %
the use case specifications in our case studies in Section~\ref{sec:evaluation}. %

Section~\ref{subsec:assumptions} summarizes our assumptions for the generation of OCL constraints. Section~\ref{subsection:ocl_algorithm} describes the constraint generation algorithm. In Section~\ref{subsec:gen}, we discuss the correctness and generalizability of the constraint generation. %

\begin{table*}[ht!]
\scriptsize
\caption{Some constraints from the BodySense and \HOD case studies in Section~\ref{sec:evaluation}, with tags generated by SRL.}
\label{table:oclExamples}
\begin{tabular}{|@{\hspace{0.02cm}}p{0.2cm}|@{\hspace{0.05cm}}p{7.8cm} |@{\hspace{0.05cm}}p{9.2cm} |}
\hline
\#&\textbf{Sentence with SRL tags}&\textbf{Corresponding OCL Constraint}\\
\hline
\SNA&
\MMREVISION{R1.18}{\{The system VALIDATES THAT\}$_{\mathit{ignored}}$}& 
$BodySense.allInstances() \rightarrow forAll(b|b.seatsensor.capacitance > 600)$\\
&\{the capacitance\}$_{A1}$ \{is\}$_{verb}$ \{above 600\}$_{AM-LOC}$&\\
\hline
\SA&
\{The system\}$_{A0}$ \{sets\}$_{verb}$ \{the occupant class for airbag& 
$BodySense.allInstances() \rightarrow forAll(b|b.occupancyStatus.occupant\-$\\
&control\}$_{A1}$ \{to Init\}$_{A2}$&\hspace{1cm}$ClassForAirbagControl = OccupantClass::Init)$\\
\hline
\SB&
\{The system VALIDATES THAT\}$_{\mathit{ignored}}$ \{the NVM\}$_{A1}$ \{is\}$_{verb}$ \{accessible\}$_{\mathit{AM-PRD}}$&
$BodySense.allInstances() \rightarrow forAll( i | i.itsNVM.isAccessible = true ) $\\
\hline
\SC&
\{The system\}$_{A0}$ \{sets\}$_{verb}$ \{temperature errors\}$_{A1}$ \{to detected\}$_{A2}$&
$TemperatureError.allInstances() \rightarrow forAll( i | i.isDetected = true ) $\\
\hline
\SE&
\{The system VALIDATES THAT\}$_{\mathit{ignored}}$ \{the build check\}$_{A1}$ \{has been passed\}$_{verb}$&
$BodySense.allInstances() \rightarrow forAll( i | i.buildCheckStatus = BuildCheckStatus::Passed)$\\
\hline
\SF&
\{The system VALIDATES THAT\}$_{\mathit{ignored}}$ \{no\}$_{\mathit{NOM-NEG}}$ &
\MMREVISION{R1.17}{}$\mathit{Errors.allInstances()}\rightarrow \mathit{select(} e | \CHANGED{\ \mathit{not}\ } \mathit{e.typeOf(VoltageError)} \mathit{and} \CHANGED{\ \mathit{not}\ }$ \\
&\{error & $\mathit{e.typeOf} \mathit{(Memory}\mathit{Error)}  ) \rightarrow forAll( i | i.isDetected = false )$\\
&\MMREVISION{R1.18}{[(except voltage errors)$_{\mathit{NOM}_{\mathit{NP}}}$ and (memory errors)$_{\mathit{NOM}_{\mathit{NP}}}$]$_{\mathit{NOM-A2}}$}&\\
&\}$_{A1}$ \{is detected\}$_{verb}$&\\
\hline
\SG&
\{The system\}$_{A0}$ \{erases\}$_{verb}$ \{the measured voltage\}$_{A1}$ &
$\mathit{BodySense.allInstances()}\rightarrow forAll( i | \mathit{i.measuredVoltage} = 0)$\\
\hline
\SH&
\{The system\}$_{A0}$ \{erases\}$_{verb}$ \{the occupant class\}$_{A1}$ \{from the &
$\mathit{AirbagControlUnit.allInstances()}\rightarrow forAll( i | \mathit{i.occupantClass} = null)$\\
&
airbag control unit\}$_{A2}$ &\\
\hline
\SL&
\{The system\}$_{A0}$ \{disqualifies\}$_{verb}$ \{temperature errors\}$_{A1}$ &
$\mathit{TemperatureError.allInstances()}\rightarrow \mathit{forAll(} i | \mathit{i.isQualified} \mathit{!=} \, true)$\\

\hline
\SM&
\{IF\}$_{ignored}$  \{some error\}$_{A1}$ \{has been qualified\}$_{\mathit{verb}}$.&
$\mathit{Error.allInstances()}\rightarrow \mathit{exists(} i | \mathit{i.isQualified} \mathit{=} \, true)$\\

\hline
\SN&
\{The system VALIDATES THAT\}$_{\mathit{ignored}}$  \{the driver\}$_{A1}$ \{put\}$_{\mathit{verb}}$ \{two hands\}$_{\mathit{A1}}$ \{on the steering wheel\}$_{\mathit{AM-LOC}}$.&
$\mathit{Hand.allInstances()}\rightarrow \mathit{select(} i | \mathit{i.onTheSteeringWheel} \mathit{=} \, \mathit{true})\rightarrow \mathit{size()} = 2$\\
\hline

\SO&
\{The system\}$_{A0}$ \{resets\}$_{verb}$ \{the counter of the watchdog\}$_{A1}$ &
$\mathit{BodySense.allInstances()}\rightarrow forAll( i | \mathit{i.watchdog.counter} = 0)$\\
\hline

\SP&
\{The system\}$_{A0}$ \{resets\}$_{verb}$ \{the watchdog counter\}$_{A1}$ &
$\mathit{BodySense.allInstances()}\rightarrow forAll( i | \mathit{i.watchdog.counter} = 0)$\\
\hline

\SQ&
\{The system VALIDATES THAT\}$_{\mathit{ignored}}$ \{the last measurement of the&
$\mathit{BodySense.allInstances()}\rightarrow forAll( i | \mathit{i.capacitanceSequence}$\\
& capacitance sequence\}$_{A1}$  \{is\}$_{verb}$ \{above 40\}$_{AM-LOC}$&$.\mathit{lastMeasurement} = 0)$\\
\hline

\end{tabular}
\\
\\
\MMREVISION{R1.18,20}{\textbf{Notes.} Sentences S1, S2, S8, S14 are taken from the \BodySense case study system, sentence S11 from \HOD; the other sentences are shared by both. In S6, different types of brackets are used to annotate phrases with nested semantic roles.}
\end{table*}%

\subsection{Working Assumptions} %
\label{subsec:assumptions}

The constraint generation is enabled by three assumptions. %

\emph{Assumption 1 (Domain Modeling).} There are domain modeling practices common for embedded systems:
\begin{enumerate}
\item[\PTWO] Most of the entities in the use case specifications are given as classes in the domain model.
\item[\PTHREE] The names of the attributes and associations in the domain model are usually similar with the phrases in the use case specifications.
\item[\PSIX] The attributes of domain entities (e.g., \emph{Watchdog.counter} in Fig.~\ref{fig:domainModel}) are often specified by possessive phrases (i.e., genitives and of-phrases such as \emph{of the watchdog} in \SO~in Table~\ref{table:oclExamples}) and attributive phrases (e.g., \emph{watchdog} in \SP) in the use case specifications.
\item[\PONE] The domain model often includes a system class with attributes that capture the system state (e.g., \emph{BodySense} in Fig.~\ref{fig:domainModel}). 
\item[\PFOUR] Additional domain model classes are introduced to group concepts that are modelled using attributes.
\item[\PFIVE] Discrete states of domain entities are often captured using either boolean attributes (e.g., \emph{isAccessible} in Fig.~\ref{fig:domainModel}), or attributes of enumeration types (e.g., \emph{BuildCheckStatus::Passed} in Fig.~\ref{fig:domainModel}). %
\end{enumerate}

To ensure that Assumption 1 holds, \M{} iteratively asks engineers to correct their models (see Section~\ref{sec:completeness}).
With Assumption 1, we can rely on string \MREVISION{R2.4}{syntactic similarity} to select the terms in the OCL constraints (i.e., classes and attributes in the domain model) based on the phrases appearing in the use case steps. 
String similarity also allows for some degree of flexibility in naming conventions.

\emph{Assumption 2 (OCL constraint pattern).} The conditions in the use case specifications of embedded systems are typically simple and capture information about the state of one or more domain entities (i.e., classes in the domain model). For instance, in \emph{BodySense}, the preconditions and condition steps describe safety checks ensuring that the environment has been properly set up (e.g., \SB{} in Table~\ref{table:oclExamples}), or that the system input has been properly obtained (e.g., \SE), while the internal steps describe updates on the system state (e.g., \SA).
They can be expressed in OCL using the pattern in Fig.~\ref{fig:bnf}, which captures assignments, equalities, and inequalities. 

The generated constraints include an entity name (\texttt{ENTITY} in Fig.~\ref{fig:bnf}), an optional selection part (\texttt{SELECTION}), and a query element (\texttt{QUERY}). 
The query element can be specified according to three distinct sub-patterns: \texttt{FORALL}, \texttt{EXISTS} and  \texttt{COUNT}. 
\texttt{FORALL} specifies that a certain expression (i.e., \texttt{EXPRESSION}) should hold for all the instances \texttt{i} of the given entity; 
\texttt{EXISTS} indicates that the expression should hold for at least one of the instances. \texttt{COUNT} is used when the expression should hold for a certain number of instances.
Examples of these three query elements are given in the OCL constraints generated for the sentences \SC, \SM, and \SN{} in Table~\ref{table:oclExamples}, respectively.
The pattern \texttt{EXPRESSION} contains a left-hand side variable (hereafter \emph{lhs-variable}), an OCL operator, and a right-hand side term (hereafter \emph{rhs-term}), which is either another variable or a literal. The lhs-variable indicates an attribute of the entity whose state is captured by the constraint, while the rhs-term captures the state information (e.g., the value expected to be assigned to an attribute). 
The optional selection part selects a subset of all the available instances of the given entity type based on their subtype; an example is given in the OCL constraint for \SF{} in Table~\ref{table:oclExamples}.

\begin{figure}
\scriptsize
\textbf{CONSTRAINT} = \textbf{ENTITY} .allInstances() \textbf{[SELECTION]} \textbf{QUERY}\\
\textbf{QUERY} = \textbf{FORALL} $|$ \textbf{EXISTS} $|$ \textbf{COUNT}\\
\textbf{FORALL} = $\rightarrow$ forAll ( \textbf{EXPRESSION} )\\
\textbf{EXISTS} = $\rightarrow$ exists ( \textbf{EXPRESSION} )\\
\textbf{COUNT} = $\rightarrow$ select ( \textbf{EXPRESSION} ) $\rightarrow$ size() \textbf{OPERATOR} \textbf{NUMBER}\\
\textbf{EXPRESSION} = i $|$ i.\textbf{LHS-VARIABLE} \textbf{OPERATOR} \textbf{RHS-TERM} \\
\textbf{LHS-VARIABLE} = \textbf{VARIABLE}\\
\textbf{RHS-TERM} =  \textbf{VARIABLE} \textbf{$|$} \textbf{LITERAL}\\
\textbf{VARIABLE} = \textbf{ATTR} \textbf{$|$} \textbf{ASSOC}  \textbf{\{} .~\textbf{ATTR} \textbf{$|$} \textbf{ASSOC} \textbf{\}}\\
\textbf{SELECTION} = $\rightarrow$ select( e $|$  \textbf{TYPESEL} \{ and \textbf{TYPESEL} \} )\\
\textbf{TYPESEL} = not e.TypeOf( \textbf{CLASS} ) \\

\textbf{Note:} This pattern is expressed using a simplified EBNF grammar~\cite{Ledgard:1980} where non-terminals are bold and terminals are not bold. \textbf{ENTITY} stands for a class in the domain model. \textbf{LITERAL} is an OCL literal (e.g., '1' or 'a'). \textbf{NUMBER} is an OCL numeric literal (e.g., '1'). \textbf{ATTR} is an attribute of a class in the domain model. \textbf{ASSOC} is an association end in the domain model. \textbf{OPERATOR} is a math operator ( -, +,  =, $\textless$, $\le$, $\ge$, $\textgreater$).
\caption{Pattern of the OCL constraints generated by \M{}.}
\label{fig:bnf}
\end{figure}

\emph{Assumption 3 (SRL).} The SRL toolset (the \CNP~tool in our implementation)  identifies all the semantic roles in a sentence that are needed to correctly generate an OCL constraint. %
Our transformation rules use the roles to correctly select the domain model elements to be used in the OCL constraint (see Section~\ref{subsection:ocl_algorithm}).

Table~\ref{table:oclExamples} reports some use case steps, the SRL roles, %
and the generated constraints. For example, in \SA, the \CNP~tool tags \textit{``The system''} with \emph{A0} (i.e., the actor performing the action), \textit{``sets''} with verb, \textit{``the occupant class for airbag control''} as \emph{A1}, and \textit{``to Init''} with \emph{A2} (i.e., the final state). We ignore the prefix \emph{``The system VALIDATES THAT''} in condition steps because it is not necessary to generate OCL constraints.

\subsection{The OCL Generation Algorithm}
\label{subsection:ocl_algorithm}

We devised an algorithm that generates an OCL constraint from a given sentence in NL (see Fig.~\ref{fig:alg:OCL}). %
We first execute the SRL toolset (the \CNP~tool) to annotate the sentence with the SRL roles (Line~\ref{algo:OCL:srl} in Fig.~\ref{fig:alg:OCL}). We select and apply the transformation rules based on the verb in the sentence (Lines~\ref{algo:OCL:select} to~\ref{algo:OCL:execute}). The same rule is applied for all the verbs that are synonyms and that belong to the same VerbNet class. %
In addition, we have a special rule, hereafter we call \emph{any-verb transformation rule}, that is shared by many verb classes.
Each rule returns a candidate OCL constraint 
with a score assessing how plausible is the constraint (Section~\ref{subsec:scoring}). %
We select the constraint with the highest score (Line~\ref{algo:OCL:returnBestOCL}).

\begin{figure}[t]
\begin{algorithmic}[1]
\scriptsize
\Require $S_{NL}$, a sentence in natural language
\Require $domainModel$, a domain model
\Ensure $\langle{}ocl,score\rangle{}$, an OCL constraint with a score

\Function{generateOcl}{$S_{NL}$, domainModel} 
\State $\mathit{S_{srl}} \gets $generate a sentence annotated with SRL roles from $S_{NL}$\label{algo:OCL:srl}
\State $\mathit{transformationRules} \gets $identify the transformation rules to apply \label{algo:OCL:select}
\State \hspace{2cm} based on the verb in $\mathit{S_{srl}}$
\For {\textbf{each}\ $\mathit{transformationRule}$}  \label{algo:OCL:For}
\State 	generate a new OCL constraint ($\mathit{constraint}$) by applying \\
\hspace{2cm} the $\mathit{transformationRule}$ \label{algo:OCL:execute}
\State 	$\mathit{constraints} \gets \mathit{constraints}  \cup  \mathit{constraint}$ \label{algo:OCL:list}
\EndFor \label{algo:OCL:ForEnd}
\State \Return the ocl constraint with the best score\label{algo:OCL:returnBestOCL}
\EndFunction

\end{algorithmic}
\caption{The OCL constraint generation algorithm.}
\label{fig:alg:OCL}
\end{figure}

For each verb, we classify the SRL roles into: (i) \emph{entity role} indicating the entity whose state needs to be captured by the constraint, and (ii) \emph{support roles} indicating additional information such as literals in the rhs-terms (see Fig.~\ref{fig:bnf}).  %
We implemented our transformation rules according to the pairs $\mathit{\langle}$\emph{entity role}, \emph{\{support roles\}}$\mathit{\rangle}$ we extracted from the VerbNet role patterns. The role patterns provide 
all valid role combinations appearing with a verb. 

Table~\ref{additionalVerbConfigs} shows the role pairs for some of our transformation rules. %
For example, the verb `to erase' has two VerbNet role patterns $\mathit{\langle}$\emph{A0}, \emph{V}, \emph{A1}$\mathit{\rangle}$ and $\mathit{\langle}$\emph{A0}, \emph{V}, \emph{A1}, \emph{A2}$\mathit{\rangle}$ where \emph{V} is the verb, and \emph{A0}, \emph{A1} and \emph{A2} are the SRL roles. The first pattern represents the case in which an object is erased (e.g., the measured voltage in \SG~in Table~\ref{table:oclExamples}), while, in the second one, an object is removed from a source (e.g., the occupant class being removed from the airbag control unit in 
\SH). %
The transformation rule for the verb `to erase' has thus two role pairs: $\mathit{\langle}$\emph{A1}, \emph{null}$\mathit{\rangle}$ and $\mathit{\langle}$\emph{A2}, \emph{\{A1\}}$\mathit{\rangle}$ 
(see Rule 4 in Table~\ref{additionalVerbConfigs}).  
Each transformation rule might be associated with multiple support roles; this is the case of the verb `to set' whose role pair 
$\mathit{\langle}$\emph{A1}, \emph{\{A2, AM-LOC\}}$\mathit{\rangle}$ appears in Rule 3 in Table~\ref{additionalVerbConfigs}.

Each transformation rule performs the same sequence of activities for each pair $\mathit{\langle}$\emph{entity role}, \emph{\{support role\}}$\mathit{\rangle}$ (see Fig.~\ref{fig:alg:transform}).
A rule first identifies the candidate lhs-variables 
(Line~\ref{algo:tranform:lhss} in Fig.~\ref{fig:alg:transform}), and
then builds a distinct OCL constraint for each lhs-variable identified (Lines \ref{algo:tranform:lhsFor} to \ref{algo:tranform:lhsForEnd}).
Finally, it returns the OCL constraint with the highest score (Line \ref{algo:tranform:returnBestOCL}).

\begin{figure}[t]
\begin{algorithmic}[1]
\scriptsize
\Require $\mathit{srl}$, a sentence annotated with the different roles identified by SRL
\Require $\mathit{systemClass}$, the main class of the system
\Require $\mathit{rolesSetPairs}$, pairs of roles sets $\langle{}$$\mathit{entity\ role}, \{\mathit{support\ role}\}$$\rangle{}$
\Ensure $\langle{}ocl,score\rangle{}$, an OCL constraint with a score

\Function{transform}{srl, systemClass, rolesSetPairs} 
\For {\textbf{each}$\ \mathit{rolesSetPairs} $ }  \label{algo:tranform:forRolesSets}
\State $\mathit{lhsVariables} \gets $process\ $\mathit{srl}$\ and\ identify\ a\ set\ of\ variables\ that\ might \label{algo:tranform:lhss}
\State \hspace{1.9cm}appear in the\ left-hand side\ of\ the\ OCL\ constraint
\For {\textbf{each}\ $\mathit{LHS\ in\ \mathit{lhsVariables}\ }$}  \label{algo:tranform:lhsFor}
\State 	$\mathit{RHS} \gets$ identify\ the\ term\ to\ put\ on\ the\ right-hand side \label{algo:tranform:rhs}
\State 	$\mathit{OP} \gets $identify the operator to use in the OCL constraint \label{algo:tranform:ops}
\State        $\mathit{SEL} \gets $if needed, build a subexpression with the selection operator \label{algo:tranform:opSEL}
\State        $\mathit{QUERY} \gets $identify the type of QUERY element \label{algo:tranform:QUERY}
\If { $\mathit{RHS} \ne \mathit{null} \ \mathit{and} \  \mathit{OP} \ne \mathit{null}$ } \label{algo:tranform:checkNull}
\State $\mathit{ocl} \gets $build the constraint using $\mathit{LHS}$, $\mathit{SEL}$, $\mathit{OP}$, $\mathit{RHS}$, and $\mathit{QUERY}$\label{algo:tranform:build}
\State $\mathit{score} \gets $calculate the score of the OCL constraint\label{algo:tranform:scoreCall}
\State $\mathit{ocls} \gets \mathit{ocls}\ \cup \{\langle{}ocl,score\rangle{}\}$
\EndIf \label{algo:tranform:checkNullEnd}
\EndFor \label{algo:tranform:lhsForEnd}
\EndFor
\State $\mathit{bestOcl} \gets$ select the constraint with the best score from the list $\mathit{ocls}$ \label{algo:tranform:findBestOCL}
\State \Return $\mathit{bestOcl}$\label{algo:tranform:returnBestOCL}
\EndFunction

\end{algorithmic}
\caption{The algorithm followed by each transformation rule.}
\label{fig:alg:transform}
\end{figure}

In Sections~\ref{subsec:lhs}~to~\ref{subsec:scoring}, we give the details of the algorithm in Fig.~\ref{fig:alg:transform}, i.e., identifying the lhs-variables (Line~\ref{algo:tranform:lhss}), selecting the rhs-terms (Line~\ref{algo:tranform:rhs}) and the OCL operators (Line~\ref{algo:tranform:ops}), identifying the types of OCL query elements (Line~\ref{algo:tranform:QUERY}), and scoring the constraints (Line~\ref{algo:tranform:scoreCall}). 
\MREVISION{R1.26}{In this paper, we only provide the pseudocode of the function to identify lhs-variables, as it is the most complex one (Fig.~\ref{fig:alg:findVar}). 
Interested readers can freely access the source code of the \UMTG OCL generation component~\cite{UMTGWeb}.}

\subsubsection{Identification of the Left-hand Side Variables}
\label{subsec:lhs}

\begin{figure}[tb]
\begin{algorithmic}[1]
\scriptsize

\Require $srl$, a sentence annotated with the different roles identified by SRL
\Require $systemClass$, the main class of the system
\Require $EntityRoles$, list of entity roles
\Require $SupportRoles$, list of support roles
\Ensure $Vars$, list of left-hand side variables

\Function{findVariables}{srl, systemClass, EntityRoles, SupportRoles} 
\For {\textbf{each}\ $\mathit{ER\ in\ \mathit{EntityRoles}\ }$} \label{algo:tranform:foreach} 
\State $\mathit{term_{ER}} \gets \mathit{preprocess(srl.get(ER))}$ \label{algo:tranform:preprocess}
\State $\mathit{attrs} \gets \mathit{findAttributes(systemClass,term_{ER})}$ \label{algo:tranform:findAttr}
\State $\mathit{Vars} \gets \mathit{Vars} \cup \mathit{attrs}$ \label{algo:tranform:findAttrAdd}

\State $\mathit{class} \gets \mathit{findClass(term_{ER})}$ \label{algo:tranform:findA1Class}

\If  {$\mathit{class} \ne \mathit{null}$ }  \label{algo:tranform:classIf}

\For {\textbf{each}\ $\mathit{role\ in\ \mathit{SupportRoles}\ }$}  \label{algo:tranform:addAttrRole}
\State $\mathit{attrs} \gets \mathit{findAttributes(class,role)}$ \label{algo:tranform:findAttr2}
\State $\mathit{Vars} \gets \mathit{Vars} \cup \mathit{attrs}$ \label{algo:tranform:findAltRoleAdd}
\EndFor \label{algo:tranform:addAttrRoleEnd}

\EndIf \label{algo:tranform:classIfEnd}

\EndFor

\For {\textbf{each}\ $\mathit{attr\ in\ \mathit{Vars}\ }$}  \label{algo:tranform:appendAttrRole}
\For {\MMREVISION{R1.23}{\textbf{each}\ $\mathit{role\ in\ \mathit{SupportRoles}\ }$}} 
\State $\mathit{attrs} \gets \mathit{extendByTraversingRelations(attr,role)}$ 
\State $\mathit{Vars} \gets \mathit{Vars} \cup \mathit{attrs}$ \label{algo:tranform:addAttrTravADD}
\EndFor 
\EndFor \label{algo:tranform:appendAttrRoleEnd}
\State \Return $\mathit{Vars}$
\EndFunction

\end{algorithmic}

\caption{The algorithm to identify lhs-variables.}
\label{fig:alg:findVar}

\end{figure}

To identify the lhs-variables, the transformation rules follow an algorithm %
using the string similarity between the names of the domain model elements (i.e., the classes, attributes and associations) and the phrases in the use case step tagged with the entity and support roles (see Fig.~\ref{fig:alg:findVar}). Based on Assumption 3, we expect that the phrase tagged with the entity role provides part of the information to identify the lhs-variable (e.g., \emph{itsNVM} in \SB~in Table~\ref{table:oclExamples}), %
while the phrase(s) tagged with the support role further characterize the variable (e.g., \emph{isAccessible} in \SB). %

\begin{table}[tb]
\scriptsize
\caption{Entity and support roles for some transformation rules in \UMTG.}
\begin{tabular}{| p{0.9cm} | p{1.0cm} | p{1.5cm} | p{3.5cm} | }
\hline
\textbf{Rule ID}&\textbf{Verb}&\textbf{Entity roles}&\textbf{Support roles}\\
\hline
1&to be& A1 & AM-PRD, AM-MNR, AM-LOC\\
\hline
2&to enable& A1& AM-MNR\\
\hline
3&to set& A1& A2, AM-LOC\\
\hline
\multirow{2}{*}{4}&\multirow{2}{*}{to erase}
& A1& \\
\cline{3-4}
& & A2 & A1\\
\hline
5&any verb&A1&AM-PRD,Verb\\
\hline
\end{tabular}
\label{additionalVerbConfigs}
\end{table}%

The algorithm is influenced by domain modeling practices (Assumption 1). 
Assumptions \PTWO and \PTHREE influence the criteria to select terms for the OCL constraint based on phrases in the use case step. Assumptions \PTWO~- \PFOUR %
influence the order in which noun phrases are processed. %

Based on \PTWO, a domain model class best matches a phrase when its name shows the highest similarity with the phrase. To identify the matching classes and phrases, we employ the Needleman-Wunsch string alignment algorithm~\cite{NeedWunsch}, which maximizes the matching between characters with some degree of discrepancy. 
\MREVISION{R2.5}{For example, it enables matching composite nouns in the presence of plural words (e.g., the entity \emph{TemperaturesDatabase} can be matched with the compound nouns \emph{temperatures database} and \emph{temperature databases}).}
However, we do not use the string alignment algorithm for attributes and associations because, 
in the context of embedded software development, attributes and associations often correspond to acronyms (e.g., RAM, ROM, and NVM) which are short and with a small alignment distance, but have different meanings. 
Based on \PTHREE, an attribute or association in the domain model best matches a given phrase (i) if it is a prefix or a postfix of the phrase, (ii) if it starts or ends with the phrase, or (iii) if it is a synonym or an antonym of the phrase. 
We ignore spaces and articles in the phrases. For each matching element, we compute a similarity score as the portion of matching characters.

The algorithm iterates over each phrase tagged with an entity role (Line~\ref{algo:tranform:foreach} in Fig.~\ref{fig:alg:findVar}).
\MREVISION{R1.24}{After the execution of \emph{findAttributes} within each iteration,}
based on \PTHREE and \PONE, we add to the list of the lhs-variables the system class attributes and associations that best match the phrase (Lines~\ref{algo:tranform:findAttr}-\ref{algo:tranform:findAttrAdd}). 
In \SB~in Table~\ref{table:oclExamples}, \emph{BodySense.itsNVM} is in the list of the lhs-variables because it terminates with \emph{NVM} tagged with \emph{A1} (i.e., the entity role).

Based on \PTWO and \PTHREE, we look for the domain model class that best matches the phrase tagged with the entity role (Line~\ref{algo:tranform:findA1Class}). 
Based on \PFOUR, we recursively traverse the associations starting from this class to identify the related attributes that best match the phrases tagged with the support roles (Lines~\ref{algo:tranform:addAttrRole}-\ref{algo:tranform:addAttrRoleEnd}). The matching attribute might be indirectly related to the starting class. We give a higher priority to the directly related attributes. Therefore, the score of the matching attribute is divided by the number of traversed associations (Line~\ref{algo:tranform:findAttr2}). %
We add the best matching attributes to the list of the lhs-variables (Line~\ref{algo:tranform:findAltRoleAdd}). %
For example, for \SA~in Table~\ref{table:oclExamples}, the lhs-variable \emph{BodySense.its\-Occupancy\-Sta\-tus.occupant\-Class\-For\-AirbagCon\-trol} is identified by traversing the association \emph{its\-Occupancy\-Sta\-tus} from the system class \emph{BodySense}. Its similarity score is 0.5 because one association has been traversed (\emph{its\-Occupancy\-Sta\-tus}) and there is an exact match between the attribute name and the noun phrase tagged with \emph{A1}.

We further refine the lhs-variables with a complex type (i.e., a class or a data type). For each attribute and association in the list of the lhs-variables, we traverse  the related associations to identify the attributes and associations that best match the phrases tagged with the support roles (Lines~\ref{algo:tranform:appendAttrRole} to~\ref{algo:tranform:appendAttrRoleEnd}). 
In \SB, \emph{BodySense.itsNVM} is refined to \emph{BodySense.itsNVM.isAccessible} since the class \emph{NVM} has a boolean attribute (\emph{isAccessible}) with a name similar to the phrase tagged with \emph{AM-PRD} (\emph{accessible}).

Based on \PSIX, when the phrase tagged with the entity role includes a possessive  
or attributive phrase, we look for attributes/associations in the domain model that reflect the relation between the  possessive/attributive phrase and the main phrase in the entity role phrase (e.g., \emph{the watchdog} and \emph{counter} in sentence \SO{} in Table~\ref{table:oclExamples}) .
We rely on the NomBank tags generated by \CNP{} to identify the main phrase and the possessive/attributive phrases.
In the domain model, the main phrase usually best matches an attribute/association that belongs to an \emph{owning entity}. The  owning entity is either (1) a class that best matches the possessive/attributive phrase or (2) 
a class referred by an attribute of the main system class that best matches the possessive/attributive phrase.
 For example, in the case of \SP, the attribute \emph{counter} (i.e., the main phrase in \SP) belongs to the entity class referenced by the attribute \emph{watchdog} (i.e., the possessive/attributive phrase) of the main system class.
Based on this observation, to identify the model elements that reflect the relation captured by a possessive/attributive phrase, we perform an \MREVISION{R1.25}{additional} execution of the function \emph{findVariables} where we treat possessive/attributive phrases as the entity role and the main phrase as a support role. 
The possessive/attributive phrase is thus used to identify the owning entity while the main phrase is used to identify the owned attribute/association.
In the case of \SP, \UMTG looks for an attribute of the system class that best matches \emph{watchdog} and then further refines the search by looking for a contained attribute that best matches \emph{counter}.

\subsubsection{Identification of the Right-hand Side Terms}
\label{subsec:rhsOCL}

The rhs-term can be a literal or a variable. %
It is identified based on the lhs-variable and on the support roles that have not been used to select the lhs-variables. 
If the lhs-variable is of a boolean or numeric type, 
the rhs-term is a boolean or numerical value derived from a phrase tagged with one of the support roles. Therefore, we look for a phrase that matches the terms 'true' and 'false' or that is a number.

When the lhs-variable is boolean and the verb is negative,
we negate the rhs-term. When the lhs-variable is boolean and %
there is no support role to identify the rhs-term, %
we use the default value \emph{true} for the rhs-term. 
For example, in \SC~in Table~\ref{table:oclExamples}, all the support roles have already been used to identify the lhs-variable and there is no support role left for the rhs-term. Therefore, we use \emph{true} for the rhs-term.

When the lhs-variable is of an enumeration type, we identify the enumeration literal in the domain model that best matches the phrases tagged with the support roles.
For instance, \CHANGED{this is the case for} \SE~in Table~\ref{table:oclExamples}, \CHANGED{which results from the application of the \emph{any-verb} transformation rule.}
\CHANGED{In \SE,} \emph{BodySense.buildCheckStatus} is the lhs-variable which is of an enumeration type, i.e., \emph{BuildCheckStatus}. 
This enumeration contains a term that matches the root of the verb in \SE~(\emph{pass}), 
\MREVISION{R1.27}{and the verb is one of the support roles in the \emph{any-verb} transformation rule.}
Therefore, the literal \emph{BuildCheckStatus::Passed} is selected as the rhs-term.

\MREVISION{R1.31}{For certain transformation rules, the rhs-term is set to a specific literal based on the meaning of the verb. This is the case for the verb \emph{erase}, used in sentences \SG~and \SH~in Table~\ref{table:oclExamples}.
For \emph{erase}, the literal to be used is selected based on the type of the lhs-variable (i.e., \texttt{0} for numeric types and \texttt{null} for complex types).}

\subsubsection{Identification of the OCL operators}
\label{subsec:identifyOCLOperators}

The OCL comparison operators are identified based on the verb in the use case step.
The operator '$=$' is used for most verbs. 
For the verb `\emph{to be}', we rely on Roy et al.~\cite{RoyViRo15} which, for example, identify the operator '$>$' for \textit{``capacitance is above 600"}. 
More precisely, we apply Roy's approach on the phrase tagged with the support-role used to identify the rhs-term.
If antonyms have been used to identify the lhs-variable and the rhs-term, we negate the selected operator, for instance, by replacing '$=$' with '$!=$' in 
\SL~in Table~\ref{table:oclExamples}.

Regarding the selection operator, we are limited to the pattern in Fig.~\ref{fig:bnf}, i.e., the exclusion of some class instances in a set. 
The selection operator is introduced when the phrase tagged with \emph{A1} contains the keyword \emph{except} (e.g., \SF~in Table~\ref{table:oclExamples}) \MREVISION{R1.28}{or any of its synonyms according to WordNet (i.e., \emph{leaving out}, \emph{excluding}, \emph{leaving off}, \emph{omitting}, and \emph{taking out})}. To identify the types of the instances to be excluded, we rely on the tags generated based on SRL NomBank. We look for the phrases tagged with \emph{A2} to identify the adverbial clause. We identify all the distinct noun phrases within the clause (e.g., \emph{voltage errors} and \emph{memory errors} in \SF).

\subsubsection{Identification of the Type of OCL Query Elements}

The query elements in our OCL pattern are used to check if an expression holds for a set of instances (see Fig.~\ref{fig:bnf}). 
The key difference among them is 
the number of instances for which the expression should hold. 
Since the number of subjects referred to in a sentence in English is specified by the determiners and quantifiers, we identify the type of the query element based on the determiners and quantifiers in the phrase tagged with an entity role. We consider entity roles because a domain entity is selected based on its similarity with the phrase tagged with an entity role.

In English, the indefinite articles \emph{a} and \emph{an} and the determiner \emph{some} refer to particular members of a group. %
Therefore, if a phrase tagged with an entity role includes an indefinite article %
or the determiner \emph{some}, we generate an OCL query following the EXISTS sub-pattern (see Fig.~\ref{fig:bnf}). 
For phrases with a quantifier referring to a certain number of members of a group (e.g., \emph{at least five}), we generate expressions following the COUNT sub-pattern. We also rely on Roy's approach~\cite{RoyViRo15} 
to generate a quantifier formula with an operator and a numeric literal.

For all other cases, we generate expressions that follow the FORALL sub-pattern. These cases include phrases with universal determiners referring to all the members of a group (e.g., \emph{any}, \emph{each}, and \emph{every}) and phrases with the definite article \emph{the}. 
The definite article is used to refer to a specific entity (e.g.,  \emph{the measured voltage} in \SG~in Table~\ref{table:oclExamples}) and typically leads to the definition of an attribute or association in the domain model (e.g., \emph{the measured voltage} matches an attribute of the system class \texttt{BodySense}). Since there might be multiple class instances with the corresponding attribute (or association), we rely on the FORALL sub-pattern to match all of them. 
\MREVISION{R1.29}{When the definite article is used to refer to a specific instance among all the instances of the same type, 
engineers should design the domain model to enable the identification of such instances.
For example, to identify the last measurement in the capacitance sequence recorded by \BodySense, which is referred to in \SQ, engineers introduced the attribute \emph{last measurement}, which is a derived attribute.}
A detailed description of how the universal determiners and the definite article influence the scoring of the generated constraints is provided in Section~\ref{subsec:scoring}.

\MREVISION{R1.21}{Our OCL constraint generation strategy does not support the generation of OCL constraints for fault tolerant devices including replicated subsystem instances (e.g., multiple instances of BodySense). 
More precisely, \UMTG cannot derive OCL constraints for subsystems with different states (e.g., one instance of BodySense with a temperature error being detected and another one with no error detected). 
This limitation is inherited from RUCM, which does not include means to capture the behavior of multiple subsystems because, in the requirements elicitation phase, analysts describe the system as a whole independently from the number of implemented instances. 
Accordingly, we use OCL constraints to capture the properties of the system as a whole, not the properties of its subsystems.
}

\subsubsection{Scoring of the OCL Constraints}
\label{subsec:scoring}

The score of an OCL constraint accounts for both the completeness and correctness of the constraint. Completeness relates to the extent to which all the concepts in the use case step are accounted for. 
Correctness relates to how similar the variable names in the constraint are to the concepts in the use case step. 

We measure the completeness of a constraint in terms of the percentage of the roles in the use case step that are used to identify the terms in the constraint. 

To compute the correctness of a constraint, we use \\ %
\begin{math}
correctness = \frac{(\mathit{lhsScore}+\mathit{rhsScore}+\mathit{matchUniversalDeterminer})}{3}
\end{math}\\
where \emph{lhsScore} and \emph{rhsScore} are the similarity scores for the lhs-variable and the rhs-term, respectively. \emph{lhsScore} is the average of the scores of all the attributes/associations in the variable. %
When the rhs-term is a boolean, numeric or enumeration literal generated from a phrase in the use case step, \textit{rhsScore} is set to one; otherwise, \textit{rhsScore} is computed like \textit{lhsScore}.
\emph{matchUniversalDeterminer} is $1$ when universal determiners (e.g., \emph{any}, \emph{every}, or \emph{no}) are properly reflected in the constraint. %
We consider a universal determiner to be properly reflected in the constraint (1) when it is not in the noun phrase tagged with an entity role and the constraint refers to a specific instance associated with the system class (e.g., \SB), and 
(2) when it is in the noun phrase tagged with an entity role and the constraint refers to all the instances of the class matching the phrase (e.g., \SF). %
Universal determiners are important to derive the correct constraints. For example,
for \SB, %
we build two constraints \emph{C2} and \emph{C3} in Table~\ref{tab:scoringExample}. We select \emph{C3} because the use case step does not explicitly indicate that all the NVM components should be considered. %

The final score is computed as the average of the completeness and correctness scores (see Table~\ref{tab:scoringExample}).

\begin{table}[t]
\scriptsize
\caption{OCL constraints and their scores for sentence \SB~of Table~\ref{table:oclExamples}.}
\begin{tabular}{|p{1mm}|@{\hspace{1mm}}p{7cm}|@{\hspace{1mm}}p{0.4cm}|}
\hline
&\textbf{Candidate OCL}&\textbf{Score}\\
\hline
C1&$\mathit{BodySense.allInstances()} \rightarrow \mathit{forAll(} \mathit{i} | \mathit{i.itsNVM} = \mathit{...} )$ 						& $-$\\
C2&$\mathit{NVM.allInstances()} \rightarrow \mathit{forAll(} i | \mathit{i.isAccessible} = \mathit{true} )$									&  $0.81$\\
C3&$\mathit{BodySense.allInstances()} \rightarrow \mathit{forAll(} i | \mathit{i.its}NVM.isAcces$ $\mathit{sible} = \mathit{true} ) $&  $0.94$	\\
\hline
\end{tabular}

\vspace{0.2cm}
\textbf{Notes on the computed scores:}

\textbf{C1} is ignored because the attribute \emph{BodySense.itsNVM} refers to a class and does not enable the identification of any rhs-term.
In \textbf{C2}, \textit{completeness} is $1$ 
 since all the roles are used to identify the terms. %
\textit{lhsScore} is $0.83$, i.e., the division of the length of the word \emph{accessible} (i.e., 10) by the length of the variable \emph{isAccessible} (i.e., 12). \textit{rhsScore} is $1$. \textit{matchUniversalDeterminer} is $0$ because the constraint refers to all the instances of class NVM although no universal determiner is used in the sentence. \textit{correctness} is calculated as $(0.83+1+0)/3 = 0.61$. 
In \textbf{C3}, \textit{completeness} is $1$. %
\textit{lhsScore} is $0.66$, which is the average of the scores for attributes \emph{itsNVM} (i.e., $0.5$) and \emph{isAccessible} (i.e., $0.83$). \textit{rhsScore} is $1$.  \textit{matchUniversalDeterminer} is $1$ because no universal determiner is used in the sentence and the constraint refers to a specific instance referenced by the system class. \textit{correctness} is calculated as $(0.66+1+1)/3=0.89$. The score is thus calculated as $(0.89+1)/2=0.94$.

\label{tab:scoringExample}
\end{table}%

\begin{table}[t]
\scriptsize
\caption{Verbs unlikely to appear in use case specifications.}
\begin{tabular}{|p{4.5cm}|p{3.5cm}|}
\hline
\textbf{Categories of Verbs for Exclusion}&\textbf{Example Verbs}\\
\hline
Verbs describing a human feeling& love, like\\
Verbs describing a human sense& smell, taste\\
Verbs describing human behaviors& wish, hope, wink, cheat, confess\\
Verbs describing body internal motion& giggle, kick\\
Verbs describing body internal states&quake, tremble\\
Verbs describing manner of speaking& burble, croak, moan\\
Verbs describing nonverbal expressions& scoff, whistle\\
Verbs describing animal behaviors& bark, woof, quack\\
Communication verbs& tell, talk\\

\hline
\end{tabular}
\label{table:excludedVerbs}
\end{table}%

\subsection{Correctness and Generalizability} %
\label{subsec:gen}

The adoption of verb-specific transformation rules may limit the correctness and generalizability of constraint generation due to the considerable number of English verbs.
For example, the \textit{Unified Verb Index}  \cite{UnifiedVerbIndex}, a popular lexicon %
based on VerbNet and other lexicons, includes 8,537 English verbs. 

\MREVISION{R1.30}{To strengthen the correctness} of constraint generation, we base our transformation rules on the VerbNet role patterns. These role patterns capture 
all valid role combinations appearing with a verb. %
The rules are applied according to the entity and support role pairs $\mathit{\langle}$\emph{entity role}, \emph{\{support roles\}}$\mathit{\rangle}$ we extracted from the role patterns (see Table~\ref{additionalVerbConfigs}). 
For example, %
the transformation rule of the verb `to erase' has two role pairs $\mathit{\langle}$\emph{A1}, \emph{null}$\mathit{\rangle}$ and $\mathit{\langle}$\emph{A2}, \emph{\{A1\}}$\mathit{\rangle}$ extracted from the VerbNet role patterns (see Rule 4 in Table~\ref{additionalVerbConfigs}). 
If the sentence contains \emph{A1} without \emph{A2}, then \emph{A1} is used to identify the entity to be erased, i.e., the attribute in the lhs-variable (see \SG~in Table~\ref{table:oclExamples}). If the sentence contains both \emph{A1} and \emph{A2}, then \emph{A2} is used to identify the entity and \emph{A1} provides some additional information (i.e., the attribute of the entity to be erased). 
As an example of the latter case, in \SH{} in Table~\ref{table:oclExamples}, we identify the entity name (\emph{AirbagControlUnit}) and the attribute in the lhs-variable (\emph{occupantClass}).
\MREVISION{R3.8}{In addition to the case studies in our evaluation (see Section~\ref{sec:evaluation}), we validated our transformation rules with sentences that include all the role patterns in VerbNet.}

To ensure the generalizability of the constraint generation, we employ three key solutions %
which prevent the implementation of hundreds of rules. First, we rely on the VerbNet classes to use a single rule targeting different verbs. 
Since all the verbs in a VerbNet class share the same role patterns,
we reuse the same rule for all the verbs in the VerbNet class that are synonyms according to WordNet.

Second, we excluded some VerbNet classes of verbs, i.e., 225 classes and 175 subclasses, that are unlikely to appear in specifications (see Table~\ref{table:excludedVerbs}). %
We manually inspected all the VerbNet classes to identify them (e.g., verbs describing human feelings). %
Our analysis results are available online~\cite{oclGenWeb}.

Third, we further analyzed the remaining classes to determine the verbs that are processed by our \textit{any-verb transformation rule}. 
This analysis shows only 33 verb-specific rules are required to process 87 classes of verbs. %

\section{Generation of Use Case Test Models}
\label{sec:testModelGeneration}

\begin{figure}[h]
  \centering
    \includegraphics{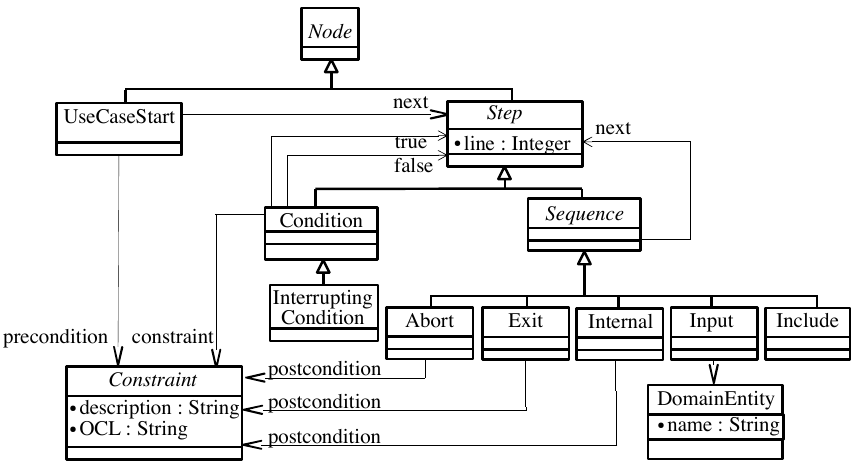}
      \caption{Metamodel for use case test models.}
      \label{fig:testingModelClasses}
\end{figure}

\M{} generates a \emph{\UCTM{}} from an RUCM use case specification along with the generated OCL constraints (Step 7 in Fig.~\ref{fig:approach}).
The model makes the implicit control flow in the use case specification explicit and maps the use case steps onto the test case steps. 
Fig.~\ref{fig:testingModelClasses} gives the metamodel for use case test models. %

A use case test model is a connected graph containing the instances of multiple subclasses of \emph{Node}. %
\emph{UseCaseStart} represents the beginning of a use case with a precondition and is linked to the first step in the use case (\emph{next}). There are two subclasses of \emph{Step}, i.e., \emph{Sequence} and \emph{Condition}. \emph{Sequence} has a single successor. \emph{Condition} is linked to a constraint (\emph{constraint}) and has two possible successors (\emph{true} and \emph{false}). 
\emph{InterruptingCondition} %
enables the execution of a global or bounded alternative flow.   
\emph{Constraint} has a condition %
and an OCL constraint generated from the condition. %

\emph{Input} indicates the invocation of an input operation %
and is linked to \emph{DomainEntity} that represents input data. 
To specify the effects of an internal step on the system state, \emph{Internal} is linked to \emph{Constraint} (i.e., \emph{postcondition}).
\emph{Exit} represents the last step of a use case flow; \emph{Abort} indicates the termination of an anomalous flow. 
\MREVISION{R1.32}{\emph{Exit} and \emph{Abort} steps have an associated \emph{Constraint} with a postcondition in textual form, which is used by \UMTG to generate oracles (see Section~\ref{sec:generationOfTestCases}). 
In the case of \emph{Exit} and \emph{Abort} steps, \UMTG does not generate an OCL constraint from the textual condition (see Section~\ref{sec:oclGeneration}).}

Output steps are not represented in use case test models because \UMTG does not rely on them to drive test generation. Although output steps might be used to select outputs to be verified by automated oracles during testing, \UMTG relies on postconditions to generate test oracles because they provide more information (e.g., the system state and the value of an output, which are not indicated in output steps).

\begin{figure}[h]
  \centering
    \includegraphics{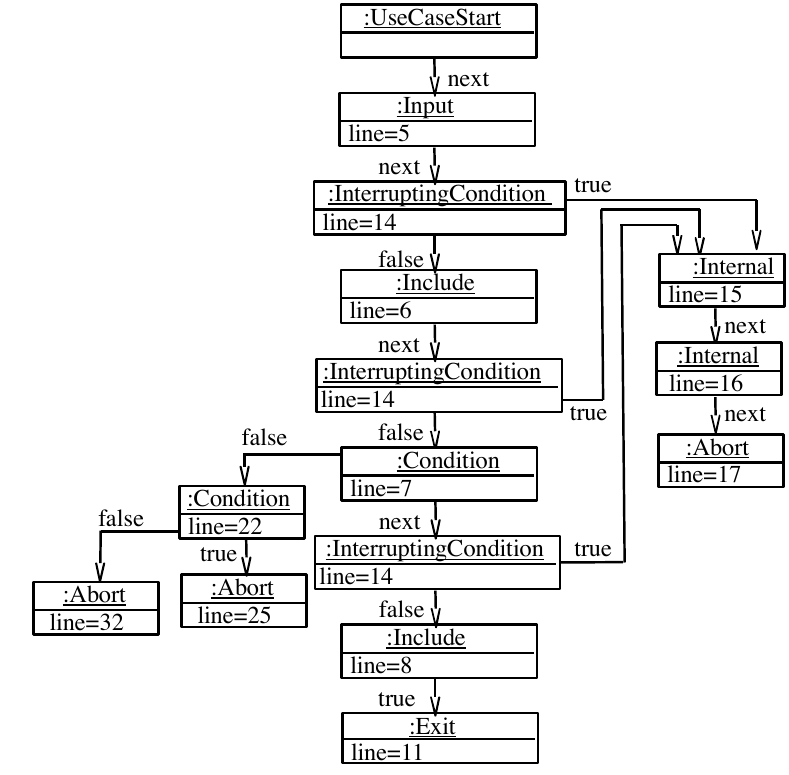}
      \caption{Part of the use case test model for \emph{Identify Occupancy Status}.}
      \label{fig:testingModelExample}
\end{figure}

Fig.~\ref{fig:testingModelExample} shows part of the use case test model generated from the use case \emph{Identify Occupancy Status} in Table~\ref{useCaseInitialOccupancy}. 
\M{} processes the use case steps annotated by the NLP pipeline in Section~\ref{sec:nlpPipeline}. For each textual element tagged with \emph{Input}, \emph{Include}, \emph{Internal}, or \emph{Condition}, a \emph{Step} instance is generated and linked to the previous \emph{Step} instance. %
For each domain entity in the use case specification, a \emph{DomainEntity} instance is created and linked to the corresponding \emph{Step} instance. 

For each specific alternative flow, a \emph{Condition} instance is generated and linked to the \emph{Step} instance that represents the first step of the alternative flow (e.g., the \emph{Condition} instance for Line~\ref{uc1:bf:validate} in Fig.~\ref{fig:testingModelExample}).
 
Global and bounded alternative flows begin with a guard condition and are used to indicate that the condition might become true at run time and trigger the execution of the alternative flow (e.g., as an effect of an interrupt being triggered). %
For each step referenced by a global or bounded alternative flow, an \emph{InterruptingCondition} instance is created and linked to the \emph{Step} instance that represents the reference flow step (e.g., the three \emph{InterruptingCondition} instances for Line 14 linked to the \emph{Step} instances for Lines 6, 7, and 8 in Fig.~\ref{fig:testingModelExample}).  

For multiple alternative flows depending on the same condition, \emph{Condition} instances are linked to each other in the order the alternative flows appear in the specification.
For an alternative flow in which the execution is resumed back to the basic flow, an \emph{Exit} instance is linked to the \emph{Step} instance that represents the reference flow step. %

\section{Generation of Use Case Scenarios and Test Inputs} %
\label{sec:testGeneration}

\M{} generates, from a use case test model along with OCL constraints, a set of use case scenarios and test inputs (Step 8 in Fig.~\ref{fig:approach}).
A scenario is a path in the use case test model that begins with a \emph{UseCaseStart} node and ends with an \emph{Abort} node or an \emph{Exit} node with the attribute \emph{next} set to null (i.e., an Exit node terminating the use case under test). 
It captures a sequence of interactions that should be exercised during testing. 
\MREVISION{R3.12}{It consists of a finite number of steps. However, it can exercise alternative flows that lead to loops in use case specifications. 
More precisely, a scenario can exercise the same use case step up to $T$ times in a loop, with $T$ being a \UMTG parameter specified by engineers.}

\UMTG needs to identify test inputs in which the conditions in the scenario hold. %
For example, to test the scenario for the basic flow of the use case \emph{Classify Occupancy Status} in Table~\ref{useCaseInitialOccupancy}, we need test inputs in which the capacitance value is above 600 (see Line~\ref{uc3:bf:capacitance} in Table~\ref{useCaseClassify}).   
We use the term \emph{path condition} to indicate a formula that conjoins all the conditions (OCL constraints) in a given scenario. 
If the path condition is satisfiable, we derive an object diagram (i.e., an instance of the domain model in which the path condition holds). 
For a given scenario, test input values are extracted from the object diagram that satisfies the path condition.

\begin{figure}[h]
\begin{algorithmic}[1]

\scriptsize
\Require $tm$, a use case test model 
\Ensure $ScenariosInputs$, a set of pairs $\langle{}Scenario,objectDiagram\rangle{}$ 
\Function{GenerateScenariosAndInputs}{tm} 
	\State $\mathit{Scenario} \gets \mathit{new} \mathit{List()}$ \textcolor{gray}{//list of Nodes} 
	\State $\mathit{pc} \gets \mathit{new} \mathit{List()}$ \textcolor{gray}{//list of Constraints}
	\State $\mathit{ScenariosInputs} \gets \mathit{new} \mathit{Set()}$ \textcolor{gray}{//set with feasible scenarios}
         \Repeat 
        \State \MMREVISION{R1.33,35}{$\mathit{ScenariosSet} \gets \mathit{new} \mathit{Set()}$ \textcolor{gray}{//set of pairs $\mathit{\langle{}Scenario,pc\rangle{}}$}}
	\State $\mathit{GenerateScenarios(tm, tm.start, Scenario, pc,}$ \label{alg:inputs:invokingGenerateScenarios}
	\Statex \hspace{3cm}$\mathit{ScenariosInputs, ScenariosSet)}$ 
	\For {\textbf{each} pair $\mathit{\langle{}Scenario,pc\rangle{}}\ in\ \mathit{ScenariosSet}$} 
		\State $\mathit{objectDiagram} \gets \mathit{Solve(Scenario,pc)}$ \label{alg:inputs:solve}
		\If {$\mathit{objectDiagram} = \mathit{null}$} \textcolor{gray}{//the scenario is infeasible} \label{alg:inputs:infeasible1}
		\State $\mathit{remove}\ \langle{}Scenario,pc\rangle{}\ \mathit{from\ ScenariosSet}$ \label{alg:inputs:infeasible2}
		\Else \textcolor{gray}{//add the scenario to the results to be returned}
		\State $\mathit{ScenariosInputs} \gets $ \\ $\hspace{2cm}\mathit{ScenariosInputs} \cup \{\langle{}\mathit{Scenario,objectDiagram}\rangle{}\}$		
		\EndIf
	\EndFor
\Until 		$\mathit{coverageSatisfied(tm,ScenariosInputs)}$ \textbf{or}  
\emph{max number of iterations T reached} \label{alg:inputs:until}
\vspace{1mm}
\If{coverage criterion is subtype coverage}
\State $\mathit{ScenariosInputs} \gets \mathit{maximizeSubTypeCoverage(ScenariosInputs)}$\label{alg:inputs:maximizeSubtype}
\EndIf
\vspace{1mm}
\State \Return $\mathit{ScenariosInputs}$
\EndFunction

\end{algorithmic}
\caption{Algorithm for generating use case scenarios and test inputs.} %
\label{alg:testInputsGen}
\end{figure}

We devise an algorithm, \emph{GenerateScenariosAndInputs} in Fig.~\ref{alg:testInputsGen}, which generates a set of pairs of use case scenarios and object diagrams from the input use case test model $tm$. %
Before calling \emph{GenerateScenariosAndInputs}, the use case test model $tm$ is merged with the use case test models of the included use cases in a way similar to the generation of interprocedural control flow graphs~\cite{Harrold-InterproceduralControlFlowGraph-ISSTA-1998}. The \emph{Include} instances in $tm$ are replaced with the \emph{UseCaseStart} instances of the included use cases; the \emph{Exit} instances of the basic flows of the included use cases are linked to the \emph{Node} instances following the \emph{Include} instances in $tm$.

To generate use case scenarios, we call the function \emph{GenerateScenarios} with \emph{tm} being a use case test model, \emph{Scenario} being an empty list, \emph{tm.start} being the \emph{UseCaseStart} instance in $tm$, %
\emph{pc} being a \emph{null} path condition, \emph{ScenariosInputs} being an empty list,
and \emph{ScenariosSet} being an empty list 
 (Line~\ref{alg:inputs:invokingGenerateScenarios} in Fig.~\ref{alg:testInputsGen}). \M{} employs the Alloy analyzer~\cite{Jackson2006} to generate an object diagram in which the path condition in OCL holds for a given scenario (Line~\ref{alg:inputs:solve}). It makes use of existing model transformation technology from OCL constraints to Alloy specifications~\cite{UML2Alloy}. %

Some of the generated scenarios may be infeasible. 
\MREVISION{R1.34}{These are the scenarios that cannot be exercised with any set of possible values such as scenarios covering auxiliary basic and alternative flows. For example, in Table~\ref{useCaseClassify}, 
this is the case for the scenario
that covers both the alternative flow that detects a low temperature error (Line~\ref{uc2:af3:start}) and 
the basic flow of Classify Occupancy Status (Line~\ref{uc3:bf}). 
Based on the condition step in Line~\ref{uc1:bf:validate}, the basic flow of Classify Occupancy Status can be executed only if no error has been detected.}
We exclude such infeasible scenarios (Lines~\ref{alg:inputs:infeasible1} and \ref{alg:inputs:infeasible2}).

We execute \emph{GenerateScenarios} multiple times until a selected coverage criterion is satisfied or the number of iterations reaches a predefined threshold (Line~\ref{alg:inputs:until}). We set the threshold to ten in our experiments.
\M{} supports three distinct coverage criteria, i.e., branch coverage, a lightweight form of def-use coverage, and a form of clause coverage that ensures each condition to be covered with different entity types. All these three coverage strategies are described in the following sections.

Section~\ref{subsec:testScenarioGeneration} describes the scenario generation algorithm (Line~\ref{alg:inputs:invokingGenerateScenarios}), while Section~\ref{subsec:testInputGeneration} provides details about the generation of object diagrams that satisfy path conditions (Line~\ref{alg:inputs:solve}).

\subsection{Generation of Use Case Scenarios} %
\label{subsec:testScenarioGeneration}

\emph{GenerateScenarios} performs a recursive, depth-first traversal of a use case test model to generate use case scenarios (see Fig.~\ref{alg:testGen}). 
It takes as input a use case test model (\emph{tm}), a node in \emph{tm} to be traversed (\inst), 
\MREVISION{R1.37}{two input lists  (i.e., \emph{Scenario} and \emph{pc}) and two input sets (i.e., \emph{Curr} and \emph{Prev}).
\emph{Scenario} is a list of traversed nodes in \emph{tm}, and 
\emph{pc} is the path condition of the traversed nodes.
\emph{Prev} is a set of feasible scenarios traversed in previous invocations of 
\emph{GenerateScenarios}
made by \emph{GenerateScenariosAndInputs}.
\emph{Prev} corresponds to \emph{ScenariosInputs} in \emph{GenerateScenariosAndInputs}.
\emph{Curr} is a set of pairs $\langle Scenario, pc \rangle$ which contain the scenarios and their path conditions identified during the current invocation
of \emph{GenerateScenarios}
made by \emph{GenerateScenariosAndInputs}.
\emph{Scenario}, \emph{pc}, and \emph{Curr} are initially empty. They are populated during recursive calls to \emph{GenerateScenarios}. 
}

\begin{figure}[t]

\begin{algorithmic}[1]

\scriptsize
\Require $tm$, a use case test model 
\Require $node$, an instance in $tm$ 
\Require $Scenario$, a linked list of node instances from $tm$
\Require $pc$, an OCL constraint of the path condition for $Scenario$
\Require $Prev$, a list of scenarios covered in previous iterations
\Ensure $Curr$, a set of pairs $\langle{}Scenario,pc\rangle$ 

\Function{GenerateScenarios}{tm, node, Scenario, pc, Prev, Curr}

		\If {($\mathit{coverageSatisfied(tm,Prev, Curr)}$)} \label{alg:ctg:allBranchesCovered}
			\State \Return
		\EndIf \label{alg:ctg:allBranchesCoveredEndIF}
\vspace{1mm}
		\If {($\mathit{unsatisfiable(pc)}$)} \label{alg:ctg:UNSAT}
			\State \Return
		\EndIf \label{alg:ctg:UNSATEndIF}
\vspace{1mm}

\If {($\mathit{node}$ is a $\mathit{UseCaseStart}$ instance)} \label{alg:ctg:initial}
	\State $\mathit{addToScenario(node, Scenario)}$ 
         \State $\mathit{pc} \gets \mathit{composeConstraints(pc, node.precondition.ocl)}$ 
         \State $\mathit{GenerateScenarios(tm, node.next, Scenario, pc, Prev, \mathit{Curr})}$ \label{alg:ctg:start_generatescen}
\EndIf \label{alg:ctg:initialEnd}

\vspace{1mm}
\If {($\mathit{node}$ is an $\mathit{Input}$ instance)} \label{alg:ctg:input}
	\State $\mathit{addToScenario(node, Scenario)}$ \label{alg:ctg:addInp}
	\State $\mathit{GenerateScenarios(tm, node.next, Scenario,  pc, Prev, \mathit{Curr})}$ \label{alg:ctg:input_generatescen}
\EndIf \label{alg:ctg:inputEnd}

\vspace{1mm}
\If {($node$ is a $Condition$ but not an $InterruptingCondition$}) \label{alg:ctg:condStart}	
	\State \textcolor{gray}{\textbf{//prepare\ for\ visiting\ the\ true\ branch}}
	\State $\mathit{Scenario_{T}} \gets \mathit{Scenario}$  \textcolor{gray}{\textbf{//create\ a\ scenario\ copy}}
	\State $\mathit{addToScenario(node, Scenario_{T})}$
	\State $\mathit{pc_{T}} \gets \mathit{composeConstraints(pc, node.constraint.ocl)}$  \label{alg:ctg:compose1}
	\State $\mathit{GenerateScenarios(tm, node.true, Scenario_{T}, pc_{T}, Prev, \mathit{Curr})}$\label{alg:ctg:followTrueGen}
	\State \textcolor{gray}{\textbf{//prepare\ for\ visiting\ the\ false\ branch}}
	\label{alg:ctg:followFalse}
	\State $Scenario_{F} \gets Scenario$  \textcolor{gray}{\textbf{//create\ a\ scenario\ copy}}\label{alg:ctg:negGB1}
	\State $\mathit{addToScenario(node, Scenario_{F})}$ \label{alg:ctg:neg1}
        \State $\mathit{nc} \gets \mathit{negateConstraint(node.constraint.ocl)}$
        \State $\mathit{pc_{F}} \gets \mathit{composeConstraints(pc, nc)}$ \label{alg:ctg:neg2}
        	\State \hspace{-1mm}$\mathit{GenerateScenarios(tm,node.false,Scenario_{F},pc_{F}, Prev, \mathit{Curr}})$ \label{alg:ctg:followFalseGen}
\EndIf \label{alg:ctg:condEnd}

\vspace{1mm}
\If {($node$ is an $InterruptingCondition$ instance)} \label{alg:ctg:intCondStart}	
	\State \textcolor{gray}{\textbf{//visit the false branch first, i.e., a scenario without any interrupt}}
	\State $Scenario_{F} \gets Scenario$  \textcolor{gray}{\textbf{//create\ a\ scenario\ copy}}
	\State $\mathit{GenerateScenarios(tm, node.false, Scenario_{F},  pc, Prev, \mathit{Curr})}$ \label{alg:ctg:followFalseGenIC}
	\State \textcolor{gray}{\textbf{//visit the true branch, i.e., a scenario with an interrupt}}
	\State $Scenario_{T} \gets Scenario$  \textcolor{gray}{\textbf{//create\ a\ scenario\ copy}}
	\State $addToScenario(node, Scenario_{T})$ 
        \State $pc_{T} \gets \mathit{composeConstraints(pc, node.constraint.ocl)}$  \label{alg:ctg:compose3}
        	\State \hspace{-1mm}$\mathit{GenerateScenarios(tm, node.true, Scenario_{T}, pc_{T}, Prev, \mathit{Curr})}$ \label{alg:ctg:followTrueGenIC}
\EndIf \label{alg:ctg:intCondEnd}

\vspace{1mm}
\If {($node$ is an $Internal$ instance)} \label{alg:ctg:internal}
	\State $\mathit{addToScenario(node, Scenario)}$ \label{alg:ctg:addInternal}
         \State $\mathit{pc} \gets \mathit{composeConstraints(pc, node.postcondition.ocl)}$ \label{alg:ctg:internalPC}
	\State $\mathit{GenerateScenarios(tm, node.next, Scenario, pc, Prev, \mathit{Curr})}$ \label{alg:ctg:generate}
\EndIf \label{alg:ctg:internalEnd}

\vspace{1mm}
\If {($\mathit{node}$ is an $\mathit{Exit}$ or $\mathit{Abort}$ instance)} \label{alg:ctg:exitBelong}
\If {($\mathit{node}$ is an $\mathit{Exit}$ instance with a non null \emph{next} association)} \label{alg:ctg:exitIncluded}
	\If {($\mathit{node.next}$ visited \MMREVISION{R1.36}{at most} T times in the current Scenario)} \label{alg:ctg:noLoop}
		\State \textcolor{gray}{\textbf{//the visit should proceed in the specified next step}}
			\State $\mathit{addToScenario(node, Scenario)}$
                           \State $\mathit{GenerateScenarios(tm, node.next, Scenario, pc, Prev, \mathit{Curr})}$ \label{alg:ctg:nextStepnodeance}
	\Else	\textcolor{gray}{\textbf{//node.next visited more than T times in the current scenario}}
	\State \Return \textcolor{gray}{\textbf{//ignore paths that traverse a same branch or loop body more}}
	 \Statex \hspace{1.2cm} \textcolor{gray}{\textbf{than T times}}
	\EndIf
\Else \label{alg:ctg:exitIncludedEnd} \textcolor{gray}{\textbf{//is the final step of the use case}}
		\State $\mathit{addToScenario(node, Scenario)}$ \label{alg:ctg:addExit}		
		\If { ($\mathit{coverageImproved( tm, Curr, Scenario)}$)} \label{alg:ctg:ifCovImproved}
			\State $\mathit{Curr} \gets \mathit{Curr}~\cup \{\langle{}Scenario, pc\rangle{}\}$  \label{alg:ctg:addTest}		
		\EndIf 
\EndIf
	
\EndIf

\State \Return
\EndFunction

\end{algorithmic}
\caption{Algorithm for generating use case scenarios.}
\label{alg:testGen}
\end{figure}

\begin{figure*}[t]
  \centering
    \includegraphics[width=18cm]{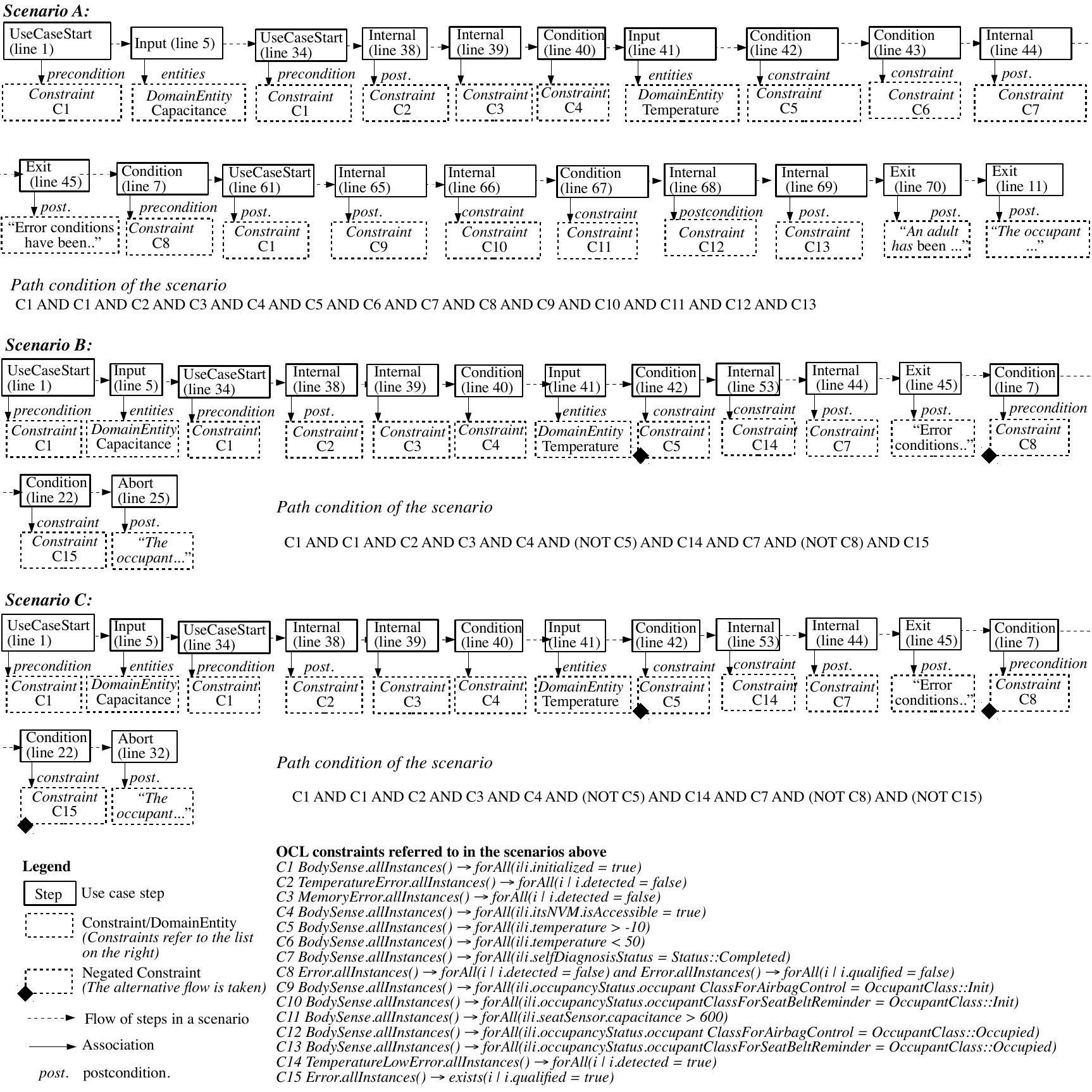}
      \caption{Three scenarios built by \M{} when generating test cases for the use case `Identify Occupancy Status' in Table~\ref{useCaseInitialOccupancy}.}
      \label{fig:scenarioExamples}
\end{figure*}

Fig.~\ref{fig:scenarioExamples} shows three scenarios generated from the use case test model in Fig.~\ref{fig:testingModelExample}. %
Scenario A covers, without taking any alternative flow, the basic flow of the use case \emph{Identify Occupancy Status} and the basic flows of the included use cases \emph{Self Diagnosis} and \emph{Classify Occupancy Status} in Table~\ref{useCaseInitialOccupancy}. Scenario B takes two specific alternative flows of the use cases \emph{Self Diagnosis} and \emph{Identify Occupancy Status}, respectively. It covers the case in which a TemperatureError has been detected (Lines~\ref{uc2:af3:start} to~\ref{uc2:af3:post} in Table~\ref{useCaseInitialOccupancy}) and some error has been qualified (Lines \ref{uc1:sat:start} to~\ref{uc1:sat:post}). Scenario C covers the case in which some error has been detected but no error has been qualified (Lines \ref{uc1:sat:startTwo} to~\ref{uc1:sat:postTwo}).

The precondition of the \emph{UseCaseStart} instance is added to the path condition for the initialisation of the test case (Lines~\ref{alg:ctg:initial}~to~\ref{alg:ctg:initialEnd} in Fig.~\ref{alg:testGen}). \textit{Input} instances do not have associated constraints to be added to the path condition (Lines~\ref{alg:ctg:input} to~\ref{alg:ctg:inputEnd}). We recursively call \emph{GenerateScenarios} for the nodes following the \emph{UseCaseStart} and \emph{Input} instances (Lines~\ref{alg:ctg:start_generatescen} and~\ref{alg:ctg:input_generatescen}). For instance, Scenario A in Fig.~\ref{fig:scenarioExamples} starts with the \emph{UseCaseStart} instance of the use case \emph{Identify Occupancy Status} followed by an \emph{Input} instance, and proceeds with the \emph{UseCaseStart} instance of the included use case \emph{Self Diagnosis}. 

For each \emph{Condition} instance which is not of the type \emph{InterruptingCondition}, we first visit the true branch and then the false branch to give priority to nominal scenarios (Lines~\ref{alg:ctg:condStart} to~\ref{alg:ctg:condEnd}). 
In the presence of a coverage-based stopping criterion, prioritizing nominal scenarios is useful to generate relatively short use case scenarios covering few alternative flows (i.e., what happens in realistic executions) instead of long use case scenarios covering multiple alternative flows.
Bounded and global alternative flows are taken in the true branch of their \emph{InterruptingCondition} instances. 
Therefore, to give priority to nominal scenarios, we first visit the false branch for the \emph{InterruptingCondition} instances %
(Lines~\ref{alg:ctg:intCondStart} to~\ref{alg:ctg:intCondEnd}). %
For each condition branch taken, 
we add to the path condition the OCL constraint of the branch (Lines~\ref{alg:ctg:compose1}, \ref{alg:ctg:neg2} and~\ref{alg:ctg:compose3}) except for the false branch of the \emph{InterruptingCondition} instances; indeed, the condition of a bounded or global alternative flow is taken into account only for the scenarios in which the alternative flow is taken.
For example, \MREVISION{R1.39}{Scenarios A, B, and C} do not cover the condition of the bounded alternative flow of the use case \emph{Identify Occupancy Status}. We call \emph{GenerateScenarios} for each branch (Lines~\ref{alg:ctg:followTrueGen}, \ref{alg:ctg:followFalseGen},~\ref{alg:ctg:followFalseGenIC} and~\ref{alg:ctg:followTrueGenIC}).

The \emph{Internal} instances represent the use case steps in which the system alters its state. Since state changes may affect the truth value of the following \emph{Condition} instances, the postconditions of the visited \emph{Internal} instances are added to the path condition (Line~\ref{alg:ctg:internalPC}). %
We call \emph{GenerateScenarios} for the nodes following the \emph{Internal} instances (Line~\ref{alg:ctg:generate}).

To avoid infinite cycles in \emph{GenerateScenarios}, we proceed with the node following an \emph{Exit} instance only if the node has 
been visited at most a number of times $T$ specified by the software engineer (Lines~\ref{alg:ctg:exitIncluded} and~\ref{alg:ctg:noLoop}). The default value for $T$ is one, which enables \UMTG to traverse loops once.
An \emph{Exit} instance followed by a node represents either a resume step or a final step of an included use case.
An \emph{Exit} or \emph{Abort} instance which is not followed by any node indicates the termination of a use case. %
For instance, Scenario A terminates with the \emph{Exit} instance of the basic flow of the use case \emph{Identify Occupancy Status}, while Scenario B terminates with the \emph{Abort} instance of the first specific alternative flow of the same use case.
When the current scenario ends, we add the scenario and its path condition to the set of pairs of scenarios and path conditions to be returned (Line~\ref{alg:ctg:addTest}), if it improves the coverage of the model based on the selected coverage criterion (Line~\ref{alg:ctg:ifCovImproved}).

The algorithm is guided towards the generation of scenarios with a satisfiable path condition (Lines~\ref{alg:ctg:UNSAT} to \ref{alg:ctg:UNSATEndIF}). 
To limit the number of invocations of the constraint solver and, consequently, to speed up test generation, the function 
\emph{unsatisfiable} (Line~\ref{alg:ctg:UNSAT}) does not execute the constraint solver to determine satisfiability but relies on previously cached results. It returns \emph{false} if the path condition has been determined to be unsatisfiable by the solver during the generation of test inputs (Line~\ref{alg:inputs:solve} in Fig.~\ref{alg:testInputsGen}); otherwise it assumes that the path condition is satisfiable and returns \emph{true}. This is why \emph{GenerateScenarios} is invoked multiple times by \emph{GenerateScenariosAndInputs} and constraint solving is executed after all the path conditions have been collected (Line~\ref{alg:inputs:solve} in Fig.~\ref{alg:testInputsGen}). 
\MREVISION{R1.38}{Multiple invocations of \emph{GenerateScenarios} might traverse already visited scenarios, except for the ones including an unsatisfiable path condition. 
Based on our experience, this does not lead to any major scalability drawback. Optimizations, 
such as avoiding the traversal of fully visited portions of the \UCTM, 
are part of our future work.}

Scenario generation terminates when the selected coverage criterion is satisfied  (Lines~\ref{alg:ctg:allBranchesCovered} to \ref{alg:ctg:allBranchesCoveredEndIF})
or the entire graph is traversed.
The depth-first traversal ensures that all edges \MREVISION{R1.40}{can be} visited at least once.
\M{} supports three coverage criteria: branch coverage, a lightweight form of def-use coverage, and a form of clause coverage that we call subtype coverage.

The \emph{branch coverage criterion} aims to maximize the number of branches (i.e., edges departing from condition steps) that have been covered. Coverage improves (Line~\ref{alg:ctg:ifCovImproved} in Fig.~\ref{alg:testGen}) any time a scenario covers a branch that was not covered yet.

Our \emph{def-use coverage criterion} aims to maximize the coverage of entities referred (i.e., used) in condition steps that present attributes defined in internal steps.
We identify \emph{definitions} from the postconditions associated to internal steps. More precisely, each postcondition generally defines one entity, i.e., one that owns the lhs-variable appearing in the postcondition; postconditions that join multiple OCL constraints using the operators \emph{and}/\emph{or} can define multiple entities. Sentence \SC{} in Table \ref{table:oclExamples} defines one entity, \emph{TemperatureError}. 
We identify \emph{uses} from condition steps' constraints; used entities are the ones that own the terms appearing in condition steps' constraints. 
More formally, our coverage criterion matches the standard definition of \emph{all p-uses}~\cite{Pezze:SWTesting} except that
we treat positive and negative evaluation of predicates as different def-use pairs.
This is done to enforce the pairing of each definition with both the true and false evaluation of the constraints in which it is used.
The def-use pairs covered by each scenario are computed by traversing the scenario and by identifying all the pairs of use case steps $\langle s_d, s_u \rangle$, where $s_d$ defines an entity and $s_u$ uses the same entity or one of its supertypes. Def-use coverage improves any time a new def-use pair is observed in a scenario. 

Def-use coverage leads to scenarios not identified with branch coverage.
In our running example, by maximizing the branch coverage criterion we ensure that both Scenario B and Scenario C in Fig.\ref{fig:scenarioExamples} are covered.  
Indeed, the entity \emph{TemperatureLowError} is defined in Line~\ref{uc2:af:resetTempLowError} and used by the constraint C15, which is referenced in the condition step in Line~\ref{uc1:af:occ2invalid}; this leads to two def-use pairs, one covered by Scenario B (C15 evaluates to true) and one by Scenario C (C15 evaluates to false). 
In addition to these two scenarios, the def-use coverage criterion also ensures that a \emph{TemperatureHighError} is covered in two additional scenarios (not shown in Fig.\ref{fig:scenarioExamples}) that pair the definition of \emph{TemperatureHighError} in Line~\ref{uc2:af:resume2} with its uses in Line~\ref{uc1:af:occ2invalid}.

The \emph{subtype coverage criterion} aims to maximize the number of entity (sub) types with an instance referenced in a satisfied condition. 
For each scenario, it leads to  a number of object diagrams such that each condition is satisfied with all the possible entity (sub) types referenced by the condition.
For example, in the presence of a condition step referring to the generic entity type \emph{Error}, it generates a number of object diagrams such that, for each subtype of \emph{Error}, there is at least one instance satisfying the condition.

To apply the \emph{subtype coverage criterion}, the algorithm \emph{GenerateScenariosAndInputs} further processes the generated scenarios with function \emph{maximizeSubTypeCoverage} (Line~\ref{alg:inputs:maximizeSubtype} in Fig.~\ref{alg:testInputsGen}). Function \emph{maximizeSubTypeCoverage} appears in Fig.~\ref{alg:maximizeSubTypeCoverage}. It identifies all the constraints appearing in condition steps that are evaluated to true (Line~\ref{alg:maximizeSubTypeCoverage:conditions} in Fig.~\ref{alg:maximizeSubTypeCoverage}). For each of these constraints, it re-executes constraint solving multiple times, once for each type \emph{sT} that is a subtype of the entity used by the constraint (Line~\ref{alg:maximizeSubTypeCoverage:findSubtypes}). Constraint solving is forced to generate a solution that contains at least one instance of the subtype \emph{sT} that satisfies the constraint (Line~\ref{alg:maximizeSubTypeCoverage:subtypes}). For each scenario, we keep only the object diagrams that contain assignments not observed before (implemented by function \emph{coverageImproved}, see Lines~\ref{alg:maximizeSubTypeCoverage:ifBetter}-\ref{alg:maximizeSubTypeCoverage:keep}).

In our example, to maximize subtype coverage, we need to cover \emph{C15} in Scenario B in Fig.~\ref{fig:scenarioExamples} with three instances of subtypes of the \emph{Error} entity: \emph{TemperatureLowError} (i.e., an instance of \emph{TemperatureLowError} should have the attribute \emph{qualified} set to true), \emph{TemperatureHighError}, and \emph{VoltageError}. 
In Scenario B, to ensure that \emph{C15} is covered with an instance of \emph{TemperatureLowError},  
the function \emph{solveWithSubType} (Line~\ref{alg:maximizeSubTypeCoverage:subtypes} in Fig.~\ref{alg:maximizeSubTypeCoverage}) extends the path condition of the scenario with the additional condition $\mathit{TemperatureLowError.allInstances()} \rightarrow \mathit{exists(}\mathit{i} | \mathit{i.qualified} = \mathit{true)}$, which is automatically derived from constraint \emph{C15} by replacing the entity type used in the constraint. The same is repeated also for \emph{TemperatureHighError} and \emph{VoltageError}. In our implementation, the subtype coverage criterion is combined with the def-use coverage criterion to ensure that constraints using input types are exercised with all the possible input subtypes. 
In our running example, the subtype coverage metric ensures that Scenario B is covered with a \emph{TemperatureLowError} being both detected and qualified.

\begin{figure}[t]
\begin{algorithmic}[1]

\scriptsize
\Require $ScenariosInputs$, a set of pairs $\langle{}Scenario,objectDiagram\rangle{}$ 
\Ensure $AugmentedInputs$, set of pairs $\langle{}Scenario,objectDiagram\rangle{}$ that maximize the subtype coverage metric
\Function{maximizeSubTypeCoverage}{ScenariosInputs} 
	\State $\mathit{AugmentedInputs} \gets \mathit{copy(SecenariosInputs)}$ 
	\For {\textbf{each} $\mathit{\langle{}Scenario,objectDiagram\rangle{}}\ in\ \mathit{ScenariosInputs}$} 
		\For {\textbf{each} $\mathit{\langle{}step\rangle{}}\ in\ \mathit{Scenario}$} 
		\If {($\mathit{node}$ is a $\mathit{Condition}$ instance evaluated to true)} \label{alg:maximizeSubTypeCoverage:conditions}
		\State $\mathit{subtypes} \gets \mathit{subtypesOf(} \mathit{identifyUsedEntity(step))}$  \label{alg:maximizeSubTypeCoverage:findSubtypes}
		\For {\textbf{each} $\mathit{sT}\ in\ \mathit{subtypes}$} 
		\State $\mathit{newObjDiagr} \gets \mathit{solveWithSubtype(Scenario,sT)}$		\label{alg:maximizeSubTypeCoverage:subtypes}
				\If {($\mathit{coverageImproved(newObjDiagr,AugmentedInputs))}$} \label{alg:maximizeSubTypeCoverage:ifBetter}
					\State $\mathit{AugmentedInputs} \gets $ \\ \hspace{2cm} $ \mathit{AugmentedInputs} \cup \langle \mathit{Scenario,newObjDiagr} \rangle$ \label{alg:maximizeSubTypeCoverage:keep}
				\EndIf %
		\EndFor	%
		\EndIf	%
		\EndFor %
	\EndFor	%

\State \Return $\mathit{AugmentedInputs}$
\EndFunction

\end{algorithmic}
\caption{Function to maximize subtype coverage.} %
\label{alg:maximizeSubTypeCoverage}
\end{figure}

\MINREV{R1.4}{The detailed \emph{time complexity analysis} for \emph{GenerateScenariosAndInputs} is provided in Appendix~\ref{app:complexity}. In the worst case, the generation of use case scenarios is at least quintic with respect to the number of nodes in the use case test model (\UCTM), i.e., $\BigO(N^{T+4})$. In the average case, we expect the complexity to be cubic, i.e., $\BigO(N^{3})$. However, we observe that, based on our empirical evaluation, the generation of use case scenarios with \UMTG is feasible in practical time. The main explanation is that the number of nodes in \UCTM{s} is usually not high since use case specifications, in practice, tend to describe high-level system behavior, e.g., system-actor interactions. 
For example, the largest \UCTM in our case studies, which has been generated for \BodySense, includes 154 nodes.
With a relatively small number of nodes, \UCTM{s} can be processed fast by using modern hardware (e.g., laptops), even in the presence of $\BigO(N^3)$ complexity.}

\subsection{Generation of Object Diagrams} %
\label{subsec:testInputGeneration}

\M{} employs the Alloy analyzer to generate an object diagram which satisfies the path condition of a given scenario (Line~\ref{alg:inputs:solve} in Fig.~\ref{alg:testInputsGen}). Test input values are later extracted from the generated object diagram (see Section~\ref{sec:generationOfTestCases}).

The Alloy analyzer does not return a solution for unsatisfiable path conditions. These path conditions are usually associated with infeasible scenarios (i.e., scenarios which cannot be tested and verified by any set of possible input values).
However, we may also observe unsatisfiable path conditions with feasible scenarios. These scenarios include an internal step that overrides the effects of the previous internal step (e.g., redefines the value of a variable).
For instance, in Scenario B in Fig.~\ref{fig:scenarioExamples}, the constraints 
\emph{C14} (i.e., {\scriptsize $\mathit{TemperatureLowError.allInstances()} \rightarrow \mathit{forAll(i} | \mathit{i.detected} = \mathit{true)}$}) and
\emph{C2} (i.e, {\scriptsize$\mathit{TemperatureError.allInstances()} \rightarrow \mathit{forAll(i} | \mathit{i.detected} = \mathit{false)}$} )
cannot be satisfied in the same path condition since \emph{TemperatureLowError} is a subclass of \emph{TemperatureError}.
The internal step with \emph{C14} (Line~\ref{uc2:af:resetTempLowError} in Table~\ref{useCaseInitialOccupancy}) sets a temperature error to \textit{detected} after the internal step with \emph{C2} (Line~\ref{uc2:af:setTempErrors} in Table~\ref{useCaseInitialOccupancy}) sets all the temperature errors to \textit{undetected}. %
We identify such feasible scenarios having unsatisfiable path conditions by incrementally re-executing the Alloy analyzer.

When the Alloy analyzer does not solve a path condition, we automatically rerun it incrementally, by following the order of the steps in the scenario, until it identifies an unsatisfiable prefix of the path condition and returns the unsat core\footnote{The unsat core is the minimal subset of OCL constraints that cannot be solved together~\cite{Torlak2008}}.
For instance, for Scenario B, \M{} runs the Alloy analyzer on (\emph{C1}), (\emph{C1} $\wedge$ \emph{C2}), 
(\emph{C1} $\wedge$ \emph{C2} $\wedge$ \emph{C3}), 
(\emph{C1} $\wedge$ \emph{C2} $\wedge$ \emph{C3} $\wedge$ \emph{C4}),
(\emph{C1} $\wedge$ \emph{C2} $\wedge$ \emph{C3} $\wedge$ \emph{C4} $\wedge$ \emph{C5}),
and (\emph{C1} $\wedge$ \emph{C2} $\wedge$ \emph{C3} $\wedge$ \emph{C4} $\wedge$ \emph{C5} $\wedge$ \emph{C14}), 
respectively. 
For (\emph{C1} $\wedge$ \emph{C2} $\wedge$ \emph{C3} $\wedge$ \emph{C4} $\wedge$ \emph{C5} $\wedge$ \emph{C14}), the Alloy analyzer returns (\emph{C2} $\wedge$  \emph{C14}) as the unsat core.

If the last OCL constraint added to the path condition (e.g., \emph{C14} in our example) belongs to an internal step $s_{i}$, we conclude that $s_{i}$ overrides the effects of previous internal steps. %
The other constraints in the unsat core can thus be safely removed from the path condition %
if they refer to the same attributes referred by this last OCL constraint.
\M{} proceeds to incrementally solve the remaining constraints in the scenario. %
Otherwise, if the last added OCL constraint does not belong to an internal step or if the  other constraints in the unsat core refer to different variables, the scenario is considered infeasible.
For instance, both \emph{C2} and \emph{C14} belong to internal steps in Scenario B; \emph{C14} is the last constraint added to the path condition before the unsat core is returned. 
Therefore, \emph{C2} is removed from 
(\emph{C1} $\wedge$ \emph{C2} $\wedge$ \emph{C3} $\wedge$ \emph{C4} $\wedge$ \emph{C5} $\wedge$ \emph{C14}). 
We incrementally rerun the Alloy analyzer by adding the remaining constraints in Scenario B to 
(\emph{C1} $\wedge$ \emph{C3} $\wedge$ \emph{C4} $\wedge$ \emph{C5} $\wedge$ \emph{C14}). %
Constraint solving proceeds until the scenario is deemed infeasible or the entire path condition excluding the removed OCL constraint(s) is solved. 
In this case, the Alloy analyzer returns an object diagram that satisfies the entire path condition for Scenario B shown in Fig.~\ref{fig:scenarioExamples} excluding \emph{C2}. The OCL constraints removed from the path condition during incremental constraint solving belong to internal steps that perform initial setups and thus do not have any effect on the identification of test inputs.

\section{Generation of Test Cases}
\label{sec:generationOfTestCases}

\begin{figure}[t]
  \centering
    \includegraphics{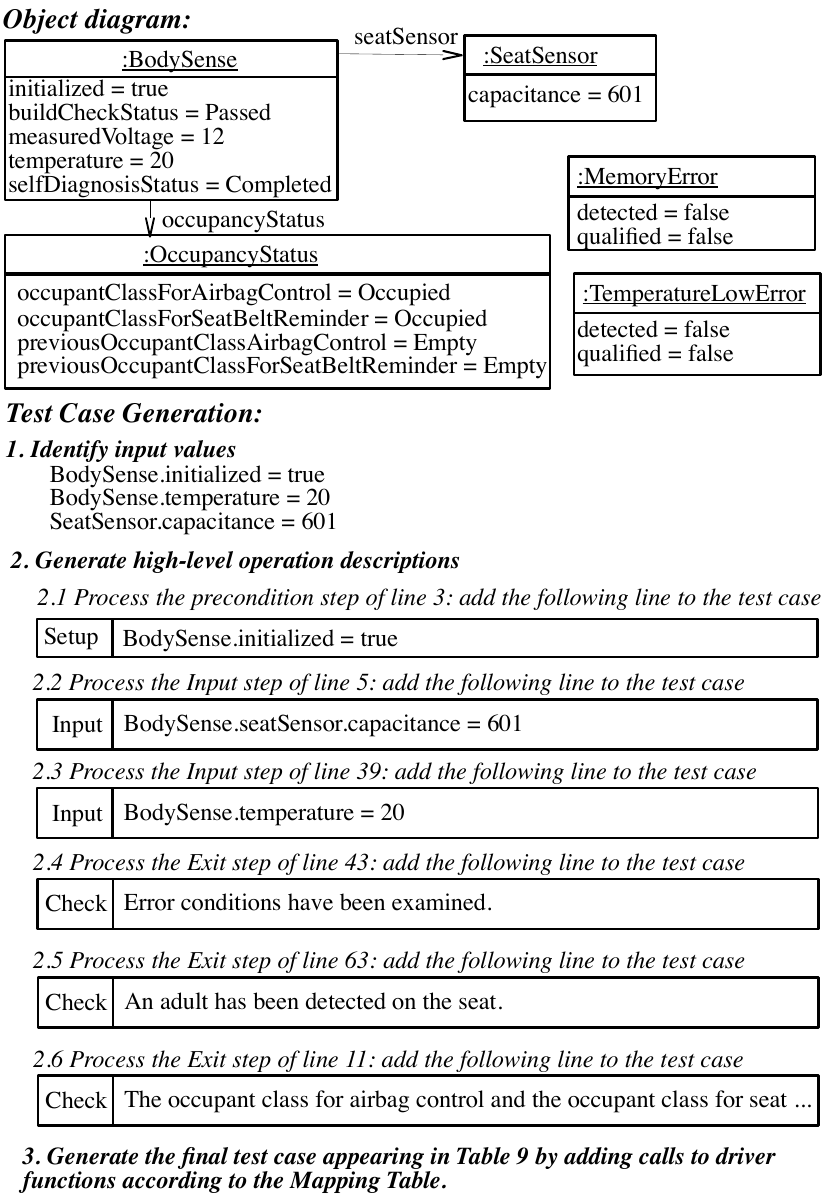}
      \caption{Activities performed to generate the test case in Table~\ref{tab:generatedTestCase}.}
      \label{fig:testCaseGeneration}
\end{figure}

\M{} uses the set of pairs $\langle{}scenario,objectDiagram\rangle$ to generate test cases (Step 10 in Fig.~\ref{fig:approach}). 
It performs three activities: %
\textit{identify input values}, \textit{generate high-level operation descriptions}, and \textit{generate test driver function calls} (see Fig.~\ref{fig:testCaseGeneration}).
Test driver function calls are tailored to the test case format in Table~\ref{tab:testCase} but can be adapted, while following similar principles, to different test case formats for different test infrastructures and hardware.

\begin{table}[h]
\scriptsize
\caption{The test case automatically generated from Scenario A in Fig.~\ref{fig:scenarioExamples}.}
\begin{center}
\begin{tabular}{|@{\hspace{0.05cm}}p{0.4cm}|@{\hspace{0.05cm}}p{2cm}|@{\hspace{0.05cm}}p{5.6cm}|}
\hline
\textbf{Line}&\textbf{Operation}&\textbf{Inputs/Expectations}\\
\hline
1&	\textit{Input}&System.initialized = true\\
\hline
2&	ResetPower&	Time=INIT\_TIME\\
\hline
3&	\textit{Input}&System.seatSensor.capacitance = 601\\
\hline
4&	SetBus&Channel=RELAY Capacitance=601\\
\hline
5&	\textit{Input}&System.temperature = 20\\
\hline
6&	SetBus&Channel=RELAY Temperature = 20\\
\hline
7&	\textit{Check}& An adult has been detected on the seat.\\
\hline
8&    ReadAndCheckBus&D0=OCCUPIED\\
&    &D1=OCCUPIED\\
\hline
9&	\textit{Check}&
The occupant class for airbag control has been sent \\
&&to AirbagControlUnit. The occupant class for seat\\
&&belt reminder has been sent to SeatBeltControlUnit.\\
\hline
10&	CheckAirbagPin&
0x010\\
\hline

\hline
\end{tabular}
\end{center}
\label{tab:generatedTestCase}
\end{table}%

Table~\ref{tab:generatedTestCase} shows the test case automatically generated from Scenario A in Fig.~\ref{fig:scenarioExamples}. 
The test cases generated by \UMTG follow the structure presented in Section~\ref{sec:context}. The odd lines
contain high-level operation descriptions followed (even line numbers) by the name of the functions that should be executed by the test driver along with the corresponding input and expected output values.
Lines 1, 3, 5, 7 and 9 are high-level operation descriptions; Lines 2, 4, 6, 8 and 10 are test driver function calls. The test case corresponds to the manually written test case in Table~\ref{tab:testCase}.

The input values are extracted from the object diagram generated from the path condition of the given scenario (Activity 1 in Fig.~\ref{fig:testCaseGeneration}). To this end, \M{} looks for the attributes that appear both
in the diagram and the path condition.

\M{} then generates the high-level operation descriptions, i.e., the \emph{Input},  \emph{Setup}, and \emph{Check} operations (Activity 2 in Fig.~\ref{fig:testCaseGeneration}).
The \emph{Input} and \emph{Setup} operations are associated with the input values (Activities 2.1 - 2.3). %
For each \emph{Input} and \emph{InterruptingCondition} step in the scenario, we generate a test line including an \emph{Input} operation. The \emph{Input} operation is associated with a set of input values %
which belong to an attribute or an entity referenced in the use case step. %
For instance, in Activity 2.2, `\emph{BodySense.seatSensor.capacitance = 601}' is selected %
because of the \emph{Capacitance} entity.
We generate a test line including the \emph{Setup} operation %
for each attribute that appears in an OCL constraint belonging to \emph{UseCaseStart} or \emph{Condition} steps but not to \emph{Input} and \emph{InterruptingCondition} steps.
For instance, in Activity 2.1, `\emph{BodySense.initialized = true}' is selected %
because the attribute \emph{initialized} is referenced in the use case precondition \emph{The system has been initialized}.
Finally, for the postcondition of each \emph{Exit} step, we create a test line including the \emph{Check} operation and the postcondition in NL (Activities 2.4 - 2.6). 
\MREVISION{R3.13}{A \emph{Check} operation is an abstract test oracle. The generated sequence of high-level operation descriptions is an \emph{abstract test case}.}

Finally, to generate the test driver function calls \CHANGED{for the executable test case}, \M{} 
processes the high-level operation descriptions by using the mapping table provided by the engineers (Activity 3 in Fig.~\ref{fig:testCaseGeneration}).
Table~\ref{tab:mappingTable} shows part of the mapping table for $BodySense$. 
The first two columns provide the operation names and the regular expressions that match the high-level operation descriptions in the test case.
The last two columns provide the test driver function calls to be added to the test case when the first two columns match the high-level operation descriptions.
For instance, the \emph{Input} operation and the expression \emph{`BodySense.seatSensor.capacitance = (.*)'} in the first row of Table~\ref{tab:mappingTable} matches the high-level operation description in Line 3 in Table~\ref{tab:generatedTestCase}, which leads to the generation of Line 4. %
We follow the Java regular expressions' syntax~\cite{JavaRegex}, which provides the grouping feature to extract char sequences from matching strings. For example, we use the grouping pattern \emph{`(.*)'} to copy the values of the high-level operation descriptions, which are extracted from the object diagram, into the test driver function calls (e.g., value 601 in Activity 2.2).

Test oracles are the test driver function calls which are generated from the high-level operation descriptions with the \emph{Check} operation (e.g., Lines 8 and 10 in Table~\ref{tab:generatedTestCase}). 

To simplify the writing of mapping tables, engineers can create them iteratively and incrementally. 
They define mappings for the first test case and then introduce (and refine) regular expressions for the following test cases. 
Moreover, when the grouping feature of regular expressions is used to copy input values generated by the OCL solver into executable test cases, 
the same mapping table can be reused for test cases generated with different coverage criteria. 
Indeed, test cases derived with different coverage criteria differ only with respect to the values assigned to inputs and the order of operations. 
Finally, mapping tables can be shared across all the systems tested on a given test infrastructure.

\begin{table}[t]
\scriptsize
\caption{Part of the mapping table for $BodySense$.}
\begin{center}
\begin{tabular}{|p{0.8cm}|p{2.5cm}|p{1.05cm}|p{2.2cm}|}
\hline
\multicolumn{2}{|c}{\textbf{Pattern to match}}&
\multicolumn{2}{|c|}{\textbf{Result}}\\
\multicolumn{2}{|c}{\textbf{(Operation and Inputs)}}&
\multicolumn{2}{|c|}{\textbf{(Operation and Inputs)}}\\
\hline
Input&BodySense.seatSensor.ca&
SetBus&Channel = RELAY\\
&pacitance = (.*)&& Capacitance = \textbackslash{}1\\
\hline
Input&System.initialized = true
& 	Reset Power&	Time=INIT\_TIME\\
\hline
Check&An adult has been de&
ReadAnd&	D0=OCCUPIED \\
&tected on the seat.&CheckBus&D1=OCCUPIED\\
\hline
\end{tabular}
\end{center}
\label{tab:mappingTable}
\end{table}%

\section{Tool Support}
\label{sec:tool}

To ease its industrial adoption, we implemented \UMTG as a toolset that extends 
IBM Doors~\cite{WebDoors} and Eclipse IDE~\cite{EclipseWeb}.
Additional information about the \M{} toolset, including executable files, download instructions and a screencast covering motivations, are available on the tool's website at \url{https://sntsvv.github.io/UMTG/}.

\begin{figure}[t]
  \centering
    \includegraphics{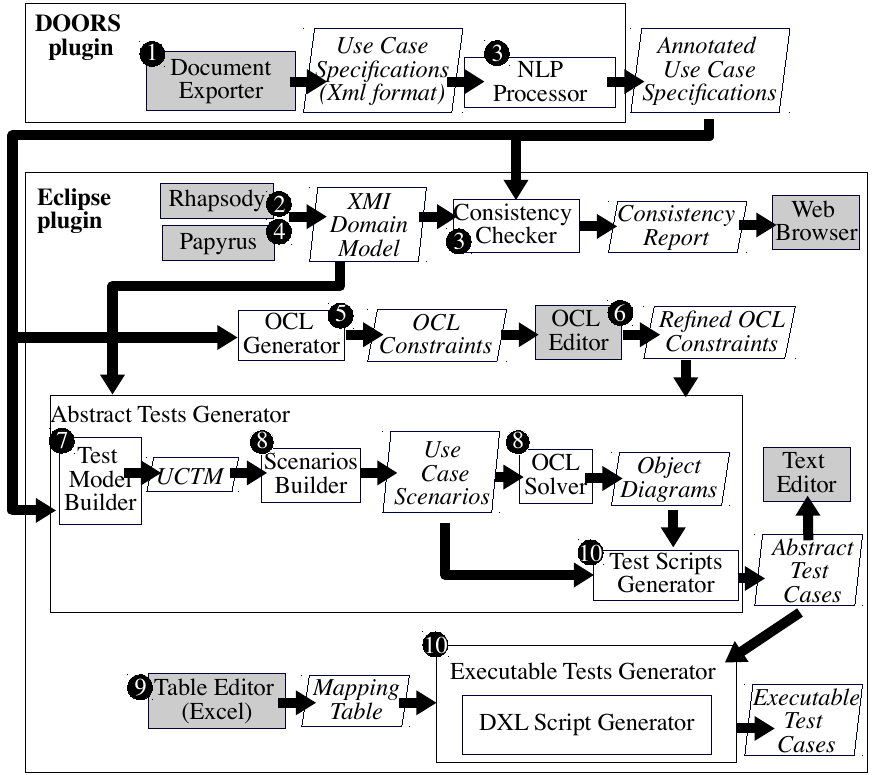}
      \caption{UMTG toolset architecture. Grey boxes show third party components, while white boxes denote \UMTG components. Black circles denote the steps in Fig.~\ref{fig:approach}.}
      \label{fig:arc}
\end{figure}

Fig.~\ref{fig:arc} shows the tool architecture. %
The main components are two plug-ins that extend IBM Doors and Eclipse IDE. 
They %
orchestrate the other components in Fig.~\ref{fig:arc}. %

The IBM Doors plug-in activates the \M{} steps for use case elicitation while the steps for test case generation are activated by the Eclipse plug-in. %
The toolset can also process use case specifications in plain text files to avoid any dependence on IBM Doors for requirements elicitation.

The Eclipse plug-in architecture enables us to rely on third party plug-ins. %
We rely on the \emph{Eclipse Web Browser} to visualize the missing entities in the domain model, the \emph{Eclipse OCL editor}~\cite{EclipseOCL} to edit the OCL constraints, the default \emph{Eclipse Table Editor} (e.g., Microsoft Excel) to edit the mapping table, and the \emph{Eclipse Project Explorer} to list the \M{} artefacts. %
To create the domain model, the user can use the IBM Rhapsody plug-in~\cite{WebRh} or the Eclipse Papyrus plug-in~\cite{WebPapyrus}. \M{} requires that the domain model be saved in the XMI format. %

The components in Fig.~\ref{fig:arc} %
automate Steps 3, 5, 7, 8, and 10 in Fig.~\ref{fig:approach}. The \textit{NLP Processor} %
contains the NL pipeline described in Section~\ref{sec:nlpPipeline} to annotate the use case specifications as required by Steps 3, 5 and 7. It is based on the GATE workbench~\cite{Cunningham2002}, an open source NLP framework. To load use case specifications from IBM Doors, we use \emph{Doors Document Exporter}, an IBM Doors API that exports the Doors content as xml files.

 \MREVISION{R1.46}{The \emph{Consistency Checker} implements Step 3 in Fig.~\ref{fig:approach}. The \emph{OCL Generator} implements Step 5;  \MREVISION{R1.47}{it relies on the \emph{CNP SRL} tool, \emph{VerbNet}, and \emph{WordNet}.}
 The \emph{Abstract Tests Generator} includes the components \emph{Test Model Builder}, \emph{Scenarios Builder}, and \emph{OCL Solver}, which implement Steps 7 and 8.
  The \emph{OCL Solver} employs the Alloy analyzer~\cite{Jackson2006} to generate object diagrams.
 The \emph{Test Scripts Generator} partially implements Step 10 by generating abstract test cases, i.e., textual test scripts with input values and high-level operation descriptions.
 The \emph{Test Scripts Generator} also filters out replicated test cases that might result from the execution of \UMTG against multiple use cases that share a subset of scenarios (e.g., the use case scenarios for unsuccessful self diagnosis in \BodySense).
The \emph{Eclipse Text Editor} is used to inspect the abstract test cases.}

The \emph{Executable Tests Generator} takes the mapping table and the abstract test cases as input, and generates the executable test cases as output (Step 10). It exploits the Door eXtension Language (DXL) to load the generated test cases into IBM Doors.
The DXL is also used to automatically generate trace links between use cases and generated test cases, an important feature to support testing of safety critical systems.

\MREVISION{R3.4}{Our toolset facilitates the adoption of \UMTG into a software development process by supporting both the generation of test cases from scratch and 
the maintenance of the generated test cases in the presence of requirements changes (i.e., when use case specifications are updated). 
In both cases, engineers simply follow the \UMTG steps in Fig.~\ref{fig:approach}.
In the case of requirements changes, the effort required by manual activities is minimal.
More precisely, in Step 4 (\textit{Refine Model}), engineers will likely revise only a limited number of missing entities because the update of requirements usually introduces a limited number of changes in the domain concepts.
In Step 5, \textit{Generate Constraints}, 
\UMTG relies on cached results to generate OCL constraints for unchanged use case steps; 
engineers manually inspect only the OCL constraints generated from new or modified use case steps.
In Step 6, \textit{Specify Missing Constraints}, engineers specify only the OCL constraints that were not generated from the updated use case sentences.
Finally, in Step 10, engineers rely on the mapping table derived for the previous version of the system. Consequently, they typically introduce only a limited number of mapping table entries.}

\section{Empirical Evaluation} 
\label{sec:evaluation}

In this section, we investigate, based on two industrial case studies in the automotive domain, the following Research Questions (RQs):

\begin{itemize}

\item \textit{\textbf{RQ1.}} \textit{\textbf{Does the approach automatically generate the correct OCL constraints?}} The generation of OCL constraints is a major building block of our approach. With this research question, we aim to measure the precision %
(i.e., the fraction of the generated OCL constraints that are correct)
and recall (i.e., the fraction of the required OCL constraints that are correctly generated) of the OCL generation algorithm in Section~\ref{sec:oclGeneration}.

\item \textit{\textbf{RQ2}} \textit{\textbf{How does the approach compare to manual test case generation based on expertise?}} In the second research question, we evaluate (i) if the approach %
generates test cases that exercise the same use case scenarios exercised by test cases manually written by experts and (ii) if it discovers additional, relevant scenarios not exercised by the manually written test cases.

\item \textit{\textbf{RQ3.}} \textit{\textbf{How does the approach compare to manual test case generation in terms of resources?}} This research question aims to compare 
our approach with manual test case generation in terms of engineers' effort and time required to generate test cases.

\end{itemize}

\MREVISION{R1.60 R3.16}{As discussed in the introduction, our empirical evaluation focuses on requirements coverage instead of code coverage and fault detection. This is because the purpose of acceptance testing is not fault detection (i.e., verification) but validation.
Earlier testing activities focus on verification. For example, ISO26262, the main international automotive standard, enforces statement, branch, and MC/DC coverage for unit testing only~\cite{ISO26262}. Therefore, analyzing the effectiveness of \UMTG in terms of fault detection is out of the scope of this paper.} 
\MREVISION{R1.4}{Further, we do not compare \UMTG with model-based approaches because their adoption in our context is not feasible (see Section~\ref{sec:context}).}

\subsection{Subjects of the Evaluation}
\label{sec:subjects}

The subjects of our evaluation are \BodySense and Hands Off Detection (\HOD), two embedded systems developed by our industry partner, IEE~\cite{IEE}.

\BodySense is a car seat occupant classification system that enables smart airbag deployment; it has been introduced in Section~\ref{sec:context}.
\HOD is an embedded system that checks if the driver has both hands on the steering wheel. It is used to enable and disable automatic braking with driver-assist features for safety. For instance, a car should not automatically brake if the driver does not have both his hands on the steering wheel. \HOD measures the capacitance between a conductive layer in the steering wheel and the electrical ground in the car body or in the seat frame. %
It notifies the autonomous driving assistance system when the driver does not have both his hands on the steering wheel. 
The complexity of these two systems lies mostly in how they deal with errors, e.g., sensor break-downs. 
Therefore, their test suites mostly concern the verification of system behavior in the presence of errors.
The two case studies are representative of automotive embedded systems and share their typical characteristics, such as handling multiple types of errors and communicating results to other components based on error status and sensor data.

\begin{table}[h]
\caption{Overview of the \BodySense and \HOD use case specifications.}
\begin{center}
\scriptsize
\begin{tabular}{|c|c|c|c|c|c|}
\hline
&\multicolumn{3}{c|}{\textbf{Use cases}}&\textbf{Use case}&\textbf{Steps}\\
&\textbf{All}&\textbf{Entry}&\textbf{Included by others}&\textbf{flows}&\\
\hline
\BodySense&7&2&5&59& 266\\
\HOD&17&11&6&  27 &  76\\
\hline
\end{tabular}
\end{center}
\label{table:caseStudiesOverview}
\end{table}%

In IEE's business context, like in many others, use cases are central development artifacts which are used for communicating requirements among stakeholders, such as customers. %
Use case specifications provide a systematic description of system-actor interactions. %
\UMTG requires use case specifications in the RUCM template. 
For \BodySense, the specifications had been initially elicited by IEE engineers following the Cockburn template~\cite{Cockburn-WritingEffectiveUseCases-Book-2000}. They were later rewritten according to the RUCM template. 
The \HOD specifications, instead, were already written by IEE engineers using RUCM with the UMTG and RUCM tools~\cite{Wang2015}.

Table~\ref{table:caseStudiesOverview} reports on the size of the use case specifications of the case studies. 
The \BodySense and \HOD specifications include 7 and 17 use cases, respectively. %
The number of use case flows is 59 for \BodySense and 27 for \HOD, which indicates that the systems must deal with multiple alternative executions (e.g., alternative flows in the case of errors). 
The difference in the number of use case flows is largely explained by \HOD specifications being less complete than \BodySense specifications.
\MREVISION{R1.50}{In general, we consider use case specifications to be \emph{complete} when their condition steps fully characterize equivalence partitions for system inputs. 
For example, in \BodySense, all error conditions are modeled by means of condition steps 
that explicitly characterize anomalous sensor data (e.g., sensor data not in a given range as in Line~\ref{uc2:TCOND} of Table~\ref{useCaseClassify}).
Consequently, \BodySense specifications include an alternative flow for every input partition leading to an error.
In contrast, use case specifications are \emph{incomplete} when such equivalence partitions are partially modeled.
This is the case of \HOD specifications, which do not include condition steps characterizing anomalous sensor data
but only describe how the system should behave in the presence of a specific error or a class of errors (e.g., Line~\ref{uc1:bf:validate} of Table~\ref{useCaseClassify}, which verifies if any error has been detected by the system).
\HOD specifications include less alternative flows than the \BodySense ones.
In this case, the descriptions of error conditions are given in other documents and the different error types appear in the domain model.
To test the system under all the possible error conditions, engineers have to carefully read multiple specification documents including the domain model.
The impact of the completeness of use case specifications on \UMTG automated test generation is discussed in the following sections.}

In Table~\ref{table:caseStudiesOverview}, 
column \emph{Entry} reports the number of use cases that are entry points for testing (i.e., the main use cases for deriving acceptance test cases). 
\MREVISION{R1.48}{Column \emph{Included by others} reports the number of use cases that are included by other use cases.}

\subsection{Results}

This section discusses the results of our case studies, addressing in turn each of the research questions.

\subsubsection{RQ1: Does the approach automatically generate the correct OCL constraints?}
\label{subsec:empirical:RQ1}

To respond to RQ1, we used \UMTG to automatically generate the OCL constraints required to drive test generation for \BodySense and \HOD. 
We compared the automatically generated OCL constraints with the OCL constraints manually written by the first author of the paper. %
The latter were written together with IEE engineers in dedicated workshops. %
The comparison was fully automated by means of string matching.

We computed precision (i.e., the fraction of automatically generated constraints that are correct) and recall (i.e., the fraction of the required constraints for test case generation that are correctly generated by \UMTG). 
A sentence may appear in multiple use case steps. To fairly compute precision and recall, we considered these sentences only once. %

Table~\ref{table:rq1} presents the results for RQ1. %
\UMTG generated 70 and 34 constraints from 86 and 36 sentences for \BodySense and \HOD, respectively. 
The number of the processed sentences is lower than the number of the use case steps because \UMTG does not need to generate constraints from input, output, and exit steps. 
Out of 70 and 34 constraints, 68 and 36 are correct, respectively. %
Precision is 0.97 for \BodySense and 1.00 for \HOD. These results are highly promising since almost each generated constraint is correct.

\begin{table}[htp]
\caption{Precision and Recall of the automated generation of OCL constraints.}
\begin{center}
\scriptsize
\begin{tabular}{|@{\hspace{0.5mm}}c@{\hspace{0.5mm}}|@{\hspace{0.5mm}}c@{\hspace{0.5mm}}|@{\hspace{0.5mm}}c@{\hspace{0.5mm}}|@{\hspace{0.5mm}}c@{\hspace{0.5mm}}|@{\hspace{0.5mm}}c@{\hspace{0.5mm}}|c@{\hspace{0.5mm}}|@{\hspace{0.5mm}}r@{\hspace{0.5mm}}|@{\hspace{0.5mm}}r@{\hspace{1mm}}|}
\hline
&\textbf{Processed}&\multicolumn{3}{c|}{\textbf{Generated constraints}}&\textbf{Not}&\textbf{Precision}&\textbf{Recall}\\
&\textbf{sentences}& \textbf{All} & \hspace{1mm} \textbf{Incorrect}\ \ & \textbf{Correct}&\textbf{Generated}&&\\
\hline
\BodySense	&86 		&70&2	&68 	& 18&	0.97&	0.77\\
\HOD		&36		&34&0	 &34 &2  &	1.00&	0.94\\
\hline
\end{tabular}
\end{center}
\label{table:rq1}
\end{table}%

For certain sentences,
the automated constraint generation fails (the columns \textit{Not Generated} and \textit{Incorrect} in Table~\ref{table:rq1}) because of imprecise writing and inconsistent terminology.  %
For example, the step `\emph{The system VALIDATES THAT the temperature is valid}' does not specify what a valid temperature is %
(i.e., a valid temperature range). %
We also cannot automatically generate a constraint from the step `\emph{The system VALIDATES THAT the occupancy status is valid}' %
because it is not straightforward to determine that `\emph{is valid}' should be translated to \texttt{<> Error}. The noun \texttt{error} is not even an antonym of the adjective \texttt{valid}.
As a result, especially for BodySense, recall is not as good as what we had hoped for. Recall is better for \HOD than for \BodySense since IEEE engineers wrote the \HOD specifications after \BodySense. 
They had more experience and were more careful in writing the \HOD specifications.

To avoid imprecise sentences and inconsistent terminology, we improved the specifications and domain models. For example, we defined the notion \emph{valid} as a derived attribute in the domain model of \BodySense. %
Our changes %
for \BodySense include eight additional derived attributes, the rewriting of
five redundant and inconsistent sentences, the renaming of two attributes in the domain model, and the refactoring of five concepts to be modelled using entities (abstract classes in two cases) instead of attributes. 
For \HOD, we introduced one derived attribute. 
This limited number of changes indicates that it is not difficult for engineers to improve their modeling practices.

\begin{table}[htp]
\caption{Precision and Recall with the Improved Specifications and Models.}%
\begin{center}
\scriptsize
\begin{tabular}{|@{\hspace{0.5mm}}c@{\hspace{0.5mm}}|@{\hspace{0.5mm}}c@{\hspace{0.5mm}}|@{\hspace{0.5mm}}c@{\hspace{0.5mm}}|@{\hspace{0.5mm}}c@{\hspace{0.5mm}}|@{\hspace{0.5mm}}c@{\hspace{0.5mm}}|c@{\hspace{0.5mm}}|@{\hspace{0.5mm}}r@{\hspace{0.5mm}}|@{\hspace{0.5mm}}r@{\hspace{1mm}}|}
\hline
&\textbf{Processed}&\multicolumn{3}{c|}{\textbf{Generated constraints}}&\textbf{Not}&\textbf{Precision}&\textbf{Recall}\\
&\textbf{sentences}&\textbf{All} &\hspace{1mm} \textbf{Incorrect}\ \ & \textbf{Correct}&\textbf{Generated}&&\\
\hline
\BodySense	&\MINREVT{R1.2}{84} 		&81 &1	&80 	& 3&	0.99&	0.95\\
\HOD		&\MINREVT{R1.2}{36}		&35 &0	 &35 &1  &	1.00&	0.97\\
\hline
\textbf{Overall}		&\MINREVT{R1.2}{120}		&116 &1	 &115 &4  &	0.99&	0.96\\
\hline
\end{tabular}
\end{center}
\label{table:rq1:clean}
\end{table}%

Table~\ref{table:rq1:clean} presents the results obtained with the improved specifications and models, which in practice can be easily obtained by carefully comparing the specifications and domain models, a task our tool now supports. Unsurprisingly, %
\UMTG achieves a particularly high recall, above 0.9 in both case studies. Overall, it achieves a precision of 0.99 and a recall of \MINREV{R1.2}{0.96}, another promising result indicating the potential usefulness of our constraint generation in practical settings. %
Among the sentences that are not correctly translated into OCL constraints, we identify (a) three sentences capturing internal behavior not useful for test input generation (e.g., `\emph{The system loads the default calibration data}'), 
and (b) one sentence referring to concepts having names that are similar to other concepts in the domain model, which, in turn, leads to the identification of the wrong entity for the constraint.

\subsubsection{RQ2: How does the approach compare to manual test case generation based on expertise?}
\label{subsec:empirical:RQ2}

To respond to RQ2, we compared the test cases automatically generated by UMTG (hereafter \emph{\UMTG test cases}) with the test cases manually written, based on extensive expertise, by IEE engineers (hereafter \emph{manual test cases}) for \BodySense and \HOD.
We focused on use case scenarios %
and test inputs used to exercise the scenarios. %

To compare the test cases based on use case scenarios,
we inspected all the manual test cases and mapped them to the scenarios in the use case test models.
To compare the test cases based on test inputs, %
we identified equivalence classes for each input domain and all the input partitions (i.e., the combination of equivalence classes) that might be considered for each scenario. We kept track of the input partitions covered by \UMTG and manual test cases.
We use the term \emph{input scenario}, %
different than \emph{use case scenario}, to refer to a use case scenario being exercised with a specific input partition.

\MREVISION{R3.16}{Because of the nature of the practical problem we address, we have to rely on NLP. Such techniques are heuristics trained on large text corpora and are therefore heuristics. Therefore no proof of soundness is possible. We can, however, empirically assess soundness and this is what we do in our two representative industrial case studies. We manually verified that the sequences of instructions in the test cases satisfying the branch and def-use coverage criteria were consistent with the use case specifications.
Furthermore, for the test cases generated with the subtype coverage criterion, we verified that the covered input partitions (i.e., the types of errors being triggered by test execution) were consistent with the use case specifications.
In our analysis, we did not discover any test case being erroneously generated in the presence of correct OCL constraints, thus providing evidence that the approach is sound.}

Table~\ref{tab:RQ2:testCases} presents the number of test cases in the test suites considered in our evaluation (i.e., the manual test suite developed by IEE and the test suites generated with three \UMTG coverage strategies).
An in-depth discussion concerning the size and the complementarity of the generated test suites is reported at the end of this section.

\begin{table}[t]
\scriptsize
\caption{Number of test cases %
in our evaluation.}
\begin{center}
\begin{tabular}{|p{2cm}|r|r|r|r|r|}
\hline
Case study&\multicolumn{1}{c|}{Manual}&\multicolumn{3}{c|}{UMTG test cases}\\
&\parbox[c]{1cm}{test cases}&\parbox[c]{1cm}{Branch}&\multicolumn{1}{c|}{Def-use}&\multicolumn{1}{c|}{Subtype}\\
\hline
\BodySense	&	134	&	63&		97&		1285	\\
\HOD		&	50	&	21&		22&		63	\\
\hline
\end{tabular}
\end{center}
\label{tab:RQ2:testCases}
\end{table}%

\begin{table}[t]
\scriptsize
\caption{Use case scenarios in the manual test cases that are also exercised by the UMTG test cases.}
\begin{center}
\begin{tabular}{|@{\hspace{1mm}}p{1cm}|r|r@{\hspace{1mm}}r|r@{\hspace{1mm}}r|r@{\hspace{1mm}}r|}
\hline
Case&\multicolumn{1}{c|}{Exercised by the}&\multicolumn{6}{c|}{Exercised also by the UMTG test cases}\\
study&\parbox[c]{1.9cm}{manual test cases}&\multicolumn{2}{c|}{Branch}&\multicolumn{2}{c|}{Def-use}&\multicolumn{2}{c|}{Subtype}\\
\hline
\BodySense	&	64	&	43& (67.19\%)&		64&		(100\%)&		64	& (100\%)\\
\HOD		&	21	&	21& (100\%)&		21&		(100\%)&		21	& (100\%)\\
\hline
\end{tabular}
\end{center}
\label{tab:RQ2:scenarios}
\end{table}%

\begin{table}[t]
\scriptsize
\caption{Input scenarios in the manual test cases that are also exercised by the UMTG test cases.}
\begin{center}
\begin{tabular}{|@{\hspace{1mm}}p{1cm}|r|r@{\hspace{1mm}}r|r@{\hspace{1mm}}r|r@{\hspace{1mm}}r|}
\hline
Case&\multicolumn{1}{c|}{Exercised by the}&\multicolumn{6}{c|}{Exercised also by the UMTG test cases}\\
study&\parbox[c]{1.9cm}{manual test cases}&\multicolumn{2}{c|}{Branch}&\multicolumn{2}{c|}{Def-use}&\multicolumn{2}{c|}{Subtype}\\
\hline
\BodySense	&	134	&	96& (71.64\%)&		131&		(97.76\%)&		134	& (100\%)\\
\HOD		&	56	&	21& (37.50\%)&		21&		(37.50\%)&		50	& (100\%)\\
\hline
\end{tabular}
\end{center}
\label{tab:RQ2:behaviors}
\end{table}%

Table~\ref{tab:RQ2:scenarios} presents the number of use case scenarios exercised by manual test cases %
and the percentage of these scenarios exercised by UMTG test cases. %
Except for branch coverage in \BodySense, the \UMTG test cases exercise all the scenarios exercised by the manual test cases.
The \UMTG test cases for branch coverage do not exercise all the scenarios covering the detection of each type of error in the presence of qualified %
errors given that branch coverage is satisfied when at least one test case covers the branch for error qualification.
Note that the \UMTG test cases cover all the branches covered by the manual test cases. %

Table~\ref{tab:RQ2:behaviors} shows the number of input scenarios exercised by manual test cases %
and the percentage of these input scenarios exercised by \UMTG test cases. %
\UMTG fares worse with branch coverage %
because the test cases do not exercise scenarios with all input partitions.
With def-use coverage, \UMTG achieves a much better result for \BodySense, i.e., 97.76\% of the manual scenarios being exercised. 
This happens because the use case specifications of \BodySense are complete %
and capture all conditions that lead to error detection. %
Therefore, the use case scenarios that can be covered only by one single input partition make def-use coverage sufficient to cover all the input scenarios. 
A different result is achieved for \HOD with def-use coverage. %
The \HOD specifications are incomplete, which leads to multiple input partitions per use case scenario. Therefore, in \HOD, subtype coverage is necessary to exercise the input scenarios covered by the manual test cases.

\begin{table}[h]
\scriptsize
\caption{Use case scenarios in the UMTG test cases that are also exercised by the manual test cases.}
\begin{center}
\begin{tabular}{|@{\hspace{1mm}}p{1cm}|r|r@{\hspace{1mm}}r|r@{\hspace{1mm}}|r@{\hspace{1mm}}r|r@{\hspace{1mm}}|r@{\hspace{1mm}}r@{\hspace{1mm}}|}
\hline
&\multicolumn{9}{c|}{Use case scenarios covered by the generated test cases}\\
Case&\multicolumn{3}{c|}{Branch}&\multicolumn{3}{c|}{Def-use}&\multicolumn{3}{c|}{Subtype}\\
study
&\multicolumn{1}{@{\hspace{0.1mm}}c@{\hspace{0.1mm}}|}{\MINREVT{R1.3a}{UMTG}}
&\multicolumn{2}{c|}{Exercised}
&\multicolumn{1}{@{\hspace{0.1mm}}c@{\hspace{0.1mm}}|}{UMTG}
&\multicolumn{2}{c|}{Exercised}
&\multicolumn{1}{@{\hspace{0.1mm}}c@{\hspace{0.1mm}}|}{UMTG}
&\multicolumn{2}{c|}{Exercised}\\
&\multicolumn{1}{c|}{}&\multicolumn{2}{c|}{by manual}&\multicolumn{1}{c|}{}&\multicolumn{2}{c|}{by manual}&\multicolumn{1}{c|}{}&\multicolumn{2}{c|}{by manual}\\
\hline
\BodySense	&	63	&	43& (68.25\%)&		97&	64&		(65.98\%)	&		97&	64&		(65.98\%)		\\
\HOD		&	22	&	21& (94.45\%)&				22&	21&		(95.45\%) &		22&	21&		(95.45\%)		\\
\hline
\end{tabular}
\end{center}
\label{tab:RQ2:UMTGscenarios}
\end{table}%

\begin{table}[h]
\scriptsize
\caption{Input scenarios in the \UMTG test cases that are also exercised by the manual test cases.}
\begin{center}
\begin{tabular}{|@{\hspace{1mm}}p{1cm}|r|r@{\hspace{1mm}}r|r@{\hspace{1mm}}|r@{\hspace{1mm}}r|r@{\hspace{1mm}}|r@{\hspace{1mm}}r@{\hspace{1mm}}|}
\hline
&\multicolumn{9}{c|}{Input scenarios covered by the generated test cases}\\
Case&\multicolumn{3}{c|}{Branch}&\multicolumn{3}{c|}{Def-use}&\multicolumn{3}{c|}{Subtype}\\
study
&\multicolumn{1}{@{\hspace{0.1mm}}c@{\hspace{0.1mm}}|}{UMTG}
&\multicolumn{2}{c|}{Exercised}
&\multicolumn{1}{@{\hspace{0.1mm}}c@{\hspace{0.1mm}}|}{UMTG}
&\multicolumn{2}{c|}{Exercised}
&\multicolumn{1}{@{\hspace{0.1mm}}c@{\hspace{0.1mm}}|}{UMTG}
&\multicolumn{2}{c|}{Exercised}\\
&\multicolumn{1}{c|}{}
&\multicolumn{2}{c|}{by manual}&\multicolumn{1}{c|}{}
&\multicolumn{2}{c|}{by manual}&\multicolumn{1}{c|}{}&\multicolumn{2}{c|}{by manual}\\
\hline
\BodySense	&	63	&	43& (68.25\%)&		97&	64&		(65.98\%)	&		867&	74&		(8.54\%)		\\
\HOD		&	22	&	18& (81.82\%)&		22&	18&		(81.82\%) &		71&	56&		(78.87\%)		\\
\hline
\end{tabular}
\end{center}
\label{tab:RQ2:UMTGbehaviors}
\end{table}%

Table~\ref{tab:RQ2:UMTGscenarios} presents the number of use case scenarios exercised by \UMTG test cases %
and the percentage of these scenarios exercised by manual test cases.
\UMTG test cases exercise more use case scenarios than manual test cases. For example, in \BodySense,
only 68\% of the scenarios for branch coverage are exercised by manual test cases.
With branch coverage, \UMTG systematically covers all the scenarios derived from bounded and global alternative flows. It is difficult to manually identify these scenarios because engineers need to create a scenario for each reference flow step which a bounded (or global) flow refers to. %
Therefore, engineers tend not to test all the scenarios that can be derived from bounded and global alternative flows. %
With def-use coverage, \UMTG systematically covers all def-use pairs. In our case studies, the def-use pairs consist of internal steps reporting the presence of specific errors and condition steps verifying the presence of these errors. 
In turn, \UMTG ensures that the systems are tested for each error in all possible combinations of condition steps verifying the presence of the error (e.g., Scenarios B and C in Fig.~\ref{fig:scenarioExamples}). %
Such systematic combination coverage is hard to achieve for engineers.
By definition, with the subtype and def-use coverage criteria, \UMTG generates the same use case scenarios.

Table~\ref{tab:RQ2:UMTGbehaviors} shows the number of input scenarios exercised by \UMTG test cases and the percentage of these input scenarios exercised by manual test cases.
\UMTG test cases exercise more input scenarios than manual test cases. 
For the branch and def-use coverage criteria, the numbers of input and use case scenarios exercised by UMTG test cases are the same (see Tables~\ref{tab:RQ2:UMTGscenarios} and~\ref{tab:RQ2:UMTGbehaviors}).
With subtype coverage, \UMTG systematically covers each condition step for all possible subtypes (e.g., for all error types in our case studies). %
In our case studies, to speed up manual test case generation, engineers do not consider all error types in condition steps.

\begin{figure}[h]
    \includegraphics[width=8.4cm]{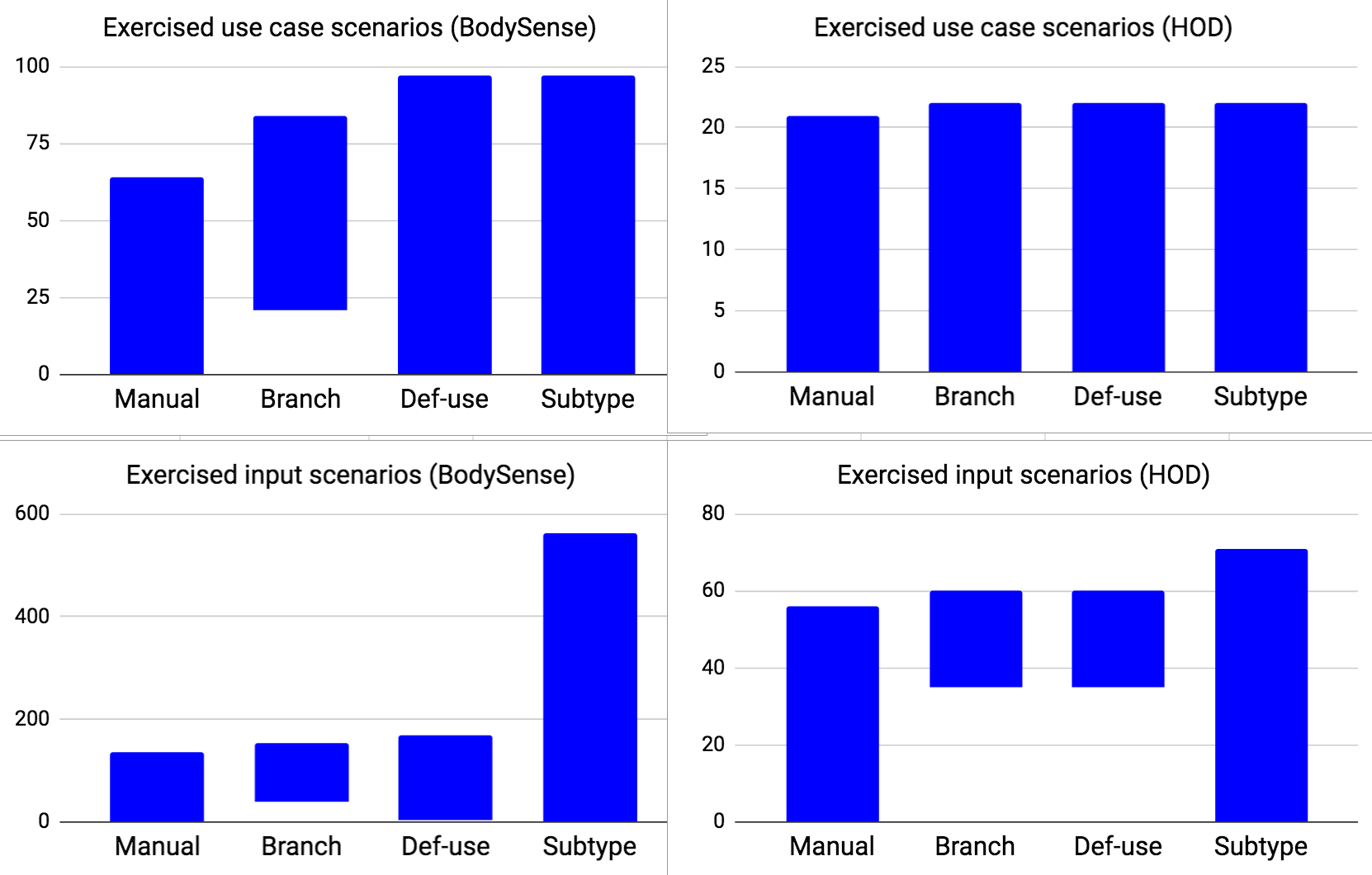}
      \caption{Use case and input scenarios exercised by the UMTG and manual test suites.}
      \label{fig:RQ2:overview}
\end{figure}

To summarize our findings,
Fig.~\ref{fig:RQ2:overview} presents four bar charts comparing all the use case and input scenarios exercised by the manual and UMTG test suites. In the charts, each data point on the vertical axis represents a use case or input scenario exercised by one or more test suites. When a scenario is exercised by a test suite, a blue horizontal line is drawn %
in the vertical axis. A blue bar indicates a group of test cases in a test suite. %
For example, the \BodySense use case scenario matching data point 50 is exercised by four test suites, while the use case scenario for data point 10 is not exercised by the branch coverage test suite. We sort the data points to simplify the visual comparison of their coverage. 

Branch coverage is the least effective one %
because it leads to test suites that do not exercise all the scenarios exercised by the manual test suite. 
For example, in \BodySense, the branch coverage test suite exercises 20 use case scenarios that are not exercised by the manual test suite (the data points between 65 and 84 are covered only by the branch coverage test suite) but it does not exercise 48 use case scenarios in the manual test suite (the data points below 49 are not covered by the branch coverage test suite).
Unsurprisingly, def-use coverage subsumes branch coverage. 
Finally, the chart shows that subtype coverage is the only criterion that exercises all the use case and input scenarios in the manual test suite
in addition to input scenarios not exercised by the manual test suite.
Unfortunately, the associated high number of test cases %
increases the cost of testing. 
On the other hand, subtype coverage is necessary only when specifications are incomplete, i.e., the \HOD case where def-use coverage misses a lot of input scenarios. 
Indeed, in \BodySense, the def-use coverage test suite exercises all the input scenarios covered by the manual test suite except three input scenarios. 
These three input scenarios exercise a use case scenario that concerns a change in the occupancy status of the seat in the presence of specific errors being both detected and qualified. 
However, \UMTG covers the same use case scenario with other types of errors not considered in the manual test suite. 
There is no specific reason for testing the above-mentioned use case scenario with a specific subset of errors rather than another.
Therefore, we conclude that def-use coverage is sufficient to generate a test suite for \BodySense that is as exhaustive as the manual one.

\MREVISION{R1.53}{Last, we observe that, for branch coverage (in \HOD) and def-use coverage (in \BodySense), \UMTG exercises more use case scenarios than manual test cases, though it generates less test cases (see Table~\ref{tab:RQ2:testCases}). 
This is due to the fact that in the case of manual test cases, engineers, to simplify the identification of test inputs, often derive test cases that focus only on the scenarios in the included use cases.
\UMTG, instead, generates test cases that start with entry point use cases and terminate in abort or exit steps of entry point use cases; consequently, each \UMTG test case may cover multiple subsequences of use case steps belonging to distinct included use cases.
For this reason, the same \UMTG test case may cover sequences of use case steps exercised by distinct manual test cases in addition to sequences of use case steps not covered by any manual test case.}
\MINREV{R1.3c,d}{This is also reflected in the results obtained for the coverage of input scenarios. Indeed, the same UMTG test case may exercise multiple input scenarios covered by distinct manual test cases; this is the reason why, for example, the 63 input scenarios in the test cases generated by UMTG with the branch coverage criterion (see Table~\ref{tab:RQ2:UMTGbehaviors}) exercise 96 input scenarios in the manual test cases (see Table~\ref{tab:RQ2:behaviors}). More precisely, we observe that engineers implemented multiple test cases to separately exercise distinct subsequences of use case steps for a specific error condition. 
Instead, a single UMTG test case exercises all such subsequences together against the same error condition.
Since the input scenarios exercised by manual test cases cover subsequences of steps exercised by UMTG test cases,
in Table~\ref{tab:RQ2:behaviors}, multiple instances of the former may be covered by a single instance of the latter, with the same input partitions.
Instead, in  Table~\ref{tab:RQ2:UMTGbehaviors}, an input scenario exercised by a UMTG test case is counted as covered by manual test cases when all the use case steps of the scenario are covered by one or more manual test cases, exercising all steps or a distinct subset, with the same input partitions.}
\MINREV{R1.3b}{Finally, we report that, in general, we generate a distinct test case for each input scenario; 
however, this is not the case for the manual test cases in HOD where the same test case covers multiple input partitions after resetting the system.
}

\subsubsection{RQ3: How does the approach compare to manual test case generation in terms of resources?}
\label{subsec:empirical:RQ3}

\begin{table}[t]
\scriptsize
\caption{Information to be manually provided when relying on \UMTG and manual test case generation.}
\begin{center}
\begin{tabular}{|@{\hspace{1mm}}p{1.2cm}@{\hspace{1mm}}|@{\hspace{1mm}}r@{\hspace{1mm}}|@{\hspace{1mm}}r@{\hspace{1mm}}||@{\hspace{1mm}}r|r@{\hspace{1mm}}|@{\hspace{1mm}}r@{\hspace{1mm}}|}
\hline
	&	\multicolumn{2}{c}{Manual testing} &\multicolumn{3}{c|}{UMTG}	\\
\parbox[c]{1cm}{Case study}			&\parbox[c]{1.4cm}{Sentences (words) in scenarios descriptions}	&\parbox[l]{1.3cm}{Executable instructions} &\parbox[l]{1cm}{Reviewed constraints}&	\parbox[l]{1.2cm}{Manually written OCLs}		&\parbox[l]{1cm}{Lines in mapping table}	\\
\hline
\BodySense	&		2,476 (18,716)&	8,886 				&	71	&	20&		67\\
\HOD		&		408 (3,890)& 	1,229					&	35	&	2&		51\\
\hline
\end{tabular}
\end{center}
\label{tab:RQ3:info}
\end{table}%

RQ3 is concerned with the effort and time required to generate test cases.
IEE engineers, for each manual test case, write a scenario including executable instructions %
and provide a descriptive comment in each executable instruction. 
In \UMTG, engineers review automatically generated OCL constraints, write missing constraints, and define a mapping table for a given test infrastructure, reusable across systems. %
We do not consider the effort required to derive use case specifications and domain models since they are typically produced during requirements analysis for other purposes~\cite{Larman-Applying-2002}. %

Table~\ref{tab:RQ3:info} presents the data characterizing the manual activities in our case studies. 
In manual test case generation, IEE engineers had to write a large number of sentences to document the test cases, i.e., 2,476 and \MREVISION{R1.55}{408} for \BodySense and \HOD, respectively. 
The test cases of the two systems include a total of 8,886 and 1,229 executable instructions and code comments. 
Thus, manual test case generation is expensive and generally takes several days (at least five working days for each case study). 

In \UMTG, engineers reviewed 71 and 35 OCL constraints for \BodySense and \HOD, respectively. %
The review was fast (on average less than two minutes per constraint) since the constraints are short and follow a well-defined pattern.
The number of constraints written by engineers is low, i.e., 20 for \BodySense and 2 for \HOD. 
The mapping tables for \BodySense and \HOD include 67 and 51 lines, respectively. 
The time taken to write the 
mapping tables is limited, i.e., 135 minutes for \BodySense and 100 minutes for \HOD. %
Mapping table generation is much faster than manual test case generation which may last days.
In the case studies, the mapping tables for the three code coverage strategies are the same. 
When use case specifications are complete, the specific values that trigger error conditions (e.g., \emph{BodySense.temperature = 100}) are generated by the constraint solver and processed in the mapping table by means of the grouping feature of regular expressions (e.g., Line 6 of Table~\ref{tab:generatedTestCase}). When use case specifications are incomplete, test case generation identifies the error condition that should hold in a scenario (e.g., \emph{TemperatureHighError.isDetected = true}) and the mapping table is used to translate such conditions into concrete values (i.e., \emph{TemperatureHighError.isDetected = true} is mapped to \emph{SetBus: Channel=RELAY Temperature = 20}). When specifications are incomplete, mapping tables become more complex. 
Incomplete use case specifications force engineers to encode system-specific, functional parameters in mapping tables, a practice that may limit the reuse of mapping tables across systems. Therefore, we recommend that engineers produce complete use case specifications.

\begin{table}[t]
\scriptsize
\caption{\UMTG test generation time and number of calls to constraint solver.}
\begin{center}
\begin{tabular}{|@{\hspace{1mm}}p{1.1cm}@{\hspace{1mm}}|@{\hspace{1mm}}r@{\hspace{1mm}}|@{\hspace{1mm}}r@{\hspace{1mm}}|@{\hspace{1mm}}r|r@{\hspace{1mm}}|@{\hspace{1mm}}r|@{\hspace{1mm}}r|}
\hline
	&\multicolumn{3}{c|}{Max test generation time}	&\multicolumn{3}{c|}{Calls to constraint solver}\\
	&\multicolumn{3}{c|}{for all the scenarios (minutes)}	&\multicolumn{3}{c|}{(total for all the scenarios)}\\
\parbox[c]{1cm}{Case study}			&\parbox[l]{8mm}{Branch}&	\parbox[l]{10mm}{Def-use}		&\parbox[l]{8mm}{Subtype}	&\parbox[c]{8mm}{Branch}	&\parbox[l]{10mm}{Def-use} &\parbox[l]{8mm}{Subtype}\\
\hline
\BodySense	&	330	&	720&		7,200&		908 &  2,762&	23,913\\
\HOD		&	2	&	2&	4& 43&		46& 	154	\\
\hline
\end{tabular}
\end{center}
\label{tab:RQ3:solving}
\end{table}%

Table~\ref{tab:RQ3:solving} presents the time, in minutes, that UMTG took to automatically generate test cases for \BodySense and \HOD. 
Since test cases for different scenarios can be generated in parallel, we report the time of the longest test case generation.
To generate test cases, \UMTG needed more time for \BodySense because its use case specifications have more alternative flows which, in turn, lead to larger use case test models to be traversed and more path conditions to be solved.
With branch and def-use coverages, it took from 1 minutes to 12 hours, which is acceptable since it is fully automated and still faster than 
manual test case generation. Manual test case generation took between two and five working days for the same case studies. With subtype coverage, %
\UMTG needed time up to 120 hours (5 days). %
Technological improvements such as parallel execution of constraint solving (we have 23,913 invocations to the constraint solver) may drastically reduce test case generation time. 
Subtype coverage is built on top of def-use coverage. Engineers can execute test cases generated based on def-use coverage and start repairing some of the faults to be detected by subtype coverage.
This can significantly speed up acceptance testing.

\MREVISION{R1.54}{The OCL constraint generation does not practically impact on the test case generation time. It takes 310 seconds for \BodySense and 110 seconds for \HOD. The OCL generation time mostly depends on the SRL execution and the network communication with the CNP server, which take between 1 and 4 seconds for each use case sentence.}

Based on our observations above, we are confident that, in practice, \UMTG requires less effort and time than manual test case generation. It can be easily integrated into the development process and provides better guarantees for achieving use case and input scenario coverage.

\subsection{Discussion}
\label{subsec:discussion}

The modeling effort required for testing a software system with \UMTG is minimized thanks to the 
automated generation of OCL constraints. 
Our results show that \UMTG automatically and correctly generated 95\% of the constraints required for testing the two systems in our evaluation.

Among all the coverage strategies in \UMTG, def-use coverage performs the best. 
Indeed, it enables the generation of test cases that exercise all the use case scenarios covered by the manual test suites; in the presence of complete specifications,
it enables engineers to automatically cover all the input scenarios considered in the manual test suites.

To cover an adequate set of input partitions (at least the ones covered by manual test case generation), engineers should select the coverage criterion based on the completeness of specifications.
Def-use coverage is sufficient when specifications are complete 
(i.e., when they capture all input partitions by means of condition steps)
while subtype coverage is needed when specifications are incomplete
(i.e., when they do not explicitly capture input partitions by means of condition steps).
Based on our experience, test case generation for these two cases can be performed overnight, which makes \UMTG appealing in industrial contexts.

\UMTG leads to significantly less test generation time than manual test case generation. The latter also consumes substantial human effort (days) whereas the former mostly requires computation time along with hours of human effort, including writing constraints and mapping tables.
This is in practice an important distinction as computation time is much more readily available than human skills.
Also, in the presence of specifications that are often updated (e.g., to reconfigure the system because it is part of a product line), automated test case generation relieves engineers from the burden of the manual identification of use case scenarios that need to be tested with newly implemented test cases.

\subsection{Threats to Validity}
\label{subsec:validity}
The main threat to validity in our study relates to generalizability, a common issue with industrial case studies. 
To deal with this threat, we consider two representative automotive embedded systems implementing very different functionalities, sold to different customers, 
with specifications written by different sets of people following different requirements analysis practices (\BodySense specifications are complete, while \HOD specifications are incomplete).

In addition, in Section~\ref{sec:oclGeneration}, we show that the OCL pattern handled by \UMTG is expressive enough for embedded systems, while the requirements specification format processed by \UMTG (use case specifications) is common in industrial settings that require precise requirements elicitation shared by multiple stakeholders.

\section{Conclusion} 
\label{sec:conclusion}
\CHANGED{In this paper, we introduced \M{}, an approach to automate acceptance testing, with a focus on embedded systems. It generates test data and executable, system-level, test cases for the purpose of validating conformance with requirements, an activity usually referred to as acceptance testing.} 
We rely on use case specifications augmented with a domain model. 
Our motivation is to achieve automated test case generation by largely relying on common practices to document requirements for
communication purposes among stakeholders in embedded system domains.

To enable the automatic identification of use case scenarios and test inputs, we combine NLP
and constraint solving.
To extract behavioral information from use case specifications by means of NLP, we rely upon a restricted use case modeling method called RUCM. %
We use an advanced NLP solution (i.e., semantic role labeling) to automatically generate OCL constraints that capture the pre and postconditions of use case steps. %
We designed an algorithm that identifies use case scenarios based on %
the branch, def-use, and subtype coverage criteria. 
It builds feasible path conditions that capture OCL constraints under which the alternative flows in each scenario are executed.
Test inputs are determined by solving the path conditions with the Alloy analyzer.

Two industrial case studies show that \UMTG effectively generates acceptance test cases for automotive sensor systems. 
\UMTG can automatically and correctly generate \MINREV{R1.2}{96}\% of the OCL constraints for test case generation. The precision of the OCL generation process is very high: 99\% of the generated constraints are correct.
The coverage criteria implemented in \UMTG enable the identification of use case scenarios missed by engineers, thus highlighting its usefulness in safety domains. 

\MREVISION{R1.57}{Our results show that the subtype coverage criterion, when specifications are incomplete, and the def-use coverage criterion, in general, generate test cases that cover all the scenarios and input partitions exercised by the test cases written by experts.}
This is achieved by automatically generated test suites that have the same size as the manually implemented ones, thus not affecting testing cost. Our experience indicates that requirements modeling, mostly limited in \M{} to RUCM use case specifications and domain modeling, is feasible in an industrial context. 
Furthermore, manual test case generation requires significantly more effort than applying \UMTG, since use case specifications and domain models are usually employed for other purposes as well. %

\MREVISION{R1.56,R1.58}{Future work includes the extension of \UMTG to deal with different types of testing problems and the identification of solutions to further the scalability of the strategy due to constraint solving for very large systems. We are currently working on UMTG-inspired approaches for security testing of Web systems~\cite{Mai2018b, Mai2019}. We also aim to address scalability issues by evaluating the feasibility of adopting alternative solving approaches, including SMT solvers~\cite{Dania:OCL2FOL:2016}, answer set programming~\cite{Potassco}, higher-order relational constraint solving~\cite{AlloyStar}, and the combination of constraint solving and search-based optimization~\cite{Soltana2017,Soltana19}.}

\appendices

\section{\MREVISION{R1.59}{Complexity} \MREVISION{R3.9}{Analysis of Test Scenarios Generation}}
\label{app:complexity}

In this appendix, we discuss the worst-case time complexity of \emph{GenerateScenariosAndInputs}, the \UMTG function that generates test scenarios. We do not discuss the time complexity of procedures implemented to generate object diagrams (i.e., test inputs) because \UMTG delegates it to the Alloy analyzer.

The time complexity of \emph{GenerateScenariosAndInputs} depends on the depth-first traversal implemented by \emph{GenerateScenarios}, which is invoked multiple times till the coverage criterion is satisfied or a max number of iterations is reached.
Also, it depends on the time complexity of function \emph{maximizeSubTypeCoverage}, which is executed to generate scenarios that satisfy the subtype coverage criterion.
 \emph{GenerateScenarios}, in turn, invokes six functions (i.e.,
 \emph{coverageSatisfied}, \emph{coverageImproved}, \emph{unsatisfiable}, \emph{addToScenario}, \emph{composeConstraints}, and \emph{negateConstraint}).

In the rest of the appendix, we discuss the time complexity of 
\emph{GenerateScenarios}, \emph{maximizeSubTypeCoverage}, and the %
functions invoked by \emph{GenerateScenarios}.
Finally, we discuss the time complexity of \emph{GenerateScenariosAndInputs}.

\subsection{Time Complexity of \emph{GenerateScenarios}}

\emph{GenerateScenarios} implements a depth-first recursive traversal of a \UCTM in which 
the same node can be visited at most $T+1$ times. 
Nodes are visited multiple times in the presence of conditional nodes with backward flows (i.e., sequences of nodes departing from the conditional node and reaching a node already visited during the traversal). Backward flows cause loops in the \UCTM. 

The execution time of \emph{GenerateScenarios} depends on (1) the number of recursive iterations performed, (2) the execution time 
of the functions being invoked at every recursive iteration
and (3) the number of operations (comparisons and assignments) performed within a single iteration of \emph{GenerateScenarios}. 

Since \emph{GenerateScenarios} does not contain loops, the number of operations performed within a single iteration of \emph{GenerateScenarios} can be considered constant and thus ignored for the worst-case time complexity analysis.

The time complexity of \emph{GenerateScenarios} can thus be computed as

\begin{equation}
\label{cost:GS:pre}
\EquationsSize
\begin{aligned}
C_{GS} &= \BigO \Big( N_v * \\
&\big(1+(C_{CS}+C_{CI}+C_{US}+C_{AS}+C_{CC}+C_{NC})\big) \Big)
\end{aligned}
\end{equation}

with $N_v$ being the number of recursive invocations of \emph{GenerateScenarios}
during the traversal of a \UCTM, 
and the terms in the innermost parenthesis (i.e., $C_{CS}$, $C_{CI}$, $C_{US}$, $C_{AS}$, $C_{NC}$, $C_{NC}$) being the
time complexity of the functions invoked by \emph{GenerateScenarios}
(i.e., \emph{coverageSatisfied}, \emph{coverageImproved}, \emph{unsatisfiable}, \emph{addToScenario}, \emph{composeConstraints}, and \emph{negateConstraint}, respectively). 
\MINREV{R1.4}{In this section, to avoid confusion with parameter $T$ of \emph{GenerateScenariosAndInputs}, we refer to the time complexity as $C$ instead of $T(n)$~\cite{Arora:ComplexityBook:2009}.}

In this subsection, we focus on the computation of $N_v$. The time complexity of the functions invoked by \emph{GenerateScenarios} is discussed in the following subsections.

Since every recursive call of \emph{GenerateScenarios} traverses only one node of the \UCTM,
for $T=0$, \emph{GenerateScenarios} visits every node once, thus having the time complexity as follows

\begin{equation}
\label{cost:GS:T1}
\EquationsSize
\begin{aligned}
C_{GS} &= \BigO \Big( N * \\
& \big(1+(C_{CS}+C_{CI}+C_{US}+C_{AS}+C_{CC}+C_{NC}) \big) \Big)
\end{aligned}
\end{equation}

with $N$ being the number of nodes in the \UCTM.

For $T>0$, \emph{GenerateScenarios} traverses every node at most $T+1$ times.
To discuss the time complexity for $T>0$, we first identify the properties of the \UCTM
that lead to the highest number of 
recursive iterations of \emph{GenerateScenarios}; we call it the worst-case \UCTM.

\begin{figure}[h]
    \includegraphics[width=8.7cm]{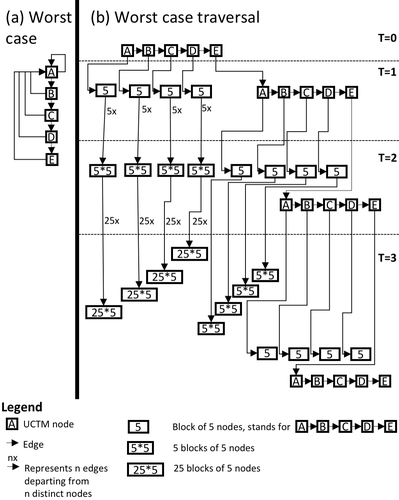}
      \caption{(a) An Example worst-case UCTM. (b) Traversal of the UCTM, showing the number of nodes visited for different values of T.}
      \label{fig:complexity:example}
\end{figure}

For $T>0$, the worst-case \UCTM is a \UCTM including only condition nodes having one forward and one backward edge, except one condition node having one backward edge and no forward edge. 
Also, it should have forward edges connecting condition nodes in a sequence, which leads to the longest possible scenario in the \UCTM.
Finally, all the backward edges should point to the root of the \UCTM; this way,   
each backward edge leads to a traversal of the root, i.e., another traversal of all the nodes of the \UCTM. 
Any other \UCTM layout would lead to a traversal of a subset of the \UCTM. An example worst-case \UCTM is shown in Fig.~\ref{fig:complexity:example}(a).

In the worst-case \UCTM, the number of nodes with a backward edge (hereafter, $N_B$) is equal to the total number of nodes of the \UCTM (i.e., $N$), while the number of leaf nodes (hereafter, $N_L$) is equal to one. In Section~\ref{sec:app:worstCase}, we discuss whether this configuration 
leads to the worst-case time complexity for \emph{GenerateScenarios}.

Leaf nodes are always Exit or Abort nodes in the \UCTM{s} generated by \UMTG. However, to simplify the discussion, we treat condition nodes with a null forward edge as leafs in Fig.~\ref{fig:complexity:example}(a) and in the rest of this appendix. Basically, we assume that \UMTG generates a new scenario when it encounters a null branch. This choice simplifies the presentation without reducing the time complexity; indeed, 
the absence of leaf nodes augments the portion of condition nodes with backward edges in a \UCTM, and, consequently, the time complexity.

Since \emph{GenerateScenarios} traverses all the nodes of the \UCTM each time the root node is visited, the value of $N_v$
can be computed as the number of visited root nodes multiplied by the number of nodes that are visited without traversing loops.
Fig.~\ref{fig:complexity:example}(b) shows, for different values of $T$, the nodes visited in the worst-case \UCTM including 5 nodes. 
Every node leads to the visit of the root node of the \UCTM, and this visit is repeated up to $T$ times.
In Fig.~\ref{fig:complexity:example}(b), for T=0, \emph{GenerateScenarios} visits 5 nodes.  
For $T=1$, it visits 30 nodes (i.e., the 5 nodes traversed for T=0 and additional 25 nodes reached from any of the nodes in the \UCTM). For $T=2$, it visits 155 nodes (i.e., 125 plus 30).
The total number of nodes visited by \emph{GenerateScenarios} is thus captured by the geometric progression $N*\sum_{i=0}^{T} N_c^{i}$, where $N_c^{i}$ represents the number of (additional) root nodes visited for an increased value of $T$. 
Because of the geometric progression, the total number of nodes visited by \emph{GenerateScenarios} can be computed as
\begin{equation}
\label{cost:NV}
\EquationsSize
\begin{aligned}
N_v = \frac{N*(1-N_B^T)}{1-N_B}
\end{aligned}
\end{equation}

However, with ${N_B} > 0$, the following condition holds
\begin{align}
\label{cost:NB:simple}
\EquationsSize
\begin{split}
\frac{1-N_B^T}{1-N_B} < {N_B}^T\\
\end{split}
\end{align}

We can thus further simplify the computation of $\BigO(N_v)$ as follows
\begin{align}
\label{cost:NV:simple}
\EquationsSize
\begin{aligned}
\BigO( N_v) = \BigO ( N*{N_B}^T )\\
\end{aligned}
\end{align}

\subsection{Time Complexity of \emph{coverageSatisfied}}

Function $\mathit{coverageSatisfied}(\mathit{tm},\mathit{Prev},\mathit{Curr})$ checks if a predefined set of coverage targets derived from $tm$ (i.e., the branches of the \UCTM and the def-use pairs of the \UCTM) is covered by the scenarios that belong to the union of the sets $Prev$ and $Curr$. 
In the following, we separately discuss the time complexity of the procedures implemented to identify coverage targets (hereafter, $C_{GT}$) and the time complexity of the procedures implemented to determine targets coverage (hereafter, $C_{DT}$). 

In Formula~\ref{cost:GS:pre}, the term $C_{CS}$ represents the time complexity of function $\mathit{coverageSatisfied}$;
after replacing $C_{CS}$ with the sum of $C_{GT}$ and $C_{DT}$ as indicated above, $C_{GS}$ can be computed as follows

\begin{equation}
\label{cost:GS2}
\EquationsSize
\begin{aligned}
C_{GS} &= \BigO \Big( N_v * (1\\
&+\big(C_{GT}+C_{DT}+C_{CI}+C_{US}+C_{AS}+C_{CC}+C_{NC}) \big) \Big)
\end{aligned}
\end{equation}

\subsubsection{Determining $C_{GT}$} 

\UMTG derives coverage targets only on the first invocation of $\mathit{coverageSatisfied}$. 
Therefore, the time complexity of the procedure for generating coverage targets 
should not be taken into account for every single recursive iteration of \emph{GenerateScenarios}. 
This leads to the following formula for $C_{GS}$

\begin{equation}
\label{cost:GS3}
\EquationsSize
\begin{aligned}
C_{GS} &= \BigO \Big( C_{GT} + N_v * \big(1\\
&+(C_{DT}+C_{CI}+C_{US}+C_{AS}+C_{CC}+C_{NC}) \big) \Big)
\end{aligned}
\end{equation}

For the branch coverage criterion, the identification of coverage targets requires that the \UCTM be traversed without traversing loops (i.e., backward edges), and thus its time complexity is $\BigO(N)$. 

For the def-use coverage criterion, we need to identify all the possible def-use pairs, which, in a \UCTM, is lower or equal to the number of unique pairs of nodes. For the def-use coverage, the number of coverage targets $N_{CT}$ can thus be computed as follows
\begin{equation}
\label{cost:NCT}
\EquationsSize
\begin{aligned}
N_{CT} = \frac{N*(N+1)}{2}
\end{aligned}
\end{equation}

Consequently, $C_{GT}$ can be computed as follows

\begin{equation}
\label{cost:C.GT}
\EquationsSize
\begin{aligned}
C_{GT} &= \BigO \Bigg(\frac{N*(N+1)}{2} \Bigg)\\
& \approx \BigO ( N^2 ) \\
\end{aligned}
\end{equation}

\subsubsection{Determining $C_{DT}$} 
\label{app:compl:det:CDT}

\UMTG implements an optimized version of $\mathit{coverageSatisfied}$ that incrementally determines if the set of coverage targets are covered by the union of $Prev$ and $Curr$.
This is done by verifying, at every iteration of  \emph{GenerateScenarios}, only the items added to $Curr$.

To check if a coverage target is covered in a scenario, it is necessary to visit, at most, every node in the scenario (e.g, to determine if the scenario contains a specific def-use pair).
By definition, the sets $Prev$ and $Curr$ can contain, at most, all the scenarios that can be derived by \emph{GenerateScenarios}. We can thus compute $C_{DT}$ as

\begin{equation}
\label{cost:C.DT}
\EquationsSize
\begin{aligned}
C_{DT} = \BigO ( N_{CT} * N_{SC} )
\end{aligned}
\end{equation}

with $N_{SC}$ being the number of nodes in all the scenarios derived by \emph{GenerateScenarios}. 

\begin{figure*}[tb]
    \includegraphics[width=18cm]{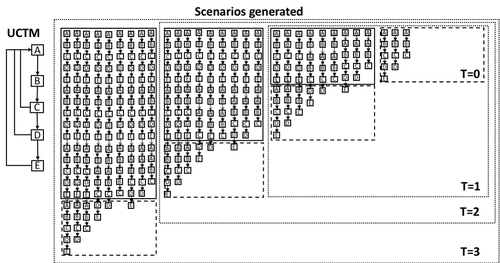}
      \caption{Example with an upper bound for the number of nodes in the scenarios generated by \emph{GenerateScenarios}, for different configurations of $T$.}
      \label{fig:complexity:example:scenarios}
\end{figure*}

To simplify the computation of $C_{DT}$, we derive an upper bound for $N_{SC}$.
To discuss the computation of the upper bound for $N_{SC}$, we rely on the example in Fig.~\ref{fig:complexity:example:scenarios}. The example includes the scenarios generated while traversing, for values of $T \leq 3$, a \UCTM that has three backward edges (departing from nodes C, D, and E) and three leaves (nodes A, B, and E).

For increasing values of $T$, \emph{GenerateScenarios} traverses additional scenarios originated from the conditional steps with backward edges. 
In each of these scenarios, we can identify a prefix (hereafter, prefix scenario) starting from the first time the root of the \UCTM is visited in the scenario until the last time the root node is visited in the scenario. We can also identify a suffix (hereafter, suffix scenario) starting from the last time the root node is visited for the scenario until a leaf is visited.
Each prefix scenario is shared by a number of scenarios that is equal to the number of leaves (i.e., $N_L$) in the \UCTM.
However, all the $N_L$ scenarios that have the same prefix scenario differ for their suffix scenario. 

Please note that worst-case \UCTM{s} contain only condition steps. Therefore, according to the formula 
($N_L = N - N_B + 1$), the number of leaves $N_L$ is given by the number of condition steps without a backward edge plus one (the last condition is always a leaf).

In Fig.~\ref{fig:complexity:example:scenarios}, boxes with black borders show the additional scenarios traversed for increasing values of $T$ (hereafter, $T_i$). The width of these boxes captures the number of the additional scenarios (i.e., $N_B*N_L$), while the height captures the length of the longest prefix scenario (i.e., $N*T_i$). The area of these boxes is thus an upper bound for the number of nodes in the prefix scenarios. It can be computed as

\begin{equation}
\label{cost:N:SC:prefix}
\EquationsSize
\begin{aligned}
Area_{\mathit{prefix}} &= \sum_{T_i=1}^{T} N_B*N_L*N*T_i\\
	&= N_B*N_L*N * \sum_{i=1}^{T_i} T_i\\
	&= N *N_B*N_L* \frac{T*(T+1)}{2} 
\end{aligned}
\end{equation}

Prefix scenarios are followed by suffix scenarios.
Since the traversal of a root node leads to the generation of a number of scenarios that is equal to the number of leaves,
the total number of suffix scenarios is equal to the number of additional root nodes traversed for the increasing values of $T$ multiplied by the number of leaves in the \UCTM (i.e., $N_B * N_L$). 
The total number of such suffix scenarios is thus $N_B * N_L * (T+1)$.
In  Fig.~\ref{fig:complexity:example:scenarios}, the maximum  number of nodes in the suffix scenarios is depicted by the area of the dashed boxes and can be calculated as 

\begin{equation}
\label{cost:N:SC:suffix}
\EquationsSize
\begin{aligned}
Area_{\mathit{suffix}} &= N * N_B * N_L * (T+1)
\end{aligned}
\end{equation}

The value of $N_{SC}$ can thus be computed as

\begin{equation}
\label{cost:NSC}
\EquationsSize
\begin{aligned}
N_{SC} &= Area_{\mathit{suffix}}  + Area_{\mathit{prefix}} \\
	&= N*N_B*N_L*(T+1) + N_B*N_L*N * \frac{T*(T+1)}{2}\\
	&= (N*N_B*N_L*T)+ (N*N_B*N_L ) \\
	&+ N_B*N_L*N * \frac{T*(T+1)}{2}\\
	&= N * N_B*N_L + N* N_B*N_L * \Bigg( T + \frac{T*(T+1)}{2} \Bigg) \\
	&= N * N_B*N_L + N* N_B*N_L * \Bigg( \frac{2T  + T^2  + T }{2} \Bigg) \\
	&= N * N_B*N_L + N* N_B*N_L * \Bigg( \frac{T^2  + 3T }{2} \Bigg) \\
	&= N * N_B*N_L  * \Bigg( 1 + \frac{T^2  + 3T }{2} \Bigg) \\
	&= N * N_B*N_L  * \frac{T^2  + 3T + 2}{2} \\
\end{aligned}
\end{equation}

The value of $C_{DT}$ can thus be computed as
\begin{equation}
\label{cost:C.DT.IN}
\EquationsSize
\begin{aligned}
C_{DT}&=
\BigO ( N_{CT} * N_{SC} )\\
&=\BigO \Bigg( \frac{N*(N+1)}{2}
* N * N_B*N_L  * \frac{T^2  + 3T + 2}{2} \Bigg)\\
&=\BigO \Bigg( \frac{(N^2+N)* N * N_B*N_L  * ( T^2  + 3T + 2)  }{4} \Bigg)\\
&=\BigO \Bigg( \frac{(N^3 * N_B*N_L  * ( T^2  + 3T + 2)  )}{4} \\
&+ \frac{(N^2 * N_B*N_L  * ( T^2  + 3T + 2)  ) }{4} \Bigg) \\
&\approx\BigO \Bigg( \frac{(N^3 * N_B*N_L  * ( T^2  + 3T + 2)  ) }{4} \Bigg) \\
\end{aligned}
\end{equation}

We use the symbol $\approx$ to indicate a simplification enabled by the big-O notation.

We discuss the time complexity for $C_{DT}$ considering $N_B=\frac{N}{2}$, which leads to 
the higher value for $N * N_B * N_L$,
and $N_B=N$, which is the worst case for $N_v$.

With $N_B=\frac{N}{2}$, $C_{DT}$ can be simplified as follows
\begin{equation}
\label{cost:C.DT.F}
\EquationsSize
\begin{aligned}
C_{DT}&=\BigO \Bigg( \frac{N^3 * N_B*(N-N_B+1)  * ( T^2  + 3T + 2)}{4} \Bigg) \\
&=\BigO \Bigg( \frac{N^3 * \frac{N}{2}*(\frac{N}{2}+1)  * ( T^2  + 3T + 2)  }{4} \Bigg) \\
&=\BigO \Bigg( \frac{(\frac{1}{4}N^2+\frac{1}{2}N)  * N^3 * ( T^2  + 3T + 2)}{4} \Bigg) \\
&=\BigO \Bigg( \frac{\Big(N^3 * \frac{1}{4}N^2  * ( T^2  + 3T + 2)  \Big)}{}\\
& \frac{+ \Big(N^3 * \frac{N}{2}  * ( T^2  + 3T + 2)  \Big) }{4} \Bigg) \\
&=\BigO \Bigg( \frac{\Big(\frac{1}{4}N^5  * ( T^2  + 3T + 2)  \Big)}{} \\
&\frac{+ \Big(\frac{1}{2}N^4  * ( T^2  + 3T + 2)  \Big) }{4} \Bigg) \\
&=\BigO \left( \frac{ \left(N^5+2N^4 \right)  * \left( T^2  + 3T + 2 \right) }{16} \right) \\
& \approx \BigO \left( \frac{\left(N^5  * \left( T^2  + 3T + 2\right)  \right) }{16} \right) \\
& \approx \BigO \left( N^5 \right) \\
\end{aligned}
\end{equation}

Instead, with $N_B=N$, $C_{DT}$ can be computed as follows
\begin{equation}
\label{cost:C.DT.2}
\EquationsSize
\begin{aligned}
C_{DT}&=\BigO \left( \frac{N^3 * N_B*N_L  * \left( T^2  + 3T + 2\right)  }{4}  \right) \\
&=\BigO \left(  \frac{N^3 * N*1  * \left( T^2  + 3T + 2\right)  }{4} \right) \\
&=\BigO \left( \frac{N^4 * \left( T^2  + 3T + 2\right)  }{4} \right)\\
& \approx \BigO \left( N^4 \right) \\
\end{aligned}
\end{equation}

Since Formula~\ref{cost:C.DT.F} leads to a higher time complexity than Formula~\ref{cost:C.DT.2}, we can conclude that
the worst-case \UCTM for $C_{DT}$ is charachterized by $N_B=\frac{N}{2}$.

Finally, since function $\mathit{coverageSatisfied}$ processes each scenario in $Prev$ and $Curr$ only once,
the time complexity $C_{DT}$ should not be taken into account for every recursive iteration of \emph{GenerateScenarios}:

\begin{equation}
\label{cost:GS4}
\EquationsSize
\begin{aligned}
C_{GS} &= \BigO\Big( C_{GT} + C_{DT} + N_v * \big(1\\
&+(C_{CI}+C_{US}+C_{AS}+C_{CC}+C_{NC})\big) \Big)
\end{aligned}
\end{equation}

\subsection{Time complexity for \emph{coverageImproved}} 

When a leaf node is reached, \emph{coverageImproved} is invoked by \emph{GenerateScenarios} to check if the new scenario improves coverage. 
To this end, the nodes of the scenario are compared with all the coverage targets.
By construction, \emph{GenerateScenarios} never generates the same scenario twice. 
Therefore, through all the recursive iterations of \emph{GenerateScenarios}, \emph{CoverageImproved} compares the coverage targets with all the nodes in the scenarios identified by \emph{GenerateScenarios}. The value of $C_{CI}$ can thus be computed as follows
\begin{equation}
\label{cost:C.CI1}
\EquationsSize
\begin{aligned}
C_{CI} = \BigO( N_{CT} * N_{SC} )
\end{aligned}
\end{equation}

Based on Formula~\ref{cost:C.DT}, the time complexity of \emph{coverageImproved} (i.e., $C_{CI}$) is thus equal to $C_{DT}$

\begin{equation}
\label{cost:C.US}
\EquationsSize
\begin{aligned}
C_{CI} = C_{DT}
\end{aligned}
\end{equation}

Since \emph{GenerateScenarios} never generates the same scenario twice, we do not need to account for the time complexity of \emph{coverageImproved} for every single recursive iteration of \emph{GenerateScenarios} but for its whole execution against a \UCTM. Consequently, we can compute $C_{GS}$ as follows

\begin{equation}
\label{cost:GS5}
\EquationsSize
\begin{aligned}
C_{GS} &= \BigO\Big( C_{GT} + C_{DT} + C_{CI} + N_v * \big(1\\
&+(C_{US}+C_{AS}+C_{CC}+C_{NC})\big) \Big)
\end{aligned}
\end{equation}

\subsection{Time complexity for \emph{unsatisfiable}} 

Function $\mathit{unsatisfiable}(\mathit{pc})$ checks if a path condition (i.e., $\mathit{pc}$) is in the set of path conditions that are known to be unsatisfiable based on the previous executions of the solver. It compares the path condition $\mathit{pc}$ with every path condition in the set. 
By definition, the set of unsatisfiable path conditions includes, at most, all the path conditions identified in the \UCTM by \UMTG, i.e., all the path conditions of all the identified scenarios. The worst-case (i.e., highest) number of path conditions identified by \UMTG is thus $N_{SC}$.
Since we assign an identifier to each path condition (i.e, their hashcode), we can assume that comparing two path conditions has unary cost.
The worst-case time complexity of $\mathit{unsatisfiable}$ (i.e., $C_{US}$) can thus be computed as follows 

\begin{equation}
\label{cost:C.US}
\EquationsSize
\begin{aligned}
C_{US} = \BigO(N_{SC})
\end{aligned}
\end{equation}

\subsubsection{Time complexity for \emph{addToScenario}, \emph{negateConstraint}, and \emph{composeConstraint}}

Functions \emph{addToScenario}, \emph{negateConstraint}, and \emph{composeConstraint} perform simple operations. Therefore, we ignore them for the computation of the time complexity.

\subsection{Worst cases for \emph{generateScenarios}} 
\label{sec:app:worstCase}

Based on the definitions above, the time complexity of \emph{GenerateScenarios} can be computed as follows
 
\begin{equation}
\label{cost:C.GS}
\EquationsSize
\begin{aligned}
C_{GS} &=  C_{GT} +  C_{DT} + C_{CI} +  \BigO\Big( N_v * \big( 1 +  C_{US} \big) \Big)\\
&=  C_{GT} +  2 * C_{DT}  +  \BigO\Big( N_v * \big( 1 +  C_{US} \big) \Big)
\end{aligned}
\end{equation}

The last term of the equation above can be simplified as follows
\begin{equation}
\label{cost:NV:red}
\EquationsSize
\begin{aligned}
\BigO( N_v * ( 1 +  C_{US} ) ) &= \bigg(N*{N_B}^T\bigg)\\
&* \left( 1+ N * N_B*N_L  * \frac{T^2  + 3T + 2}{2} \right)\\
&=N*{N_B}^T \\
&\ \ \ \ + N*{N_B}^T * N * N_B*N_L  \\
&\ \ \ \ * \frac{T^2  + 3T + 2}{2}\\
& \approx \frac{T^2  + 3T + 2}{2} * N^2*{N_B}^{T+1} *N_L\\
\end{aligned}
\end{equation}

The highest value of the term above depends on the values of $T$, $N_B$, and $N_L$. 
We discuss the default case for \UMTG (i.e., $T=1$) and the general case (i.e., $T>1$).
For $T=1$, and $N_B=\frac{N}{2}$, 
we obtain the following

\begin{equation}
\label{cost:NB_T1}
\EquationsSize
\begin{aligned}
N_L=N-\frac{N}{2}+1=\frac{N}{2}+1
\end{aligned}
\end{equation}

\begin{equation}
\label{cost:C.GS}
\EquationsSize
\begin{aligned}
\BigO( N_v * ( 1 +  C_{US} ) ) &= 
\frac{T^2  + 3T + 2}{2} * N^2*{N_B}^{T+1} *N_L \\
&= 3* N^2*\left({\frac{N}{2}}\right)^{1+1} * \left(\frac{N}{2}+1\right)\\
&= 3* N^2*{\frac{N^{2}}{4}} * \left(\frac{N}{2}+1\right) \\
&= 3* {\frac{N^{4}}{4}} * \frac{N}{2} + 3* N^2*{\frac{N^{2}}{4}} \\
&= \frac{3}{8} * N^{5} +  \frac{3}{4} * N^{4} \\
& \approx N^{5} \\
\end{aligned}
\end{equation}

With $T=1$, and $N_B=N$, 
we obtain the following

\begin{equation}
\label{cost:C.GSno}
\EquationsSize
\begin{aligned}
\BigO( N_v * ( 1 +  C_{US} ) ) &= 
\frac{T^2  + 3T + 2}{2} * N^2*{N_B}^{T+1} *N_L \\
&= 3* N^2*N^{1+1} * 1\\
&= 3* N^4\\
& \approx N^4\\
\end{aligned}
\end{equation}

Based on Formula~\ref{cost:C.GS} and Formula~\ref{cost:C.GSno}, we can conclude that the higher value for $\BigO( N_v * ( 1 +  C_{US} ) )$, with $T=1$, is thus given by $N_B=\frac{N}{2}$.

With $T\geq1$, and $N_B=\frac{N}{2}$, we obtain the following

\begin{equation}
\label{cost:NV:no}
\EquationsSize
\begin{aligned}
\BigO( N_v * ( 1 +  C_{US} ) ) &= 
\frac{T^2  + 3T + 2}{2} * N^2*{N_B}^{T+1} *N_L \\
&= \frac{T^2  + 3T + 2}{2} * N^2*\frac{N^{T+1}}{2^{T+1}} *\left(\frac{N}{2}+1\right) \\
&\approx \frac{T^2  + 3T + 2}{2} * \frac{N^3}{2}*\frac{N^{T+1}}{2^{T+1}} \\
&\approx \frac{T^2   * N^{T+4}}{2^{T+3}} \\
&\approx N^{T+4}\\
\end{aligned}
\end{equation}

With $T\geq1$, and $N_B=N$, we obtain the following

\begin{equation}
\label{cost:NV:Tabove}
\EquationsSize
\begin{aligned}
\BigO( N_v * ( 1 +  C_{US} ) ) &= \frac{T^2  + 3T + 2}{2} * N^2*{N_B}^{T+1} *N_L \\
& \approx T^2 * N^2 * N^{T+1} * 1 \\
& \approx N^{T+3} \\
\end{aligned}
\end{equation}

With $T\geq1$, the highest value for $\BigO( N_v * ( 1 +  C_{US} ) )$ is thus given by ($N_B=\frac{N}{2}$).

Since Formula~\ref{cost:C.GT} does not refer to $N_{B}$, while Formulas~\ref{cost:C.DT.F} and~\ref{cost:NV:no} show that the worst-case time complexity is given by $\left(N_B=\frac{N}{2}\right)$, we can conclude that the worst-case time complexity for ${GenerateScenarios}$ is thus given by 
$\left(N_B=\frac{N}{2}\right)$.

\subsection{Time complexity for \emph{maximizeSubTypeCoverage}}

For each condition node in the scenario generated by \emph{GenerateScenarios}, function \emph{maximizeSubTypeCoverage} invokes the Alloy solver $S$ times, where $S$ is the number of subtypes of the domain entity used in the condition node. 
$S$ may vary for condition nodes. However, we may assume a constant value that is lower than the number of nodes in the \UCTM (i.e., $S < N$). 
In the computation of the worst-case time complexity for \emph{maximizeSubTypeCoverage},
we ignore the time complexity of the Alloy solver.

The total number of nodes in the scenarios generated by \emph{GenerateScenarios} is given by Formula~\ref{cost:NSC}. The worst-case time complexity for \emph{maximizeSubTypeCoverage}  (hereafter, $C_{MS}$) can thus be computed as follows

\begin{align}
\label{cost:subType}
\EquationsSize
\begin{split}
C_{MS}  & = \BigO( S * N_{SC} )
\end{split}
\end{align}

For ($N_B=\frac{N}{2}$), Formula~\ref{cost:subType} leads to

\begin{align}
\label{cost:subTypeX}
\EquationsSize
\begin{split}
C_{MS}  & = \BigO\left( S*N*\frac{N}{2}*\left(\frac{N}{2}+1\right)*\frac{T^2+3T+2}{2} \right)\\
& \approx \BigO\left( S*T^2*\frac{N^3}{4}\right)\\
& \approx \BigO\left( N^3 \right)\\
\end{split}
\end{align}

For ($N_B=N$), Formula~\ref{cost:subType} leads to

\begin{align}
\label{cost:subTypeNo}
\EquationsSize
\begin{split}
C_{MS}  & = \BigO\left( S*N^2*\frac{T^2+3T+2}{2} \right)\\
& \approx \BigO\left( T^2N^2 \right)\\
\end{split}
\end{align}

The worst-case time complexity for \emph{maximizeSubTypeCoverage} is thus given by $\left(N_B=\frac{N}{2}\right)$.

\subsection{Time complexity for \emph{GenerateScenariosAndInputs}} 

In this subsection, we discuss the time complexity of \emph{GenerateScenariosAndInputs}.
\emph{GenerateScenariosAndInputs} invokes \emph{GenerateScenarios} a number of  times equal to $\mathit{MaxIt}$.
Also, in the case of subtype coverage, \emph{GenerateScenariosAndInputs} invokes function \emph{maximizeSubTypeCoverage}.

The time complexity of \emph{GenerateScenariosAndInputs} can thus be computed as follows
\begin{align}
\label{cost:C.GSI}
\EquationsSize
\begin{split}
C_{GSI}  & = C_{GT} + \mathit{MaxIt} * \big( 2 * C_{DT} +  N_v * ( 1 +  C_{US} ) \big) + C_{MS} \\
\end{split}
\end{align}

In Formula~\ref{cost:C.GSI}, we do not multiply $C_{GT}$ by $\mathit{MaxIt}$ because the number of coverage targets does not change across different invocations of \emph{GenerateScenarios}. 

Based on 
Formulas~\ref{cost:C.GT},~\ref{cost:C.DT.F},~\ref{cost:C.GS},~and~\ref{cost:subTypeX}, the time complexity of \emph{GenerateScenariosAndInputs} can be computed as follows

\begin{align}
\label{cost:C.GSI:one}
\EquationsSize
\begin{split}
C_{GSI}  & = \BigO\Big( N^2 + \mathit{MaxIt} * \big( 2 * N^5  + N^{T+4} \big) + N^3 \Big)\\
&\approx \BigO\big( N^{T+4} \big)\\
\end{split}
\end{align}

Formula~\ref{cost:C.GSI:one} shows that,
in the worst-case \UCTM{s}
 the generation of use case scenarios has at least quintic time complexity, for $T\geq1$.

Despite the high time complexity, we observe that, based on our empirical evaluation, the generation of use case scenarios with \UMTG is feasible in practice. This has two reasons which we detail below.

First, both the number of condition steps, which may potentially lead to loops, and the number of leaves are always lower than half of the total number of use case steps.
In our evaluation, the number of use case steps in the \UCTM{s} for \BodySense and \HOD is 272 and 136, respectively. 
The number of condition steps leading to backward edges in these \UCTM{s} is 48 and 25.
The number of leaves in the \UCTM{s} is 16 for both case studies.
In our case studies, the number of condition steps leading to backward edges is, on average, 18\% of the total number of steps, while the number of leaves is 8\% of the total number of steps.

Based on our case studies, we may assume that $\left(N_B=\frac{1}{4}N\right)$ and $\left(N_L=\frac{1}{12}N\right)$. 
Because of their low value, the impact of $N_B$ and $N_L$ on the computation time for \emph{GenerateScenariosAndInputs} is minimal. 
When we ignore $N_B$ and $N_L$, 
for $T=1$ and $\mathit{MaxIt}=10$,
 we can compute the value of $C_{GSI}$ using 
Formulas~\ref{cost:C.GT},~\ref{cost:C.DT.IN},~and~\ref{cost:NV:red} as follows

\begin{align}
\label{cost:C.GSI:final}
\EquationsSize
\begin{split}
C_{GSI}  & = C_{GT} + \mathit{MaxIt} * \bigg( 2 * C_{DT} +  N_v * \left( 1 +  C_{US} \right) \bigg) + C_{MS} \\
& = \BigO \Bigg( N^2 + 10 * \bigg( 2 * \frac{N^3 * N_B*N_L  * ( T^2  + 3T + 2)}{4} \\
&= + \frac{(T^2  + 3T + 2) * N^2*{N_B}^{T+1} *N_L}{2}  \bigg) + N^3\Bigg)\\
& \approx \BigO\Bigg( N^2  + 10 * \bigg( 2 * \frac{\left(N^3   * 6  \right) }{2}  + \frac{6}{2} * N^2 \bigg) + N^3\Bigg)\\
& = \BigO\Bigg( N^2 + 10 *\bigg(  \left(6 * N^3 \right) + 3 * N^2 \bigg) + N^3\Bigg)\\
& \approx \BigO\Bigg( 60 * N^3 \Bigg) \\
\end{split}
\end{align}

Finally, the total number of nodes in the \UCTM{s} for testing the real systems in our case studies is not high. 
This is mostly due to the fact that the use case specifications are manually written and describe the high-level behavior of the systems. 
For example, the largest \UCTM in our case studies, which has been generated for \BodySense, includes 154 nodes.
With a relatively small number of nodes, \UCTM{s} can be processed fast by using modern computing systems (e.g., laptops), even in the presence of $\BigO(N^3)$ complexity.

\ifCLASSOPTIONcompsoc
  \section*{Acknowledgments}
\else
  \section*{Acknowledgment}
\fi

We gratefully acknowledge funding from FNR and IEE S.A. Luxembourg, the grant FNR/P10/03, the Canada Research Chair program, and the European Research Council (ERC) under the European Union’s Horizon 2020 research and innovation programme (grant agreement No 694277).

\ifCLASSOPTIONcaptionsoff
  \newpage
\fi

\bibliographystyle{IEEEtran}
\bibliography{references}

\end{document}